\documentclass[aps,physrev,twocolumn,superscriptaddress]{revtex4-2}
\usepackage{amsmath}
\usepackage{amssymb} 
\usepackage{graphicx}
\usepackage{hyperref}
\hypersetup{hidelinks}
\usepackage{mathtools}
\usepackage{float}
\usepackage{xurl}
\usepackage[anythingbreaks]{breakurl}
\usepackage[T1]{fontenc}
\usepackage[utf8]{inputenc}

\newcommand{\Teff}{T_{\mathrm{eff}}}

\begin{document}

\title{Energetics of stochastic limit-cycle oscillators: when does coupling reduce dissipation?}

\author{Anton F. Burnet}
\email[]{anton.burnet@lmu.de}

\author{Vansh Kharbanda}

\affiliation{Fakultät Physik, Technische Universität Dortmund, 44227 Dortmund, Germany}

\affiliation{
 Faculty of Physics and Center for NanoScience, Ludwig-Maximilians-Universit\"at M\"unchen, 80752 Munich, Germany
}\affiliation{
 Department of Veterinary Sciences, Ludwig-Maximilians-Universit\"at M\"unchen, 80752 Munich, Germany
}

\author{David Tobias}

\affiliation{
 Faculty of Physics and Center for NanoScience, Ludwig-Maximilians-Universit\"at M\"unchen, 80752 Munich, Germany
}

\author{Benedikt Sabass}

\email[]{benedikt.sabass@tu-dortmund.de}
\affiliation{Fakultät Physik, Technische Universität Dortmund, 44227 Dortmund, Germany}
\affiliation{
 Faculty of Physics and Center for NanoScience, Ludwig-Maximilians-Universit\"at M\"unchen, 80752 Munich, Germany
}\affiliation{
 Department of Veterinary Sciences, Ludwig-Maximilians-Universit\"at M\"unchen, 80752 Munich, Germany
}

\date{November 24, 2026}

\begin{abstract}
Non-linear oscillators occur in a wide range of biological and engineered systems. In noisy environments, the steady-state entropy production rate (EPR) quantifies time irreversibility and, under a thermodynamic interpretation, the associated energy dissipation. We study an idealized, analytically tractable family of stochastic circular limit-cycle oscillators and examine how mutual coupling in pairs and populations alters dissipation. The dynamics are characterized by a dimensionless effective temperature given by the ratio of radial relaxation and diffusion timescales. For identical oscillators, we derive analytical results for the EPR under radial, phase (Kuramoto-like), and Cartesian couplings. The phase-coupling analysis also considers nonidentical intrinsic frequencies. Coupling can increase or decrease EPR through changes in the radial mean and variance and in probability currents, with the response depending on effective temperature and system size. In particular, weak attractive Cartesian coupling lowers the EPR of identical oscillators at any effective temperature. At sufficiently low effective temperature, full Cartesian coupling produces a finite-coupling minimum below the uncoupled EPR throughout the oscillator family considered. Reduced EPR can also accompany improved individual and collective temporal coherence.
\end{abstract}

\maketitle

\section{Introduction}

\begin{figure}
    \centering
    \includegraphics[bb=0 0 1361.00000000 1161.00000000,width=1.\linewidth]{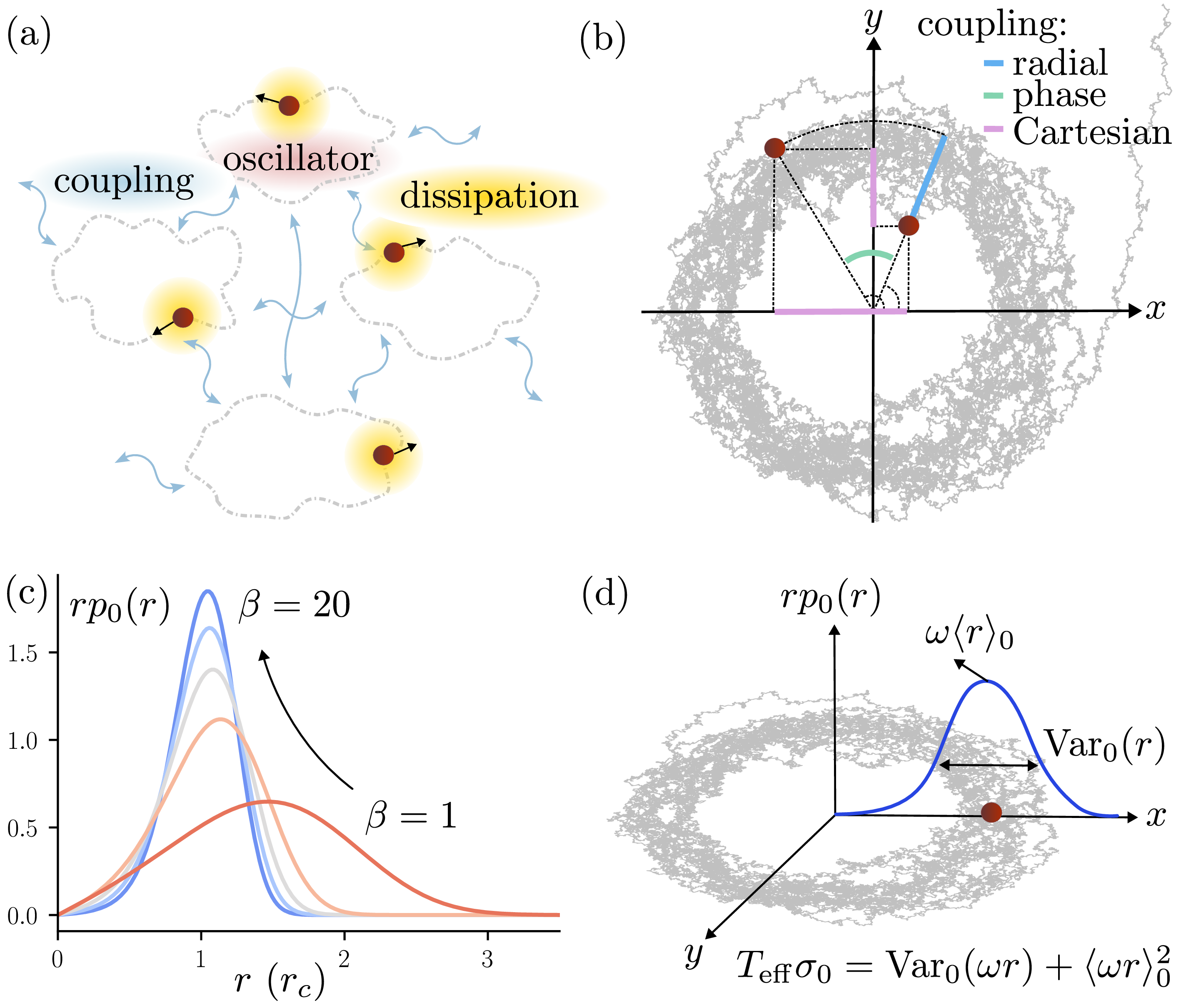}
    \caption{(a) A stochastic oscillator (red particle) fluctuates around a deterministic periodic orbit in phase space (gray dashed line). The self-sustained oscillations are inherently out of equilibrium and, when the dynamics admit a thermodynamic interpretation, require continuous energy dissipation (yellow halo). For a system of stochastic oscillators, there can be an arbitrary coupling among them (blue arrows). (b) A sample trajectory of the stochastic oscillator described by Eq.~(\ref{eq:1}) with $m=n=1$, shown in gray. In this work, two particles following such trajectories can couple via their radial, angular, and Cartesian coordinates, illustrated by the blue, green, and pink colors, respectively. (c) Radial probability distribution $rp_0$ of a single oscillator (Eq.~({\ref{eq:general-radial-density}}) {with $m=n=1$}) for various $\beta$. (d) {The radial distribution is shown over a sample trajectory. The squared mean radius and the radial variance contribute to the steady-state EPR according to Eq.~(\ref{eq:epr single}).} Sample trajectory in (b) and (d) simulated with $\beta=50$, $\omega=1\tau_r^{-1}$, and simulation time $100\tau_r$.}
    \label{fig:1}
\end{figure}

Stochastic oscillations play a crucial role in biological systems, including hair bundles~\cite{martin2021mechanical}, biochemical reaction networks~\cite{novak2008design}, {and macroscopic brain activity~\cite{deco2017dynamics}, and also arise in engineered systems such as driven colloidal rotors~\cite{kotar2013optimal}}. In many cases, the underlying deterministic dynamics admit a stable limit cycle, which is an isolated periodic orbit in phase space to which trajectories are attracted over time. Stochastic fluctuations, resulting, for instance, from thermal noise, molecular activity, or environmental noise, continually perturb the periodic orbit. The resulting noisy oscillations are shaped by the competition between deterministic attraction back to the limit cycle and stochastic forcing that displaces the state.
\\
\indent
When the dynamics admit a thermodynamic interpretation, maintaining a circulating probability current around a limit cycle requires continuous energy dissipation, quantified by the steady-state entropy production rate (EPR)~\cite{seifert2012stochastic}.  Under this interpretation, the EPR provides a natural measure when considering the ongoing energetic cost of sustaining coherent oscillations, and is thus a central aspect of the energy-accuracy or energy-coherence trade-offs that constrain biological oscillations~\cite{cao2015free, shiraishi2023entropy, santolin2025dissipation, kolchinsky2025comment}. In the auditory system, for example, spontaneous hair-bundle oscillations that underlie amplification and sharpen frequency discrimination have been used to infer non-zero EPR in bullfrog hair cells~\cite{ghosal2022inferring,roldan2021quantifying}. {In macroscopic stochastic models with nonthermal noise, such as those describing whole-brain activity, EPR quantifies time irreversibility, and its relation to physical dissipation requires an explicit thermodynamic description~\cite{NARTALLOKALUARACHCHI20261,korbel2026quo}.}
\\
\indent
In typical settings, oscillators occur in ensembles, making the collective dynamics of interacting oscillators a central theme of interest. The Kuramoto model and its variants have served as paradigmatic descriptions of weakly coupled deterministic phase oscillators, providing insights into synchronization and entrainment~\cite{kuramoto2003chemical,strogatz2000kuramoto,nakao2016phase, hong2007entrainment}. These ideas have naturally been explored in the context of stochastic oscillators~\cite{nandi2007effective, callenbach2002oscillatory, gupta2014kuramoto,buendia2025mesoscopic, perez2023universal, kreider2025q}. Several studies have examined how synchronization affects the EPR of coupled phase oscillators~\cite{sasa2015collective, imparato2015stochastic, izumida2016energetics, lee2018thermodynamic}.
For example, Ref.~\cite{izumida2016energetics} showed that the odd part of the coupling function contributes to reduced dissipation upon frequency synchronization, while the contribution from the even part depends on the system parameters. Synchronization-dependent work output and its reliability have also been investigated in an inertial-like Stuart-Landau dimer with Cartesian coupling~\cite{ryu2021stochastic}. Synchronization can also reduce the EPR in driven Potts models~\cite{meibohm2024small, meibohm2024minimum}.
\\
\indent
Retaining radial fluctuations allows coupling to affect dissipation through both amplitude statistics and radius-dependent angular diffusion. In this work, we compare radial, phase (Kuramoto-like), and Cartesian interactions within a common overdamped circular-oscillator model, resolving how radial redistribution and phase probability currents determine their distinct EPR responses, see Fig.~\ref{fig:1}. We derive analytical expressions, validate them numerically, and examine how these mechanisms depend on the radial relaxation and diffusion timescales and the number of coupled oscillators. We also compare EPR with individual and collective temporal coherence to assess how the energetic changes accompany changes in oscillatory performance.
\\
Section~\ref{sec: 2} introduces the oscillator model and EPR framework. Sections~\ref{sec: 3}-\ref{sec: 5} examine radial, phase, and Cartesian coupling, respectively. {Section~\ref{sec:coherence} compares dissipation and oscillatory coherence, followed by the Summary and Discussion in} Sec.~\ref{sec: 6} and Methods in Sec.~\ref{sec:7}.

\section{Background}\label{sec: 2}
\subsection{Single oscillator model}
We consider a family of stable deterministic circular limit cycles of radius $r_c$, with radial dynamics $\dot{r}{=C}(r_c^m-r^m)r^n$, {where $C>0$,} $m>0$, and $n\geq1$. These dynamics have a stable fixed point at $r=r_c$ and a repelling fixed point at $r=0$. For instance, the isochronous Stuart-Landau oscillator corresponds to $m=2$ and $n=1$ {(Appendix~\ref{app: SL}).} With additive noise in the Cartesian plane, the overdamped Langevin equations in polar coordinates read
\begin{equation}\label{eq:1}
   \begin{split}
        \dot{r}&=C(r_c^m-r^m)r^n {+\frac{D}{r}} + \sqrt{2D}\eta_r,\\
         \dot{\theta}&=-\omega  + \frac{\sqrt{2D}}{r}\eta_\theta,\\
    \end{split}
\end{equation}
{where we use the It\^o convention for the stochastic equations. The coordinate noises are independent Gaussian white noises with} $\langle\eta_i(t)\rangle=0$ and $\langle\eta_i(t)\eta_j(t')\rangle={\delta_{ij}}\delta(t-t')$. {The geometric drift $D/r$ follows from transforming additive Cartesian noise to polar coordinates.} $D$ is the diffusion constant and the damping coefficient is set to unity. The radial relaxation timescale is $\tau_r:=(mCr_c^{m+n-1})^{-1}$. Without noise, the motion on the limit cycle is $(x(t),y(t))=(r_c\cos\omega t,-r_c\sin\omega t)$ up to an arbitrary phase, driven by the intrinsic angular frequency $\omega$. Figure~\ref{fig:1}(b) shows a stochastic trajectory for $m=n=1$.

Scaling $t\to t/\tau_r$ and $r\to r/r_c${, and defining $F(r):=(1-r^m)r^n/m$, gives}
\begin{subequations}\label{eq:3}
   \begin{align}
        \dot{r}&={F(r)}  {+\frac{T_\mathrm{eff}}{r}} + \sqrt{2T_\mathrm{eff}}\eta_r,\\
         \dot{\theta}&=-\omega  + \frac{\sqrt{2T_\mathrm{eff}}}{r}\eta_\theta,\label{eq:3b}
    \end{align}
\end{subequations}
where $\tau_r$ has been absorbed into $\omega$ and we have defined the dimensionless effective temperature
$T_{\mathrm{eff}}:={D\tau_r/r_c^2}={D/(mCr_c^{m+n+1})}=\tau_r/\tau_D$,
with $\tau_D:=r_c^2/D$, which characterizes the ratio of the radial relaxation and angular diffusion timescales. Throughout, we will denote the inverse effective temperature as $\beta:=1/T_{\mathrm{eff}}$.

Rotational symmetry gives a stationary distribution $p(r,\theta)=p_0(r)/(2\pi)$ with vanishing radial current and
\begin{equation}\label{eq:general-radial-density}
    p_0(r) = \frac{1}{{\mathcal{I}_0}}
    \exp\!\left[\frac{\beta}{m}\left(\frac{r^{n+1}}{n+1} - \frac{r^{m+n+1}}{m+n+1}\right)\right],
\end{equation}
{where the normalization and radial moments are determined by
\begin{equation}\label{eq:general-radial-moments}
\mathcal{I}_q(\beta):=\int_0^\infty r^{q+1}
\exp\!\left[\beta\int_0^r F(s)\,\mathrm{d}s\right]\mathrm{d}r,
\quad q\geq -1.
\end{equation}}
{$p(r,\theta)$ is a density with respect to the area measure $r\,\mathrm{d}r\,\mathrm{d}\theta$, and $\int_0^\infty p_0(r)r\,\mathrm{d}r=1$. Thus, the radial probability density with respect to $\mathrm{d}r$ is $rp_0(r)$.}

{For the quantitative comparisons, we choose $m=n=1$, giving $F(r)=r(1-r)$. Keeping the intrinsic radial dynamics fixed provides a common basis for comparing the coupling mechanisms. Unless stated otherwise, the figures and numerical thresholds below use this choice. We retain $F$ and its stationary density $p_0$ in the general analytical formulation.} For $m=n=1$, $\mathcal{I}_q$ has the closed-form expression given in Appendix~\ref{app:moments}. Figure~\ref{fig:1}(c) displays $rp_0$ for various $\beta$. Unless stated otherwise, all probability densities in the following are time independent corresponding to the steady-state distributions. The moments of $p_0$ read $\langle r^q\rangle_{0,\beta}:=\int_0^\infty \mathrm{d}r \, r \,p_0(r) r^q=\mathcal{I}_q(\beta)/\mathcal{I}_0(\beta)$. Throughout this work, unless necessary to be explicit, we relax the notation $\langle r^q\rangle_{0,\beta}=\langle r^q\rangle_{0}$.

\subsection{Entropy production rate}
For the dynamics considered here, the steady-state EPR is expressed in terms of the steady-state probability density $p(\mathbf{x})$ and probability current $\mathbf{J}(\mathbf{x})$, satisfying ${\nabla}\cdot\mathbf{J}(\mathbf{x})=0$, as~\cite{seifert2012stochastic}
\begin{equation}\label{eq: epr}
    \sigma = \int \frac{\mathbf{J}D^{-1}\mathbf{J}^\top}{p}\mathrm{d}\mathbf{x},
\end{equation}
where $D$ is the diffusion matrix. With the local mean velocity $\mathbf{v}=\mathbf{J}/p$, this becomes $\sigma=\langle\mathbf{v}^{\top}D^{-1}\mathbf{v}\rangle$~\cite{sekizawa2024decomposing}.
From Eq.~(\ref{eq:3}) and {Eq.~(\ref{eq:general-radial-density})}, the radial current vanishes $J_r=0$ and the angular current is $J_\theta = -\omega p_0 \cdot (2\pi)^{-1}$ with $D_{\theta\theta}=T_\mathrm{eff}/r^2$,
so that the entropy production rate for a single oscillator is given by
\begin{equation}\label{eq:6}
 \sigma_0=\frac{\omega^2}{T_\mathrm{eff}}\langle r^2\rangle_0,
\end{equation}
{where $\langle\cdot\rangle_0$ is the average with respect to $rp_0$. The local mean velocity is purely tangential, with component $v_{\mathrm{tan}}=rJ_\theta/p=-\omega r$. Hence $T_\mathrm{eff}\sigma_0=\langle v_{\mathrm{tan}}^2\rangle_0$, which separates into the variance and squared mean of $v_{\mathrm{tan}}$,}
\begin{equation}\label{eq:epr single}
\begin{split}
T_\mathrm{eff}\sigma_0
&={\mathrm{Var}_0(v_{\mathrm{tan}})
  +\langle v_{\mathrm{tan}}\rangle_0^2}\\
&={\omega^2\bigl[\mathrm{Var}_0(r)+\langle r\rangle_0^2\bigr]}.
\end{split}
\end{equation}
{At fixed $\omega$, the two contributions are determined by the radial variance and squared mean radius, respectively (Fig.~\ref{fig:1}(c, d)).} 

As $T_\mathrm{eff}\to0$, $\mathrm{Var}_0(r)\to0$ and $\langle r\rangle_0\to1$, {so $T_\mathrm{eff}\sigma_0\to\omega^2$, recovering the fixed-radius EPR $\omega^2/T_\mathrm{eff}$} (Appendix~\ref{app:kuramoto}).

\subsection{\texorpdfstring{Interactions and EPR}{Interactions and stationary EPR}}\label{sec: conservative interaction}
{We now consider the EPR for oscillators interacting through a coupling force
$\boldsymbol f_{\mathrm c}$. The intrinsic radial force is conservative, and the
intrinsic rotational drift $(\boldsymbol f_{\mathrm{rot}})_i=(\omega_i y_i,-\omega_i x_i)$
is divergence-free.}

{Let $\langle\cdot\rangle_k$ denote an average in the coupled stationary
state. Under the stationary boundary conditions, integration by parts against
the divergence-free current gives $\langle\mathbf v\cdot\nabla\chi\rangle_k=0$
for scalar functions $\chi$. With $D=T_{\mathrm{eff}}I$, the local mean
velocity is $\mathbf v=\mathbf A-T_{\mathrm{eff}}\nabla\ln p$, where
$\mathbf A$ is the total Cartesian drift. Equation~(\ref{eq: epr}) then gives
(Appendix~\ref{app:conservative-epr})}
\begin{equation}\label{eq:stationary-epr-general}
\begin{aligned}
{\sigma(k)}
&{=\beta\langle(\boldsymbol f_{\mathrm{rot}}+\boldsymbol f_{\mathrm c})\cdot\mathbf v\rangle_k}\\
&{=\beta\langle(\boldsymbol f_{\mathrm{rot}}+\boldsymbol f_{\mathrm c})\cdot\mathbf A\rangle_k
+\langle\nabla\cdot\boldsymbol f_{\mathrm c}\rangle_k.}
\end{aligned}
\end{equation}

{For conservative coupling, $\boldsymbol f_{\mathrm c}=-\nabla V$, so stationarity
gives $\langle\boldsymbol f_{\mathrm c}\cdot\boldsymbol v\rangle_k=0$.
Only the rotational term remains in the first line of Eq.~(\ref{eq:stationary-epr-general}).
Since the intrinsic radial force is perpendicular to the rotational drift, the stationary EPR reduces to}
\begin{equation}\label{eq:stationary-conservative-epr}
\begin{aligned}
{\sigma(k)}
&{=\beta\sum_{i=1}^N\omega_i^2\langle r_i^2\rangle_k-\mathcal R(k),}\\
{\mathcal R(k)}
&{=-\beta\langle\boldsymbol f_{\mathrm{rot}}\cdot\boldsymbol f_{\mathrm c}\rangle_k\geq0.}
\end{aligned}
\end{equation}
{The first term is set by the coupled radial second moments. As in Eq.~(\ref{eq:epr single}), each second moment is the sum of the radial variance and the squared mean radius, which is a useful decomposition when coupling-induced changes in these two quantities compete. This radial contribution bounds the EPR
from above~\cite{dechant2023upper}, since
$\mathcal R=\beta\langle|\boldsymbol f_{\mathrm{rot}}-\boldsymbol v|^2\rangle_k\geq0$
(Appendix~\ref{app:conservative-epr}). A positive $\mathcal R$ means that the coupling force
opposes rotation on average, as we find for single-coordinate coupling in
Sec.~\ref{sec:single-coordinate}. With fluctuating radii, the sinusoidal phase coupling of
Sec.~\ref{sec: 4} is nonconservative in Cartesian coordinates, so the general identity
in Eq.~(\ref{eq:stationary-epr-general}) applies (Appendix~\ref{app: EPR phase}).}

{If $V$ is unchanged when each oscillator rotates at its own intrinsic
angular velocity, then $\boldsymbol f_{\mathrm{rot}}\cdot\nabla V=0$ and
$\mathcal R=0$. Equivalently,} $\mathbf{J}_0\cdot\nabla V=0$, where
$\mathbf{J}_0$ is the uncoupled stationary current. {This holds for radial
coupling at arbitrary intrinsic frequencies. For identical frequencies, the
oscillators rotate through the same angle and preserve their pairwise
distances, so it also holds for full Cartesian coupling. In these cases, the
stationary density has the Gibbs-reweighted form}
\begin{equation}\label{eq: dist gen}
    p(\{r_j\},\{\theta_j\}) = \frac{1}{\mathcal{Z}}\left(\prod\limits_{i=1}^N \frac{p_0^i}{{2\pi}} \right) e^{-\frac{1}{T_\mathrm{eff}}V(\{r_j\},\{\theta_j\})},
\end{equation}
where $p_0^i:=p_0(r_i)$ and $\mathcal{Z}=\langle e^{-{\beta} V}\rangle_0$.
{$\langle\cdot\rangle_0$ is the normalized uncoupled average with measure $\prod_i[r_i p_0(r_i)\,\mathrm{d}r_i\,\mathrm{d}\theta_i/(2\pi)]$. For phase-independent integrands, the angular integrals equal unity and this reduces to the radial average defined above.} Coupled averages are given by
\begin{equation}\label{eq: expect k}
 \langle f\rangle_k={\mathcal{Z}^{-1}}\langle f e^{-{\beta} V}\rangle_0.
\end{equation}
{For a $k$-independent observable $f$, at fixed $\beta$,}
\begin{equation}\label{eq: deriv k}
      \frac{\mathrm{d}}{\mathrm{d}k}    \left\langle f \right \rangle_k = -\beta\, \mathrm{Cov}_k(f, \partial_k V(\{r_j\},\{\theta_j\})),
\end{equation}
where $\mathrm{Cov}_k(f,g)=\langle fg\rangle_k-\langle f\rangle_k\langle g\rangle_k$ denotes the covariance. 

{For rotationally symmetric conservative coupling, Eq.~(\ref{eq: deriv k}) gives sufficient conditions for the EPR to
decrease with coupling} (Appendix~\ref{app: gibbs dist}). {For radial
and full Cartesian coupling, the covariance sum that determines the EPR
gradient can change sign as $k$ increases.}

In the following, we compare the analytical results with EPR estimates from Cartesian Langevin trajectories using the Stratonovich drift integral~\cite{sekimoto1998langevin}, as described in Sec.~\ref{app: sim details}.

\section{Radial coupling}\label{sec: 3}
{Radial coupling acts on the oscillator amplitudes and leaves the angular drift unchanged.
We consider $N$ oscillators with the Hookean interaction $V=\tfrac{k}{2}\sum_{i<j}c_{ij}(r_i-r_j)^2$,
where $c_{ij}=c_{ji}\geq0$, $c_{ii}=0$, and $kc_{ij}$ is the strength of the link between oscillators $i$ and $j$.
For $k>0$, each link penalizes differences between the two radii.}

{This reciprocal interaction permits an explicit stationary distribution with vanishing radial currents, providing a minimal setting for isolating how coupling modifies the EPR through radial statistics. Nonreciprocal coupling can additionally sustain radial probability currents and constitutes a distinct extension of the present model.}
{The Gibbs construction of Sec.~\ref{sec: conservative interaction} applies to any conservative radial interaction.
For identical intrinsic frequencies $\omega_i=\omega$, the EPR is determined by the sum of the stationary radial second moments.}

The radial dynamics obey
\begin{equation}\label{eq:7}
    \dot{r}_i={F(r_i)+\frac{T_{\mathrm{eff}}}{r_i}
    +k\sum_{j=1}^N c_{ij}(r_j-r_i)}
    +\sqrt{2T_{\mathrm{eff}}}\,\eta_{r_i}.
\end{equation}
{The angular dynamics are given by Eq.~(\ref{eq:3b}) for each oscillator.}

{We first consider a coupled pair to examine how changes in the radial distribution, particularly its mean and variance, affect the EPR.
We then ask how the EPR response depends on population size and coupling weights.}

\subsection{Two coupled systems}
{For a pair with unit coupling weight $c_{12}=1$, Eq.~(\ref{eq: dist gen}) gives the stationary density with $V=\tfrac{k}{2}(r_1-r_2)^2$.}

\subsubsection{Entropy production rate}

{The only nonvanishing currents are the angular currents $J_{\theta_i}=-\omega p$, so the pair EPR is}
\begin{equation}\label{eq:12}
    \sigma=\frac{\omega^2}{T_\mathrm{eff}}
    \sum_{i=1}^2\langle r_i^2\rangle_k.
\end{equation}
Coupling therefore changes the EPR through the radial second moment, $\langle r_i^2\rangle_k=\mathrm{Var}_k(r_i)+\langle r_i\rangle_k^2$. The expansion in moments of the free system is given in Appendix~\ref{app:radial-expansion}.

In Fig.~\ref{fig:2}(a), we show the EPR versus coupling strength displaying three distinct behaviors for different $\beta$: a monotonic decrease, an increase with a small local minimum, and a monotonic increase. 

\begin{figure}[!tp]
    \centering    \includegraphics[bb=0 0 447.36089472 768.48153696,width=1.\linewidth]{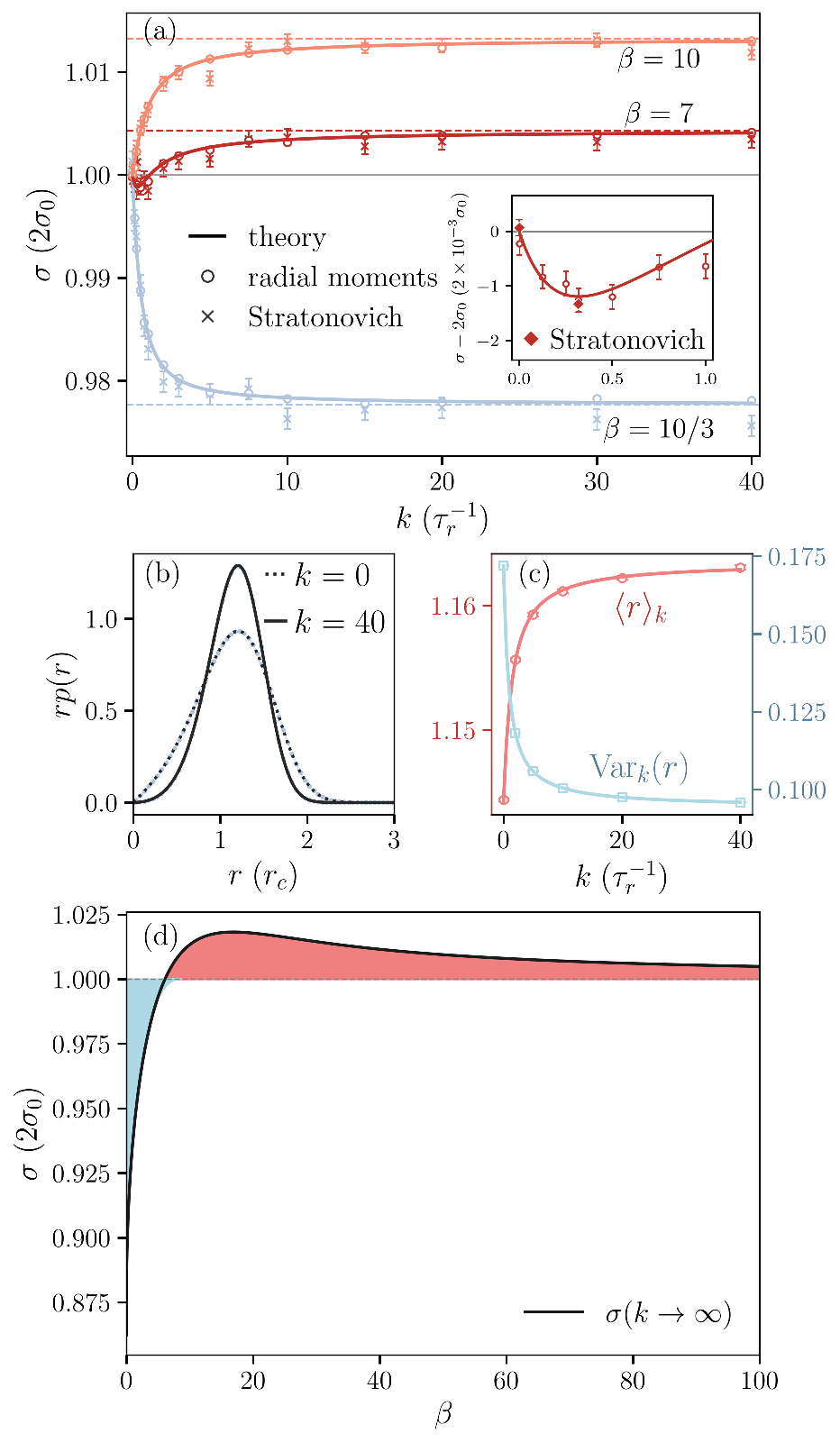}
    \caption{(a) The steady-state entropy production rate (EPR) $\sigma$ versus coupling strength $k$ of radial coupling for inverse effective temperatures $\beta\in\{{10/3},7,10\}$. The EPR is plotted in units of the EPR of the free system $2\sigma_0$ (Eq.~(\ref{eq:6})). The solid lines show Eq.~(\ref{eq:12}) evaluated using the stationary distribution. Open circles show the same expression evaluated using radial moments measured from simulations of {Eq.~(\ref{eq:7}) for $N=2$ and $c_{12}=1$}{; crosses show Stratonovich estimates (Eq.~(\ref{eq:numerical-epr-integral})) from the same trajectories. Horizontal dashed lines show the strong-coupling limits. Inset: enlargement of the $\beta=7$ minimum; diamonds show Stratonovich estimates from longer trajectories.} (b) The marginal radial distribution for coupling strengths $k\in\{0,40\}\tau_r^{-1}$ for $\beta={10/3}$. Lines show the stationary marginal densities; dots and shaded bands show simulation estimates and one standard error. (c) The average radius and radial variance versus coupling strength of the marginal distribution in (b). The distribution variation is qualitatively the same for $\beta\in\{{10/3},7,10\}$. (d) The limiting EPR $\sigma(k\to\infty)$ (solid black line) as a function of inverse effective temperature $\beta$ in units of the free system EPR. The blue and red regions represent the attainable fraction of the free system EPR under finite coupling below and above unity, respectively. Error bars show one standard error.}
    \label{fig:2}
\end{figure}

\subsubsection{EPR change with coupling strength}

{Using Eq.~(\ref{eq: deriv k}), we obtain the initial response to coupling (Appendix~\ref{app:EPR grad zero rad}):}
\begin{equation}\label{eq:18}
    \begin{split}
        \frac{\mathrm{d}\sigma}{\mathrm{d}k}\bigg|_{k=0}
        &=-\omega^2\beta^2\bigl[\mathrm{Var}_0((r-\langle r\rangle_0)^2)\\
        &\qquad\qquad +2\langle r\rangle_0{\mu_3}\bigr],
    \end{split}
\end{equation}
{where $\mu_3:=\langle(r-\langle r\rangle_0)^3\rangle_0$. At the onset of coupling, the radial variance decreases, whereas the mean radius can increase or decrease according to the skewness of the free radial distribution. For the evaluated $m=n=1$ model,} $\mu_3>0$ for $\beta\lesssim1.8$ and $\mu_3<0$ for $\beta>1.8$. Above this threshold, suppression of radial fluctuations competes with an increasing mean radius, allowing either an increase or a decrease in EPR.

With the relative radius $\delta:=r_1-r_2$ and the center-of-mass radius $\bar{r}:=(r_1+r_2)/2$, {Eq.~(\ref{eq:12}) becomes}
\begin{equation}\label{eq: epr bar r del}
     \sigma= \frac{2\omega^2}{T_\mathrm{eff}}\left\langle \bar{r}^2 \right \rangle_k +  \frac{\omega^2}{2T_\mathrm{eff}}\left\langle \delta^2 \right \rangle_k. 
\end{equation}
Using the identity in Eq.~({\ref{eq: deriv k}}), the rate of change of the EPR with respect to the coupling strength reads
\begin{equation}\label{eq:17}
    \frac{\mathrm{d}\sigma}{\mathrm{d}k}
    =-\omega^2\beta^2\mathrm{Cov}_k(\bar{r}^2,\delta^2)
    -\frac{\omega^2\beta^2}{4}\mathrm{Var}_k(\delta^2).
\end{equation}
The relative-radius contribution decreases monotonically as the radii align. The center-of-mass contribution can increase when $\mathrm{Cov}_k(\bar{r}^2,\delta^2)<0$, competing with this reduction. For the cases shown in Fig.~\ref{fig:2}(a), the negative covariance accompanies an increasing mean radius (Appendix~\ref{app: dist radial}). Figures~\ref{fig:2}(b, c) illustrate the associated changes in the radial distribution.

{For weak repulsive coupling, reversing the sign of $k$ reverses the nonzero linear EPR correction in Eq.~(\ref{eq:18}). In Appendix~\ref{app:repulsive-coupling}, Fig.~\ref{fig:repulsive-coupling}(a) compares finite attractive and repulsive couplings.}

\subsubsection{Different regimes}

To gain an overview of how $\beta$ affects the system response to coupling, we can study the strong-coupling limit EPR in Eq.~(\ref{eq: epr bar r del}). In the limit of strong coupling, the radii align and so the relative-radius contribution vanishes (Appendix~\ref{app: radial-strong}) giving
\begin{equation}\label{eq:16}
\begin{split}
    \sigma(k\to\infty)&= \lim_{k\to\infty}\frac{2\omega^2}{T_\mathrm{eff}}\left\langle \bar{r}^2 \right \rangle_k  = \frac{2\omega^2}{T_\mathrm{eff}}\frac{\langle r^3  \rangle_{0,2\beta}}{\langle r  \rangle_{0,2\beta}}.
\end{split}
\end{equation}
Figure~\ref{fig:2}(d) compares the above strong-coupling value with the uncoupled value $2\sigma_0{(\beta)}$ {at the same $\beta$. A ratio} $\sigma(k\to\infty)/(2\sigma_0)$ {below (above) unity means that the strongly coupled system dissipates less (more) than the uncoupled system. For $m=n=1$ and} ${0<}\beta<\beta_1^*\approx6.117$, the ratio is below unity; above this threshold it is slightly greater than unity. As $\beta\to\infty$, the ratio approaches unity and radial coupling has no effect in the fixed-radius limit.

Furthermore, the rate of change of the EPR with respect to coupling strength in Eq.~(\ref{eq:18}) satisfies  $\mathrm{d}\sigma/\mathrm{d}k|_{k=0}\leq0$ for $\beta<\beta^*_2\approx8.617>\beta_1^*$, which implies the existence of a regime by which a local minimum emerges. Indeed, a local minimum emerges for $5\lesssim\beta< \beta_2^*$. This local minimum represents only a small deviation from unity (Fig.~\ref{fig:2}(d)).  An example curve is shown in Fig.~\ref{fig:2}(a) for $\beta=7$, with the minimum enlarged in the inset. For ${\beta\geq\beta_2^*}$, numerical evaluation gives a monotonic EPR increase with coupling. The strong-coupling ratio approaches its minimum in the {high-noise limit} $\beta\to0$,  $\sigma(k\to\infty)/(2\sigma_0)\to\frac{\Gamma[2/3]\Gamma[5/6]}{\sqrt{\pi}}\approx 0.86$ {(Eq.~(\ref{app eq: general moment high noise}), Appendix~\ref{app:moments})}, where $\Gamma$ denotes the Gamma function.

\begin{figure}
    \centering    \includegraphics[bb=0 0 360.96072192 519.36103872,width=1.\linewidth]{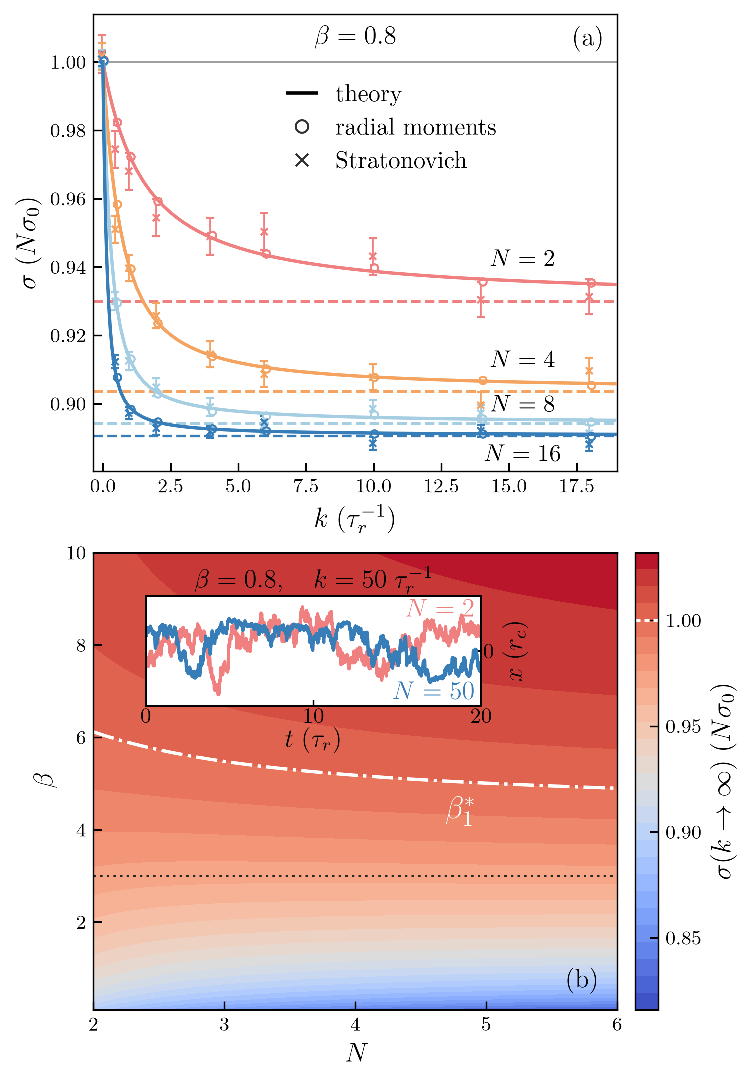}
    \caption{(a) Steady-state EPR $\sigma$ versus coupling strength $k$ of radial coupling for inverse effective temperature $\beta=0.8$ and system sizes $N\in\{2,4,8,16\}$. EPR is normalized by the uncoupled value $N\sigma_0$ (Eq.~(\ref{eq:6})). {Solid lines show the EPR $\beta\omega^2\sum_{i=1}^N\langle r_i^2\rangle_k$ evaluated using the stationary distribution. Open circles show the same expression evaluated using radial moments measured from} numerical simulations of {Eq.~(\ref{eq:7}) with unnormalized all-to-all coupling.} {Crosses show Stratonovich estimates (Eq.~(\ref{eq:numerical-epr-integral})) from the same trajectories}. The dashed lines denote the strong-coupling limit given by Eq.~(\ref{eq:24}). (b) A contour plot of the limiting EPR $\sigma(k\to\infty)$ as a function of inverse effective temperature $\beta$ and system size $N$ in units of the free system EPR. The white dotted-dashed line shows $\beta_1^*$ as a function of $N$. The black dotted line indicates $\beta\approx3$, where for $\beta$ above and below this line, increasing $N$ increases and decreases {$\sigma(k\to\infty)/(N\sigma_0)$}, respectively. {Inset: sample $x$-coordinate trajectories at} $k=50\,\tau_r^{-1}$, for $N\in\{2,50\}$, $\beta=0.8$, and $\omega=0.1\,\tau_r^{-1}$. Error bars show one standard error.}
    \label{fig:3 N}
\end{figure}

\subsection{N coupled systems}\label{sec: n coup rad}
{At fixed coupling weights, the initial EPR slope for $N$ oscillators is $W$ times the unit-weight pair slope in Eq.~(\ref{eq:18}),
where $W=\sum_{i<j}c_{ij}$ is the total edge weight (Appendix~\ref{app:EPR grad zero rad}).
Increasing $W$ therefore amplifies either an initial increase or an initial decrease in EPR.
For an individual oscillator $i$, the initial changes in its mean radius and radial variance are proportional to its weighted degree $s_i=\sum_j c_{ij}$, with $W=\tfrac12\sum_i s_i$.
For binary weights, $W$ counts the edges and $s_i$ counts the neighbors of oscillator $i$.}

{The unnormalized all-to-all population calculations represent finite clusters with fixed-strength pairwise interactions. $c_{ij}=1$ for $i\neq j$, $k$ is the per-edge strength, and $K=(N-1)k$ is the total incident coupling weight per oscillator. Increasing $N$ at fixed $k$ therefore also increases $K$, affecting the coupling scale of the approach to the strong-coupling regime. To compare system sizes at fixed total incident coupling strength $K$, one can instead use degree-normalized weights $c_{ij}=1/(N-1)$ for $i\ne j$. Each oscillator then has weighted degree $s_i=1$, the total edge weight is $W=N/2$, and the pairwise coupling strength is $K/(N-1)$.}

{The two conventions give identical dynamics under the substitution $k=K/(N-1)$. This rescales the coupling axis, preserving the existence and depth of EPR minima while changing their locations to $K_{\min}=(N-1)k_{\min}$. In Sec.~\ref{sec:coherence}, we hold $K$ fixed as $N$ varies to examine coherence without simultaneously increasing the total incident coupling weight. This equivalence between the two all-to-all conventions holds for every coupling mechanism considered here.}

Defining $\bar{r} = \frac{1}{N}\sum_{i=1}^N r_i$ and $\Delta^2:= \sum_{i=1}^N(r_i-\bar{r})^2$, the total EPR can be expressed in terms of the center-of-mass and relative mode components
\begin{equation}\label{eq:radial-N-decomposition}
    \sigma = \omega^2 \beta  [N\langle \bar{r}^2 \rangle_k + \langle \Delta^2 \rangle_k],
\end{equation}
which generalizes Eq.~(\ref{eq: epr bar r del}). {With $Q_r:=\sum_{i<j}c_{ij}(r_i-r_j)^2$,} the gradient thereof reads
\begin{equation}\label{eq:radial-N-gradient}
    \begin{split}
        \frac{\mathrm{d}}{\mathrm{d}k}\sigma
        &= -\frac{\omega^2\beta^2}{2}[N\mathrm{Cov}_k(\bar{r}^2,{Q_r})\\
        &\qquad\qquad + {\mathrm{Cov}_k(\Delta^2,Q_r)}].
    \end{split}
\end{equation}
{For unnormalized all-to-all coupling, $Q_r=N\Delta^2$, so the second covariance is $N\mathrm{Var}_k(\Delta^2)\geq0$. The relative-radius contribution therefore decreases with $k$. The total EPR can nevertheless rise if the collective contribution $N\beta\omega^2\langle\bar r^2\rangle_k$ grows enough to outweigh this decrease.}

{For fixed finite $N$ and fixed symmetric weights whose nonzero edges form} a connected graph, the strong-coupling limit {$k\to\infty$} is given by (see Appendix~\ref{app: radial-strong})
\begin{equation}\label{eq:24}
    \sigma({k\to\infty})  = \frac{N\omega^2}{T_\mathrm{eff}} \frac{\langle r^{N+1}\rangle_{0,N\beta}}{\langle r^{N-1}\rangle_{0,N\beta}}.
\end{equation}
This equation reveals a scaling of the effective temperature with system size, $N\beta$. {More generally, the strong-coupling EPR exceeds the uncoupled value at sufficiently low noise throughout the family $m>0$, $n\geq1$ (Eq.~(\ref{app eq: general moment low noise}), Appendix~\ref{app:moments}).} Figure~\ref{fig:3 N}(a) illustrates that an increase of the system size leads to a reduction of the normalized EPR for coupled noisy oscillators with $\beta=0.8$. However, an increase of the system size can also increase the limiting EPR when $\beta\gtrsim 3<\beta^*_1$, see Fig.~\ref{fig:3 N}(b). In particular, for ${4.40}<\beta\lesssim 6.117=\beta_1^*(N=2)$, increasing $N$ can change the limiting EPR from below to above the free system EPR value. The inset shows a sample trajectory of $N\in\{2,50\}$ strongly coupled oscillators, where the larger system visibly constrains the amplitude fluctuations more effectively. For the general family, the {high-noise limit} of Eq.~(\ref{eq:24}) relative to the free system EPR $\sigma({k=0})$ {follows from Eq.~(\ref{app eq: general moment high noise}) in Appendix~\ref{app:moments}:}
\begin{equation}
\begin{split}
     \frac{\sigma({k\to\infty})}{\sigma({k=0})}(\beta\to0)
     &=N^{-2/{(m+n+1)}}\\
     &\quad\times\frac{\Gamma[\frac{2}{{m+n+1}}]\Gamma[\frac{N+3}{{m+n+1}}]}
     {\Gamma[\frac{4}{{m+n+1}}]\Gamma[\frac{N+1}{{m+n+1}}]}.
\end{split}
\end{equation}
{Taking $N\to\infty$ after the strong-coupling and high-noise limits gives}
\begin{equation}
\begin{split}
     &\lim_{N\to\infty}\frac{\sigma({k\to\infty})}{\sigma({k=0})}(\beta\to0)\\
     &\qquad={(m+n+1)^{-2/(m+n+1)}}
     \frac{\Gamma[\frac{2}{{m+n+1}}]}{\Gamma[\frac{4}{{m+n+1}}]}.
\end{split}
\end{equation}
{For $m=n=1$, this limit is approximately} $73\%$ of the EPR of the uncoupled systems.

\subsection{Summary}

Radial coupling among oscillators can either increase or decrease the EPR, depending on effective temperature. The different regimes result from competition between reduction in radial fluctuations and increase in average radius under coupling, due to how the coupling interaction deforms the radial distributions.

\mbox{}
\section{Phase coupling}\label{sec: 4}
{Phase coupling acts on the angular dynamics and leaves the radial dynamics of Eq.~(\ref{eq:3}a) unchanged. For $N$ oscillators with intrinsic frequencies $\omega_i$, we consider the Kuramoto-like interaction}
\begin{equation}\label{eq:29}
    {\dot{\theta}_i=-\omega_i+k\sum_{j=1}^{N}c_{ij}\sin(\theta_j-\theta_i)
        +\frac{\sqrt{2T_{\mathrm{eff}}}}{r_i}\eta_{\theta_i},}
\end{equation}
{with symmetric weights $c_{ij}=c_{ji}\geq0$ and $c_{ii}=0$, as in Sec.~\ref{sec: 3}. In the low-temperature limit, the radii approach unity and the phase dynamics reduce to the Kuramoto model (Appendix~\ref{app:kuramoto}). With fluctuating radii, the sine coupling is nonconservative in Cartesian coordinates, so the Gibbs-reweighted stationary-density construction of Sec.~\ref{sec: conservative interaction} does not apply.}

We first consider a pair to examine how synchronization and radial fluctuations affect the EPR. We then use the pair result to determine how frequency differences across the links of a network affect its EPR.

\subsection{Two coupled systems}
{For a pair with $c_{12}=1$, let $\psi=(\theta_1+\theta_2)/2$ and $\varphi=\theta_2-\theta_1$ denote the center-of-mass and relative phases, with $\bar\omega=(\omega_1+\omega_2)/2$ and $\Delta\omega=\omega_2-\omega_1$. Invariance under a common phase shift makes $\psi$ uniform at stationarity. The two radii evolve independently, so their joint marginal remains $p_0(r_1)p_0(r_2)$. The stationary density therefore has the form}
\begin{equation}\label{eq:phase-pair-density}
    {p(r_1,r_2,\psi,\varphi)=\frac{p_0(r_1)p_0(r_2)}{2\pi}
        p_\varphi(\varphi\mid r_1,r_2).}
\end{equation}
Substitution into the stationary Fokker-Planck equation gives
\begin{equation}\label{eq:78}
    \begin{split}
        {0}&{=\mathcal L_r p_\varphi-\partial_\varphi j_\varphi,}\\
        {j_\varphi}&{=A_\varphi p_\varphi-D_{\varphi\varphi}\partial_\varphi p_\varphi,}
    \end{split}
\end{equation}
{where $A_\varphi=-\Delta\omega-2k\sin\varphi$ is the relative-phase drift and $D_{\varphi\varphi}=\Teff(r_1^{-2}+r_2^{-2})$ its diffusivity. The radial generator is $\mathcal L_r f=\sum_{i=1}^2\frac{\Teff}{r_i p_0(r_i)}\partial_{r_i}[r_i p_0(r_i)\partial_{r_i}f]$. We solve Eq.~(\ref{eq:78}) numerically in $r_1$, $r_2$, and $\varphi$.}

\subsubsection{\texorpdfstring{Entropy production rate}{Entropy production rate}}
{In the fixed-radius Kuramoto model, the EPR separates into independent center-of-mass and relative-phase contributions, with the latter vanishing at strong coupling (Appendix~\ref{app:kuramoto}). With fluctuating radii, however, the angular noise strengths depend on the radii, and the corresponding decomposition contains mixed contributions that obscure the distinct effects of coupling (Appendix~\ref{app: EPR phase}). Although the general stationary EPR identity of Sec.~\ref{sec: conservative interaction} provides an exact expression in terms of stationary observables, we seek a decomposition that more transparently distinguishes the different effects of coupling. To this end, we introduce the radially weighted mean phase velocities $\Omega_i=\langle r_i^2 v_{\theta_i}\rangle_k /\langle r^2 \rangle_0$, where $v_{\theta_i} = -\omega_i + k\sin(\theta_j-\theta_i) +(-1)^{i+1} \frac{\Teff}{r_i^2}\partial_\varphi\ln p_\varphi$. Note that, by construction, $\Omega_i(0)=-\omega_i$.} The total EPR is found to be~(Appendix~\ref{app: EPR phase})
\begin{equation}\label{eq: EPR mean-variance}
    \begin{split}
        \sigma &= \frac{2\bar{\omega}^2}{\Teff}\langle r^2 \rangle_0 + {\frac{\langle r^2 \rangle_0}{2\Teff}(\Omega_2-\Omega_1)^2 }\\
     &{+  \frac{1}{\Teff} \sum_{i=1}^2 \langle r_i^2 (v_{\theta_i}-\Omega_i)^2\rangle_k}\\
     &+ \sum_{i=1}^2 \Teff \left\langle \langle (\partial_{r_i} \ln p_\varphi)^2 \rangle_\varphi \right\rangle_0\\
     &=: \sigma_\psi +{ \sigma_{\Delta\Omega} + \sigma_{\mathrm{var}}} + \sum_{i=1}^2\sigma_{r_i\varphi},
    \end{split}
\end{equation}
where $\langle\ldots\rangle_\varphi$ denotes the expectation with respect to the distribution $p_\varphi$. {In this decomposition, $\langle\cdot\rangle_0$ denotes the normalized average over the two radii, after the conditional phase average has been taken.} {Since $(\Omega_1+\Omega_2)/2=-\bar\omega$,} $\sigma_\psi$ is the coupling-independent center-of-mass {contribution. The term $\sigma_{\Delta\Omega}$ characterizes the difference between the radially weighted mean phase velocities. It vanishes for identical intrinsic frequencies and tends to zero under strong coupling for detuned oscillators. At zero coupling, only these two terms remain.}

{The contribution $\sigma_{\mathrm{var}}$ captures coupling-induced fluctuations of the local phase velocities. For fixed radii and $\Delta\omega\neq0$, the initial increase in $\sigma_{\mathrm{var}}$ is outweighed by the decrease in $\sigma_{\Delta\Omega}$, giving a net EPR reduction, with both contributions vanishing as $k\to\infty$ (Appendix~\ref{app:kuramoto}). Radial fluctuations contribute to $\sigma_{\mathrm{var}}$ in Eq.~(\ref{eq: EPR mean-variance}) by continually changing the angular noise strengths $T_{\mathrm{eff}}/r_i^2$. Coupling transmits phase disturbances between oscillators with unequal noise strengths, generating fluctuations in their local phase velocities. The terms} $\sigma_{r_i\varphi}$ {are $T_\mathrm{eff}$ times radial averages of the} Fisher information {of $p_\varphi$ with respect to $r_i$, and we refer to them as} radial Fisher contributions (Appendix~\ref{app:FI}). {A radial displacement changes the angular noise strength without directly changing the relative phase. An outward excursion can therefore carry poorly aligned phases into conditions where weaker angular noise favors tighter locking. Radial motion continually mixes configurations with different phase statistics, and angular dynamics readjust their relative phases. Radial Fisher information measures how sensitively the conditional relative-phase distribution depends on radius. Its EPR contribution quantifies the radial-current dissipation associated with the continual readjustment of these statistics under radial fluctuations, despite zero net radial marginal current.}
\newline
{All terms in Eq.~(\ref{eq: EPR mean-variance}) are nonnegative, so $\sigma(k)\geq\sigma_\psi$. For $\Delta\omega=0$, $\sigma_\psi=2\sigma_0$, and the EPR is therefore at least as large as that of the uncoupled pair. However, when $\Delta \omega\neq 0$, $\sigma(0) = \sigma_\psi + (\Delta \omega )^2\langle r^2 \rangle_0/ (2 T_\mathrm{eff})$ and so it follows that
\begin{equation}
    \sigma(0)- \sigma(k) \leq \frac{(\Delta \omega )^2}{2 T_\mathrm{eff}}\langle r^2 \rangle_0.
\end{equation}
This places a bound for how much the EPR can decrease in the case $\Delta\omega\neq 0$. In the fixed-radius Kuramoto model, synchronization suppresses the relative-phase contribution, and the corresponding bound is saturated at strong coupling (Appendix~\ref{app:kuramoto}).}

{For the phase-coupled pair, $A_\varphi(\varphi+\pi;-k)=A_\varphi(\varphi;k)$. Since the radial dynamics and diffusion coefficients are independent of $\varphi$, reversing the coupling sign translates the stationary density and currents by $\pi$ in relative phase. The EPR integrand therefore has identical values at corresponding phase-shifted configurations, so integration over a full phase period gives $\sigma(-k)=\sigma(k)$ (Appendix~\ref{app:repulsive-coupling}, Fig.~\ref{fig:repulsive-coupling}(b)).}

\begin{figure}[!htp]
	\centering    \includegraphics[bb=0 0 518.40103680 1026.96205392,width=0.98\linewidth]{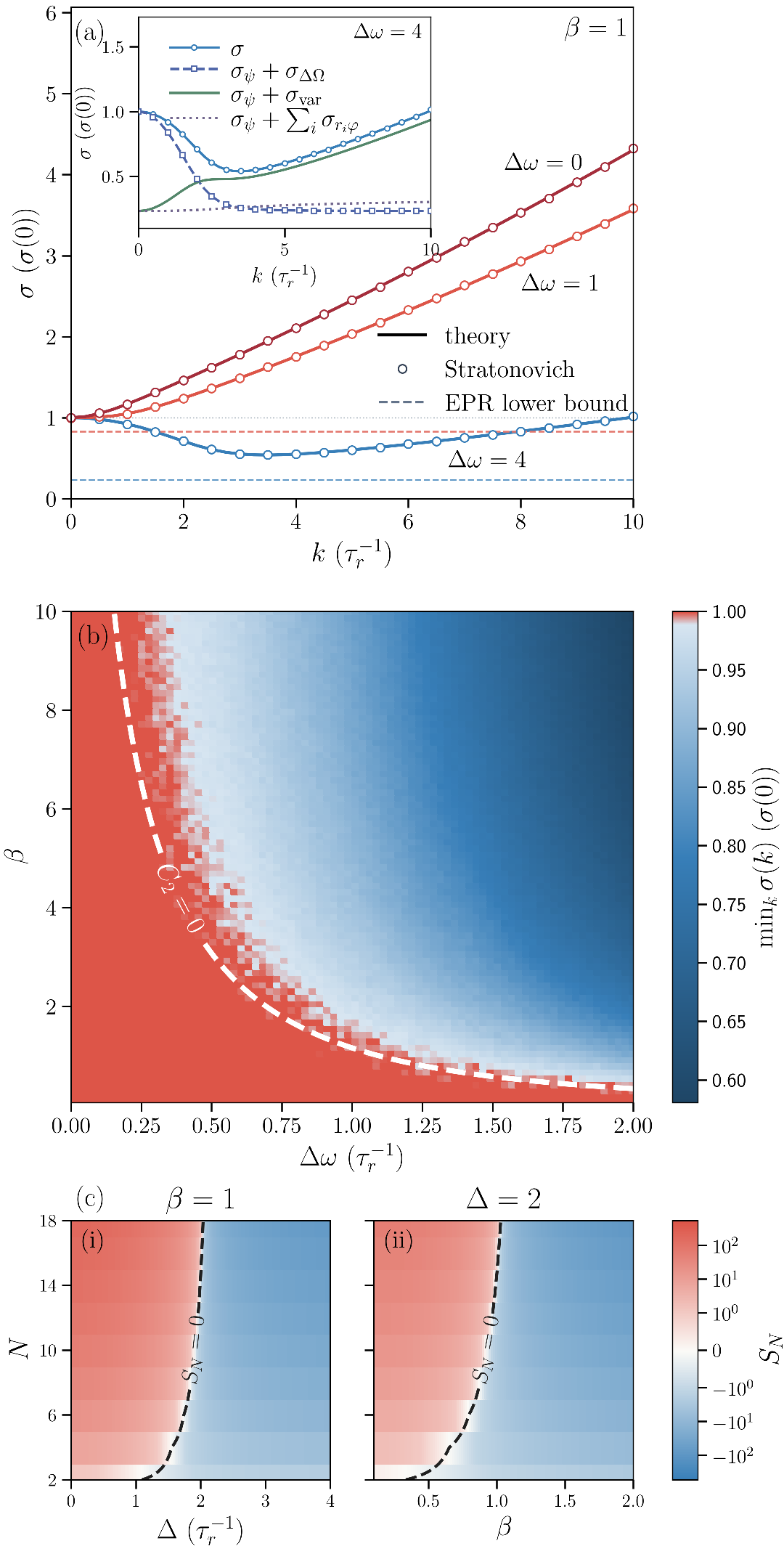}
	\caption{(a) Steady-state entropy production rate (EPR) $\sigma$ versus coupling strength $k$ of phase coupling for two oscillators {when $\beta=1$, $\Delta\omega\in\{0,1,4\}\,\tau_r^{-1}$, and $\bar\omega=1.1\tau_r^{-1}$. Open circles show Stratonovich estimates from simulations of Eqs.~(\ref{eq:3}a) and~(\ref{eq:29}) for $N=2$ and $c_{12}=1$. EPR is normalized by its uncoupled value $\sigma(0)$ at the same intrinsic frequencies and effective temperature. Solid lines show Eq.~(\ref{eq: EPR mean-variance}) evaluated with the numerical solution to Eq.~(\ref{eq:78}). Horizontal dashed lines show lower bounds $\sigma_\psi/\sigma(0)$. Inset ($\Delta\omega=4\tau_r^{-1}$): total EPR and its components, each component offset by $\sigma_\psi$. (b) Minimum normalized EPR over the simulated coupling grid versus $\beta$ and $\Delta\omega$. The white dashed curve $C_2=0$ (Eq.~(\ref{eq: C_2 pair})) separates positive and negative initial quadratic EPR responses. (c) The second order perturbation coefficient $S_N$ (Eq.~(\ref{eq: SN})) of a bi-frequency population with unnormalized all-to-all coupling, with fixed $\beta=1$ in (i) and fixed $\Delta=2\tau_r^{-1}$ in (ii). Error bars show one standard error.} }
	\label{fig:3}
\end{figure}

In Fig.~\ref{fig:3}(a), we show the EPR as a function of coupling strength (Eq.~(\ref{eq: EPR mean-variance})) together with numerical simulation results for {$\Delta\omega\in\{0,1,4\}\tau_r^{-1}$}. Unlike with the Kuramoto model, radial fluctuations result in the EPR increasing monotonically with moderate-to-strong coupling{. However, this increase, dominated by $\sigma_\mathrm{var}$, competes at weak coupling with the decrease in $\sigma_{\Delta\Omega}$, shown in Fig.~\ref{fig:3}(a) inset for the case $\Delta \omega = 4$. Evidently, $\Delta \omega\neq0$ does not guarantee a decrease in EPR under weak coupling. Indeed, a weak coupling expansion of $p_\varphi$ reveals
\begin{equation}
    \sigma(k)- \sigma(0) =   C_2 k^2 +  \mathcal{O}(k^4),
\end{equation}
where the quadratic coefficient is
\begin{equation}\label{eq: C_2 pair}
    C_2 = \underbrace{\frac{\langle r^2\rangle_0}{T_\mathrm{eff}} -  2 \langle  u\rangle_0}_{=:\mathcal F\geq0} - \underbrace{\frac{\Delta\omega}{T_\mathrm{eff}}\langle a \,
    w\rangle_0}_{=:\mathcal S\geq0},
\end{equation}
where $a=r_1^2+r_2^2$ and $u +\mathrm{i}w=h$ is the first Fourier response of
$p_\varphi$, determined by
$(D_{\varphi\varphi}-\mathcal L_r-\mathrm{i}\Delta\omega)h=1$ from Eq.~(\ref{eq:78}) (Appendix~\ref{app: phase weak coupling}). This coefficient characterizes whether the EPR decreases or increases around the onset of coupling. In terms of the decompositions in Eq.~(\ref{eq: EPR mean-variance}), 
$\sigma_\psi + \sigma_{\Delta\Omega} = \sigma(0)-\mathcal S k^2 + \mathcal{O}(k^4)$ and $\sigma_{\mathrm{var}} + \sum_{i=1}^2\sigma_{r_i\varphi}= \mathcal F k^2 + \mathcal{O}(k^4)$. Therefore, an initial decrease with coupling occurs when the relative mean velocity contribution decreases sufficiently quickly relative to the coupling-induced gain. In Fig.~\ref{fig:3}(b), we show the line $C_2=0$ with numerical simulation results characterizing the minimum intrinsic frequency difference $\Delta\omega_\mathrm{min}$ required for an EPR decrease upon the onset of coupling.}

{Under strong coupling, to leading order the EPR is given by~(see Appendix~\ref{app sec: strong coupling pair})
\begin{equation}\label{eq: strong coupling pair main }
\begin{split}
     \sigma(k) &=  \frac{k}{2}\left\langle \left(\frac{r_1}{r_2}-\frac{r_2}{r_1}\right)^2\langle \cos\varphi\rangle_\varphi \right\rangle_0 + o(k)\\
    &=\frac{\langle r^2\rangle_0 \, p_0(0)}{2} \, k\ln \frac{k}{T_\mathrm{eff}} + \mathcal{O}(k).
\end{split}
\end{equation}
The leading EPR growth thus arises from differences between the fluctuating radii. Phase coupling leaves these fluctuations unsuppressed, and the resulting unequal angular diffusivities $T_{\mathrm{eff}}/r_i^2$ sustain currents. This leading growth is carried by $\sigma_{\mathrm{var}}$, whereas the radial Fisher contribution is subleading.} Such an effect is analogous to the energy flow between two coupled particles each sitting in a different thermal bath~\cite{sekimoto1998langevin}.

{The same leading contribution takes a radially weighted phase-Fisher form, expressing the dissipative cost of maintaining a tightly confined relative phase under unequal angular noise. Stronger coupling narrows the relative-phase distribution and increases its sensitivity to phase shifts. The radial weighting accounts for the noise imbalance (Appendix~\ref{app sec: strong coupling pair}). The leading logarithmic term originates from the region $r_*\ll r_i\ll1$, with $r_*\sim\sqrt{T_{\mathrm{eff}}/k}$ setting an effective lower cutoff for the logarithmic contribution. In Appendix~\ref{app sec: strong coupling pair}, we show Eq.~(\ref{eq: strong coupling pair main }) against simulation results for strong coupling along with the subleading $\mathcal{O}(k)$ term.}

\protect\leavevmode{\subsection{N coupled systems}
For $N$ oscillators, the independent radial dynamics give a stationary density $P_0 p_{\{\theta_j\}}$, where $P_0=\prod_i p_0(r_i)$ and $p_{\{\theta_j\}}=p(\{\theta_j\}\mid\{r_j\})$ is the conditional phase density. It satisfies $0=\mathcal L_r p_{\{\theta_j\}}-\sum_{i=1}^N\partial_{\theta_i}j_{\theta_i}$, with $\mathcal L_r$ summed over the $N$ radii and $j_{\theta_i}=J_{\theta_i}/P_0$. As for the pair, the EPR separates into contributions from the radially weighted mean phase velocities, local fluctuations about those means, and the radial probability currents (Appendix~\ref{app: EPR phase N}),
\begin{equation}\label{eq: EPR mean-variance N}
\begin{split}
\sigma_N
&=\frac{N\langle r^2\rangle_0}{\Teff}\bar\Omega^2
+\frac{\langle r^2\rangle_0}{N\Teff}
\sum_{i<j}(\Omega_i-\Omega_j)^2\\
&\quad+\frac{1}{\Teff}\sum_{i=1}^N
\langle r_i^2(v_{\theta_i}-\Omega_i)^2\rangle_k\\
&\quad+\Teff\sum_{i=1}^N
\langle(\partial_{r_i}\ln p_{\{\theta_j\}})^2\rangle_k\\
&=: \sigma_{\bar\Omega} + \sigma_{\Delta\Omega,N} + \sigma_{\mathrm{var},N} + \sum_{i=1}^N\sigma_{r_i,\{\theta_j\}}.
\end{split}
\end{equation}
where $\bar\Omega=N^{-1}\sum_{i=1}^N\Omega_i$. The four terms correspond, in the same order, to those in the pair decomposition in Eq.~(\ref{eq: EPR mean-variance}).}

{Expanding the stationary EPR about $k=0$ gives a sum over links. Each link contributes the pair coefficient $C_2$ from Eq.~(\ref{eq: C_2 pair}), evaluated at the frequency difference of the oscillators it connects,
\begin{equation}\label{eq: SN}
    \begin{split}
        \sigma_N(k)-\sigma_N(0)&=k^2 S_N+\mathcal O(k^3),\\
        S_N&=\sum_{i<j}c_{ij}^2 C_2(\omega_j-\omega_i).
    \end{split}
\end{equation}
The EPR initially decreases if $S_N<0$. Its initial response can change sign as the relative numbers of links connecting equal and unequal frequencies change. As a simple illustration, consider an even population with $N/2$ oscillators at each intrinsic frequency $\bar\omega\pm\Delta/2$. For unnormalized all-to-all coupling, $c_{ij}=1$ for $i\neq j$, the sum is then given by
\begin{equation}
S_N=
\frac{N(N-2)}{4}\,C_2(0)
+\frac{N^2}{4}\,C_2(\Delta).
\end{equation}
For $N=2$, the only edge connects oscillators with
different frequencies. As $N$ increases, equal-frequency edges account for a growing fraction of the network, so their nonnegative contribution $C_2(0)$ competes with $C_2(\Delta)$ when the latter is negative. A change in the sign of $S_N$ therefore reflects a change in pair composition, not merely an increase in the number of interactions. 
\newline 
Figure~\ref{fig:3}(c) shows how $S_N$ varies with $N$ and $\Delta$ at fixed $\beta$ in (i), and with $N$ and $\beta$ at fixed $\Delta$ in (ii). For some of the parameters shown, $S_N$ changes from negative to positive as $N$ increases, so weak coupling lowers the EPR in smaller populations and raises it in larger ones.
\newline
For strong, unnormalized all-to-all coupling ($c_{ij}=1$ for $i\neq j$), the local phase-velocity variance in Eq.~(\ref{eq: EPR mean-variance N}) carries the leading EPR growth (Appendix~\ref{app sec: strong coupling N}),
\begin{equation}
\begin{split}
    \sigma_N 
    &= (N-1) \frac{ \langle r^2 \rangle_0}{2} p_0(0)k \ln \frac{k}{\Teff} +  \mathcal{O}(k)\\
    &= \sigma_{\mathrm{var},N} + \mathcal{O}(k).
\end{split}
\end{equation}\mbox{}}

\subsection{Summary}

{Phase coupling leaves the radial fluctuations
unsuppressed, while generating phase-velocity
fluctuations and radial-phase probability currents. For two oscillators with identical intrinsic frequencies, these contributions cannot reduce the EPR below its uncoupled value. For detuned oscillators, a reduction in the weighted mean-velocity difference competes with this additional dissipation. Small-radius excursions
produce a leading $k\ln(k/T_{\mathrm{eff}})$ growth
carried by the phase-velocity variance contribution. Consequently, for sufficiently different intrinsic frequencies, the initial decrease is followed by at least one finite-coupling minimum below the uncoupled EPR. For networks, the initial response is determined by the weighted sum of the corresponding pair coefficients.}

\section{Cartesian coupling}\label{sec: 5}
In Secs.~\ref{sec: 3} and \ref{sec: 4}, we considered the effect of coupling radial and phase coordinates of oscillators on the steady-state EPR, respectively. Finally in this section, we will consider coupling Cartesian coordinates, where the previously considered normal coordinates are coupled simultaneously.

\subsection{Full coordinate coupling}

{Full Cartesian coupling penalizes the distance between oscillators in the $(x,y)$ plane.
Writing $z_i=x_i+\mathrm{i}y_i$, where $\mathrm{i}=\sqrt{-1}$ is the imaginary unit, we consider the Hookean potential $V=\tfrac{k}{2}\sum_{i<j}c_{ij}|z_i-z_j|^2$.
As in Sec.~\ref{sec: 3}, the weights satisfy $c_{ij}=c_{ji}\geq0$ and $c_{ii}=0$, and $kc_{ij}$ is the strength of the link between oscillators $i$ and $j$.}

{To isolate changes in EPR caused by radial redistribution at fixed mean rotation frequency, we consider identical intrinsic frequencies $\omega_i=\omega$.
The intrinsic rotation then preserves every pairwise distance, so the Gibbs construction of Sec.~\ref{sec: conservative interaction} gives the stationary density.
The equations of motion are}
\begin{equation}\label{eq:41}
    \begin{split}
        \dot{x}_i &= {\frac{F(r_i)}{r_i}}x_i+{\omega}y_i
        +{k\sum_{j=1}^N c_{ij}(x_j-x_i)}+\sqrt{2T_\mathrm{eff}}\,\xi_{x_i},\\
        \dot{y}_i &= {\frac{F(r_i)}{r_i}}y_i-{\omega}x_i
        +{k\sum_{j=1}^N c_{ij}(y_j-y_i)}+\sqrt{2T_\mathrm{eff}}\,\xi_{y_i},
    \end{split}
\end{equation}
{where the noises are independent standard Gaussian white noises.
We first consider a coupled pair to examine how radial and phase alignment affect the EPR.
We then ask how population size and coupling weights influence its approach to the strong-coupling limit.}

\subsubsection{Two coupled systems}
\begingroup
For a pair with unit coupling weight $c_{12}=1$, the interaction potential separates into two terms,
\begin{equation}\label{eq:51}
    V=\frac{k}{2}(r_1-r_2)^2+kr_1r_2\bigl[1-\cos(\theta_1-\theta_2)\bigr].
\end{equation}
For attractive coupling, the first term favors equal radii.
The second favors phase alignment, with a strength proportional to $r_1r_2$.
Substituting this potential into Eq.~(\ref{eq: dist gen}) gives the joint stationary density.
As for radial coupling, the EPR is determined by the radial second moments,
\begin{equation}\label{eq:46}
 \sigma=\frac{\omega^2}{T_\mathrm{eff}}\sum_{i=1}^2\langle r_i^2\rangle_k.
\end{equation}
The phase-integrated density and the EPR expansion in uncoupled moments are given in Appendix~\ref{app:cartesian-expansion}.

The initial EPR gradient is (Appendix~\ref{app: EPR grad zero cart})
\begin{equation}\label{eq:cartesian-pair-weak-gradient}
 \left.\frac{\mathrm{d}\sigma}{\mathrm{d}k}\right|_{k=0}
 =-\omega^2\beta^2\mathrm{Var}_0(r^2)\leq0.
\end{equation}
Thus, for nonzero intrinsic frequency, weak attractive Cartesian coupling decreases the EPR. The average radius initially decreases, whereas the radial variance can initially increase; for $F(r)=r(1-r)$, this occurs for $\beta\gtrsim8.6$ (Appendix~\ref{app: dist cartes}).
\endgroup

{For the pair with identical nonzero intrinsic frequencies, repulsive coupling instead increases the EPR above its uncoupled value for every finite $k<0$, see Appendix~\ref{app:repulsive-coupling} with representative examples in Fig.~\ref{fig:repulsive-coupling}(c).}

\begin{figure}
    \centering
    \includegraphics[bb=0 0 437.76087552 768.48153696,width=1.\linewidth]{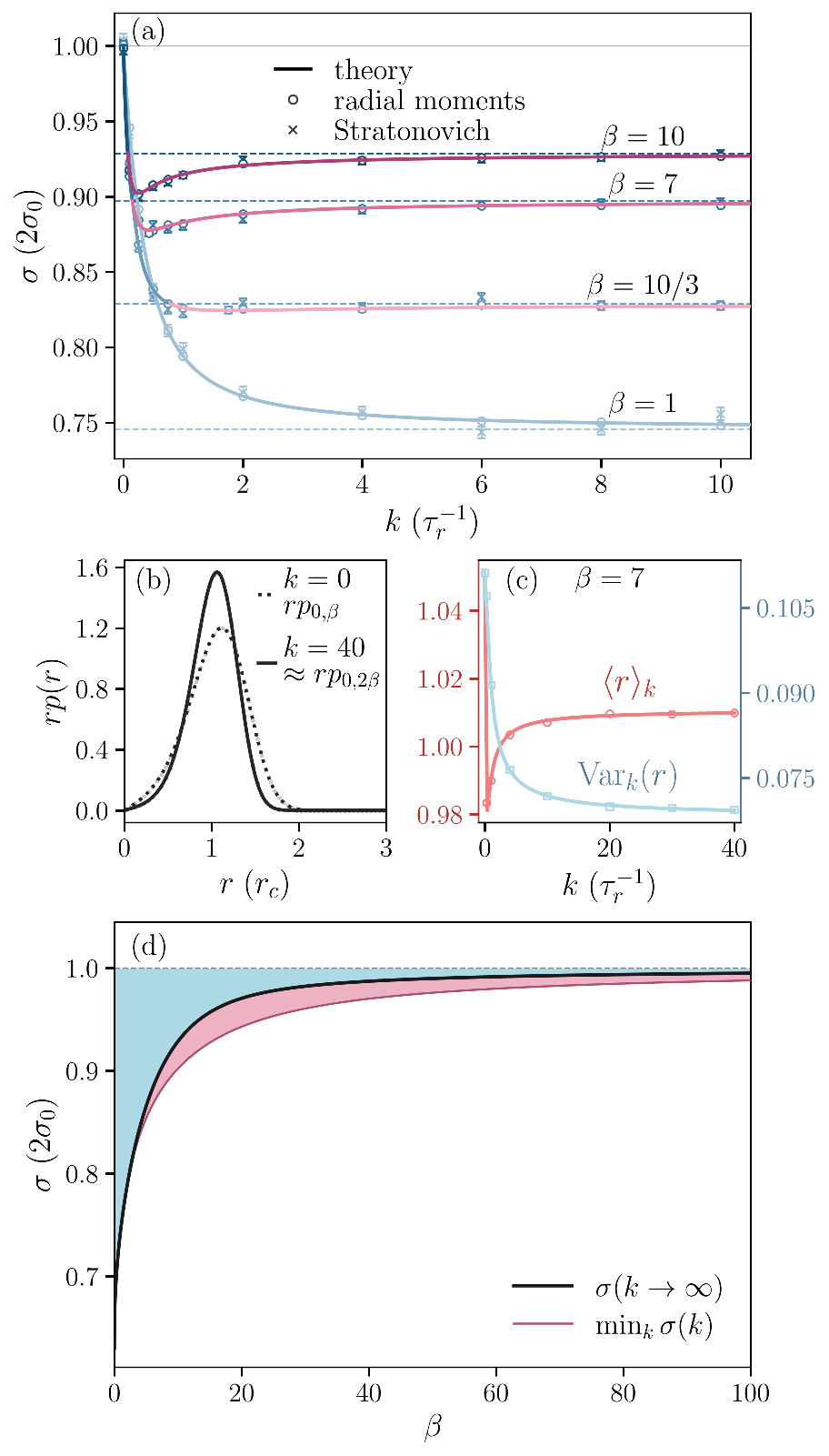}
    \caption{(a) The steady-state entropy production rate (EPR) $\sigma$ versus coupling strength $k$ of full Cartesian coupling for inverse effective temperatures $\beta\in\{1,{10/3},7,10\}$. The EPR is plotted in units of the EPR of the free system $2\sigma_0$ (Eq.~(\ref{eq:6})). The solid lines show Eq.~(\ref{eq:46}) evaluated using the stationary distribution. Open circles show the same expression evaluated using radial moments measured from numerical simulations of {Eq.~(\ref{eq:41}) for $N=2$ and $c_{12}=1$}{; crosses show Stratonovich estimates (Eq.~(\ref{eq:numerical-epr-integral})) from the same trajectories. Horizontal dashed lines show the strong-coupling limits.} Blue and pink segments indicate EPR values above and below the strong-coupling limit, respectively. (b) The marginal radial distribution for coupling strengths $k\in\{0,40\}\tau_r^{-1}$ for $\beta=7$. The strong-coupling limit distribution is ${r}p_{0,2\beta}(r)$. Lines show the stationary marginal densities; dots and shaded bands show simulation estimates and one standard error. (c) The average radius and radial variance versus coupling strength of the marginal distribution for the case $\beta=7$. (d) The limiting EPR $\sigma(k\to\infty)$ (solid-black line) as a function of inverse effective temperature $\beta$ in units of the free system $2\sigma_0$. The blue and pink regions represent the attainable fraction of the free system EPR under finite coupling, above and below the strong-coupling limit, respectively. Error bars show one standard error.}
    \label{fig:4}
\end{figure}

{For large $k$, the leading gradient is (Appendix~\ref{app: limit cart})
\begin{equation}\label{eq:cartesian-pair-strong-gradient}
\begin{split}
 \frac{\mathrm{d}\sigma}{\mathrm{d}k}&\approx\frac{\omega^2}{k^2}\left[-\frac{\beta}{2}\mathrm{Cov}_{0,2\beta}
 \left(r^2,F'(r)+\frac{F(r)}r\right)-1\right].
\end{split}
\end{equation}
The covariance term describes a change in the center-of-mass EPR contribution, while the $-1$ term reflects suppression of relative-coordinate fluctuations.
A positive bracket, together with the negative initial slope, implies a finite-coupling minimum. For $m=n=1$, numerical evaluation of the stationary distribution and Langevin simulations show this minimum for sufficiently large $\beta$ (Fig.~\ref{fig:4}(a)). At $\beta=7$, the average radius initially decreases and then increases, while the radial variance decreases (Fig.~\ref{fig:4}(b, c)). The associated radial redistribution is examined through gradient distribution plots in Appendix~\ref{app: dist cartes}.}

In the strong-coupling limit, the radii and phases align and relative-radius fluctuations vanish, so that (Appendix~\ref{app: limit cart})
\begin{equation}
     \sigma(k\to\infty)=\frac{2\omega^2}{T_\mathrm{eff}}\langle r^2\rangle_{0,2\beta}.
\end{equation}
{This EPR results from the center-of-mass motion and equals the free single-oscillator EPR at noise intensity $T_\mathrm{eff}/2$. Relative to the uncoupled pair, the ratio is $\langle r^2\rangle_{0,2\beta}/\langle r^2\rangle_{0,\beta}$. For $F(r)=r(1-r)$, the stationary mean-square radius decreases as the noise intensity decreases, so the strong-coupling EPR lies below the uncoupled value (Fig.~\ref{fig:4}(d)).} {Equation~(\ref{eq:cartesian-pair-strong-gradient}) predicts a finite-coupling minimum for $\beta>1.896$, whose} depth below the strong-coupling limit is shown in pink in Fig.~\ref{fig:4}(d). For the general family, the {high-noise limit} {follows from Eq.~(\ref{app eq: general moment high noise}) in Appendix~\ref{app:moments}:}
\begin{equation}
 \frac{\sigma(k\to\infty)}{\sigma(k=0)}(\beta\to0) = 2^{-2/{(m+n+1)}},
\end{equation}
{For $m=n=1$, this is $2^{-2/3}\approx0.63$, or} $63\%$ of the uncoupled EPR. Note that this is a greater decrease compared to the radial coupling of $86\%$ of the free system EPR.

\subsubsection{N coupled systems}
{At fixed coupling weights, the initial EPR slope for $N$ oscillators is $W$ times the unit-weight pair slope in Eq.~(\ref{eq:cartesian-pair-weak-gradient}),
where $W=\sum_{i<j}c_{ij}$ is the total edge weight (Appendix~\ref{app: EPR grad zero cart}).
Thus, for $\omega\ne0$, weak attractive coupling decreases the EPR for any network with $W>0$.}

\begin{figure}
    \centering
    \includegraphics[bb=0 0 360.96072192 519.36103872,width=1.\linewidth]{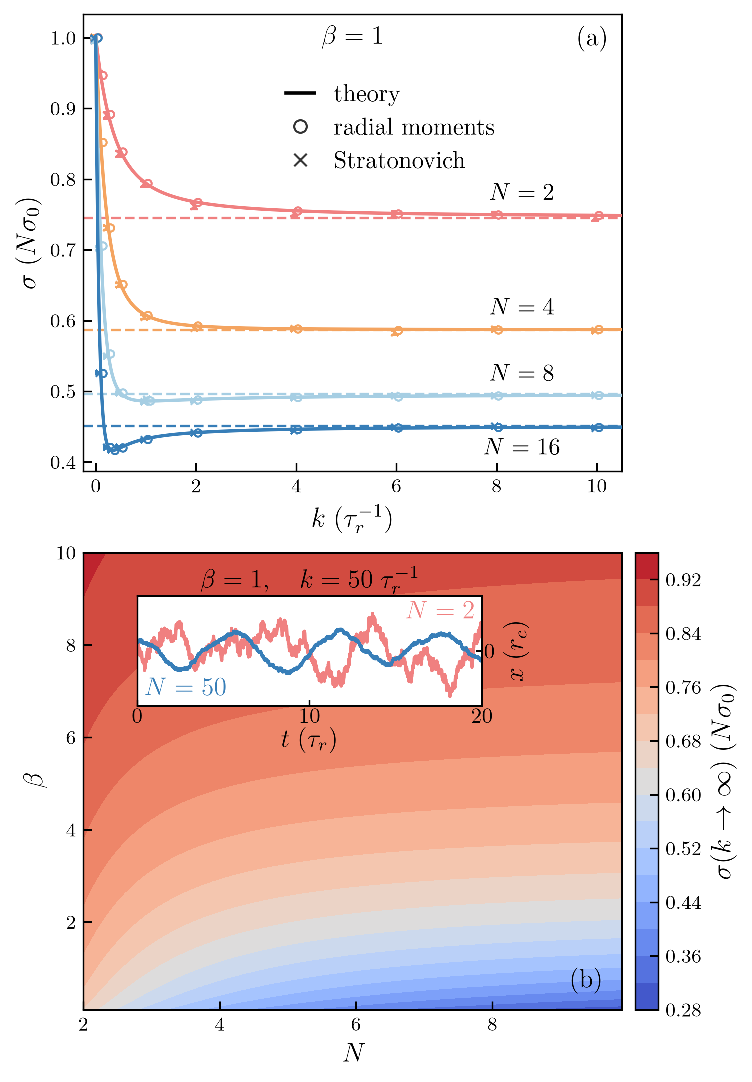}
    \caption{(a) The steady-state entropy production rate (EPR) $\sigma$ versus coupling strength $k$ of full Cartesian coupling for inverse effective temperature $\beta=1$ and system sizes $N\in\{2,4,8,16\}$. The EPR is plotted in units of the EPR of the free system $N\sigma_0$ (Eq.~(\ref{eq:6})). {Solid lines show the EPR $\beta\omega^2\sum_{i=1}^N\langle r_i^2\rangle_k$ evaluated using the stationary distribution. Open circles show the same expression evaluated using radial moments measured from} numerical simulations of {Eq.~(\ref{eq:41}) with unnormalized all-to-all coupling. Crosses show Stratonovich estimates (Eq.~(\ref{eq:numerical-epr-integral})) from the same trajectories}. The dashed lines denote the strong-coupling limit given by Eq.~(\ref{eq:63}). (b) A contour plot of the limiting EPR $\sigma(k\to\infty)$ as a function of inverse effective temperature $\beta$ and system size $N$ in units of the free system EPR. The inset displays two sample trajectories along the $x$-coordinate with large coupling strength, $k=50\,\tau_r^{-1}$, for system sizes $N\in\{2,50\}$, $\beta=1$, and $\omega={1}\tau_r^{-1}$. Error bars show one standard error.}
    \label{fig:5N}
\end{figure}

{At strong coupling, the approach to the limiting EPR also depends on how the links are arranged.
For fixed finite $N$, fixed $\beta$, and a connected network, the leading gradient is (Appendix~\ref{app: limit cart})
\begin{equation}\label{eq:cartesian-N-strong-gradient}
\begin{split}
 \frac{\mathrm{d}\sigma}{\mathrm{d}k}&\approx
 \frac{\omega^2}{k^2}\left(\sum_{\alpha=1}^{N-1}\lambda_\alpha^{-1}\right)\\
 &\quad\times\left[-\frac{N\beta}{2}\mathrm{Cov}_{0,N\beta}
 \left(r^2,F'(r)+\frac{F(r)}r\right)-2\right],
\end{split}
\end{equation}
where $\lambda_\alpha$ are the nonzero eigenvalues of the weighted graph Laplacian. For unnormalized all-to-all coupling, $\lambda_\alpha=N$. The covariance term describes a change in the center-of-mass EPR contribution, while the $-2$ term reflects suppression of relative-coordinate fluctuations. A positive bracket again implies a finite-coupling minimum. For $m=n=1$, the leading gradient changes sign at $N\beta\approx3.792$, so increasing $N$ at fixed $\beta$ can produce such a minimum, as observed in Fig.~\ref{fig:5N}(a). As $k\to\infty$, the oscillators align in both radius and phase. Their relative displacements vanish, leaving only the center-of-mass motion to contribute to the limiting EPR.}

{For fixed finite $N$ and fixed symmetric weights whose nonzero edges form a connected graph, the strong-coupling limit $k\to\infty$ is given by} (Appendix~\ref{app: limit cart})
\begin{equation}\label{eq:63}
    \sigma({k\to\infty})  =  \frac{N\omega^2}{T_\mathrm{eff}} \langle r^2\rangle_{0,N\beta}.
\end{equation}
{The total limiting EPR equals that of a single oscillator at noise intensity $T_\mathrm{eff}/N$. At sufficiently low noise (large $\beta$), this strong-coupling EPR lies below the uncoupled value when $m+2n<4$, and above it when $m+2n>4$ (Eq.~(\ref{app eq: general moment low noise}), Appendix~\ref{app:moments}). Throughout the family, the initial EPR slope is negative and the strong-coupling slope becomes positive at sufficiently low noise, implying a finite-coupling minimum (Appendix~\ref{app: limit cart}). For $F(r)=r(1-r)$, increasing $N$ lowers the strong-coupling EPR relative to the uncoupled value $N\sigma_0$.} Figure~\ref{fig:5N}(a) shows that increasing system size reduces the normalized EPR for coupled noisy oscillators with $\beta=1$. {The relative reduction is stronger at small $\beta$,} see Fig.~\ref{fig:5N}(b). The inset shows a sample trajectory of $N\in\{2,50\}$ strongly coupled oscillators, where the larger system visibly dampens fluctuations resulting in coherent oscillations. The {high-noise limit} of Eq.~(\ref{eq:63}) in units of the free system {follows from Eq.~(\ref{app eq: general moment high noise}) in Appendix~\ref{app:moments}:}
\begin{equation}
     \frac{\sigma({k\to\infty})}{\sigma({k=0})}(\beta\to0) = N^{-2/{(m+n+1)}}.
\end{equation}
{For $m=n=1$, this reduces to $N^{-2/3}$. Throughout the family, the normalized EPR tends to zero as $N\to\infty$ after the strong-coupling and high-noise limits.}

\subsection{Single coordinate coupling}\label{sec:single-coordinate}
{We next couple identical oscillators through their $x$ coordinates, retaining the radial drift $F(r)$ and symmetric coupling weights $c_{ij}$ introduced above.} The Langevin equations read
\begin{equation}\label{eq:64}
    \begin{split}
        \dot{x}_i &= {\frac{F(r_i)}{r_i}}x_i +\omega y_i + k \sum_{j\neq i}^N{c_{ij}}(x_j-x_i) + \sqrt{2T_\mathrm{eff}}\xi_{x_i},\\
        \dot{y}_i &= {\frac{F(r_i)}{r_i}}y_i -\omega x_i + \sqrt{2T_\mathrm{eff}}\xi_{y_i}.
    \end{split}
\end{equation}

\begin{figure}[h!]
    \centering
    \includegraphics[bb=0 0 246.00000000 342.00000000,width=1.\linewidth]{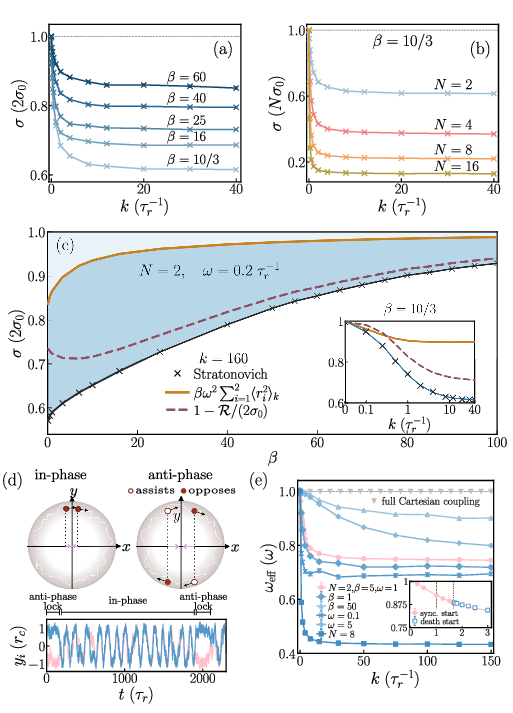}
    \caption{{Single-coordinate Cartesian coupling with unnormalized all-to-all interactions. (a, b) EPR versus coupling strength $k$ for the indicated inverse effective temperatures $\beta$ at $N=2$ (a), and system sizes $N$ at $\beta=10/3$ (b), normalized by the uncoupled value $N\sigma_0$ (Eq.~(\ref{eq:6})). (c) EPR at the large finite coupling $k=160\tau_r^{-1}$ versus $\beta$ for $N=2$. Orange shows the radial contribution $\beta\omega^2\sum_i\langle r_i^2\rangle_k/(2\sigma_0)$. The inset also shows the normalized stationary EPR (blue) versus $k$ at $\beta=10/3$. The dashed burgundy curves subtract $\mathcal R/(2\sigma_0)$ from unity in the main panel and from the normalized uncoupled Stratonovich estimate in the inset. Crosses in (a-c) show Stratonovich estimates; $\omega=0.2\tau_r^{-1}$ throughout (a-c). (d) Top: in-phase and anti-phase configurations of two $x$-coupled oscillators. In the anti-phase sketch, coupling assists rotation for the white pair and opposes it for the red pair. Bottom: sample $y$-coordinate trajectories showing intermittent anti-phase locking for $k=50\tau_r^{-1}$, $\omega=0.05\tau_r^{-1}$, and $\beta=100$. (e) Effective frequency $\omega_\mathrm{eff}/\omega$ versus $k$. The pink reference curve has $\omega=1\tau_r^{-1}$, $\beta=5$, and $N=2$; other curves change one parameter as indicated. The horizontal line at unity denotes full Cartesian coupling. The inset shows slowing near the deterministic existence and stability thresholds, $k=1$ and $k=5/3$, for $\omega=1$ (Appendix~\ref{app: single-od}). Circles and open squares show synchronized and stable anti-phase stationary initial conditions, respectively. Error bars and narrow uncertainty bands show one standard error.}}
    \label{fig:5}
    \label{fig:7}
\end{figure}

{Single-coordinate coupling breaks rotational symmetry, so
the preceding stationary-density construction no longer applies.
Nevertheless, the coupling remains conservative. Applying the EPR identity
of Sec.~\ref{sec: conservative interaction} to Eq.~(\ref{eq:64}) gives
\begin{equation}\label{eq:single-coordinate-epr}
    \begin{split}
        \sigma(k)
        &=\beta\omega^2\sum_{i=1}^N\langle r_i^2\rangle_k-\mathcal R(k),\\
        \mathcal R(k)
        &=-\beta\big\langle\boldsymbol f_{\mathrm{rot}}\cdot\boldsymbol f_{\mathrm c}\big\rangle_k\\
        &=\beta\omega k\sum_{i<j}c_{ij}
        \big\langle(x_i-x_j)(y_i-y_j)\big\rangle_k\geq 0.
    \end{split}
\end{equation}
A positive $\mathcal R$ means that
the coupling force $\boldsymbol f_{\mathrm c}$ opposes the intrinsic rotational drift on average.
We return below to the relative orientations that give rise to this cross correlation.}
	\newline
	{At the onset of coupling, rotational symmetry of the uncoupled
	system gives an initial EPR gradient equal to half that for full
	Cartesian coupling with the same per-coordinate strength
	(Appendix~\ref{app:single-coordinate-gradient}):
	\begin{equation}\label{eq:single-coordinate-weak}
		\left.\frac{\mathrm{d}\sigma}{\mathrm{d}k}\right|_{k=0}
		=-\frac{\omega^2\beta^2 W}{2}
		\mathrm{Var}_0(r^2)\leq0,
	\end{equation}
	where $W=\sum_{i<j}c_{ij}$ is the total edge weight
	defined in Sec.~\ref{sec: n coup rad}, with
	$W=N(N-1)/2$ for unnormalized all-to-all coupling.
	The above implies that $\mathcal R(k)=\mathcal O(k^2)$,
	so the linear decrease in EPR is entirely due to radial
	redistribution. The initial slopes of the mean radius and radial variance
	are likewise half those for full Cartesian coupling
	(see Eq.~(\ref{eq:  m v deriv 0}) for the pair).}
	\newline
	{For the numerical comparisons, we use $F(r)=r(1-r)$ and
	unnormalized all-to-all coupling, $c_{ij}=1$ for $i\ne j$.}
	We employ numerical simulations and show in {Fig.~\ref{fig:7}(a)}
	the effect of coupling on the EPR. The EPR decreases with coupling
	strength {over the parameters studied. Its value at large finite
	coupling is lower than for full Cartesian coupling in these
	comparisons (Fig.~\ref{fig:4}). Moreover, increasing system size $N$
	further reduces the normalized EPR (Fig.~\ref{fig:7}(b)).
	In Fig.~\ref{fig:7}(c), the light blue region above the orange curve
	shows the EPR reduction due to changes in the radial second moments
	at large finite coupling, and the darker blue region below the curve
	estimates the additional reduction $\mathcal R$, both in units of $2\sigma_0$. The inset shows the corresponding
	terms as functions of coupling strength.}

\begingroup

{To illustrate how the coupling can oppose rotational motion, consider
the in-phase and anti-phase pair configurations in Fig.~\ref{fig:7}(d).
Near the $y$-axis, the intrinsic angular driving moves an in-phase pair in the same $x$-direction, so the coupling cannot oppose their center-of-mass motion. For an anti-phase pair, the upper oscillator is driven to the right and the lower one to the left.
Rotation therefore increases the $x$-separation of the red pair and decreases that of the white pair.
The red and white pairs illustrate positive and negative contributions to $\mathcal R$ in Eq.~(\ref{eq:single-coordinate-epr}), respectively.
Anti-phase configurations in which coupling opposes rotation therefore contribute to the positivity of $\mathcal R$.}
\endgroup

{To make this picture quantitative, consider a deterministic anti-phase pair, $(x_2,y_2)=-(x_1,y_1)$, with $F(r)=r(1-r)$ and $c_{12}=c_{21}=1$.
Writing $x_1=r\cos\phi$ and $y_1=-r\sin\phi$ gives
$\dot\phi=\omega+k\sin(2\phi)$. For $0<k<\omega$, the phase continues
to wind, with mean angular rate $\sqrt{\omega^2-k^2}<\omega$.
For $k\geq\omega$, the coupling can balance
the rotational drive at a fixed angle.}

{At strong coupling, this balance gives $x_i/y_i\approx\omega/(2k)$,
and radial force balance gives $r_i\to1$ (Appendix~\ref{app: single-od}).
The two oscillators thus approach stationary positions near $x=0$,
one at $y\approx+1$ and the other at $y\approx-1$. These anti-phase fixed points
correspond to oscillation death when they are stable in the full two-oscillator
system~\cite{zakharova2013death}. For the parameters in
Fig.~\ref{fig:7}(d), the deterministic states are stable, and the noisy
trajectory shows intermittent anti-phase locking.}

To characterize the net circulation, we consider the effective frequency $\omega_\mathrm{eff}:= \langle \Delta\phi/\Delta t\rangle$, {where $\Delta\phi$ is the net clockwise phase change accumulated over a measurement interval of duration $\Delta t$, and the brackets denote an average over oscillators and trajectories.} {Figure~\ref{fig:7}(e)} shows the effective frequency reducing with coupling for various parameter values of $\omega$, $\beta$, and $N$. {For $N=2$ and $\omega=1$, the inset shows frequency reduction below both the existence ($k=1$) and stability ($k=5/3$) thresholds of the nonzero stationary states. The smooth decrease across these thresholds is consistent with the combined effects of slower angular motion and intermittent locking at finite noise.} {At fixed $\beta=5$, increasing $\omega$ weakens the coupling relative to the intrinsic angular drive and reduces the fractional slowing. For the pair at $\omega=1\,\tau_r^{-1}$, the weaker slowing at larger $\beta$ may reflect smaller noise-driven departures from the synchronized orbit, on which the coupling force vanishes.} {Notably, for the parameters shown in Fig.~\ref{fig:7}(e), the $N=8$ curve exhibits a larger reduction in $\omega_\mathrm{eff}$ than the $N=2$ comparison curve. This is consistent with an oscillator temporarily escaping an anti-phase locked configuration without bringing the whole population into an in-phase state.}
 
For identical oscillators, radial and full Cartesian coupling can reduce the EPR exclusively by modifying the radial distribution, while preserving the intrinsic frequency of the oscillators $\omega$. {For single-coordinate Cartesian coupling, Eq.~(\ref{eq:single-coordinate-epr}) separates radial redistribution from the additional reduction $\mathcal R$ due to the coupling force's average opposition to the intrinsic rotational drift.}

\subsection{Summary}

{For identical intrinsic frequencies, full Cartesian coupling reduces the EPR at the onset of coupling. For $F(r)=r(1-r)$, its strong-coupling limit also lies below the uncoupled value. More generally, at sufficiently low effective temperature, a finite-coupling minimum below the uncoupled EPR exists throughout the oscillator family considered.}

{Single-coordinate Cartesian coupling also lowers the EPR over the parameter range studied. Its initial EPR slope is half that for full Cartesian coupling. This initial decrease arises entirely from radial redistribution. At stronger coupling, the coupling force further reduces the EPR by opposing rotation on average. Angular slowing and intermittent anti-phase locking illustrate this opposition (Fig.~\ref{fig:7}(d,e)).}

\begin{figure}[h!]
    \centering
    \includegraphics[bb=0 0 255.00000000 364.00000000,width=1.\linewidth]{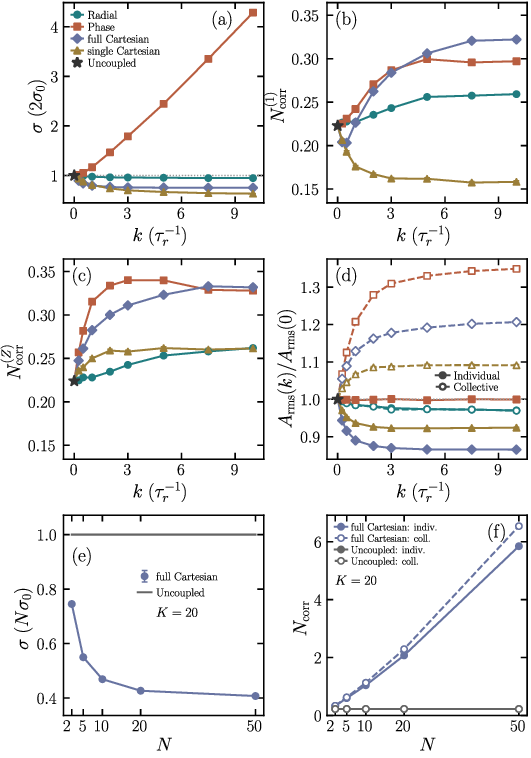}
    \caption{{Dissipation and oscillatory coherence. All panels use $T_\mathrm{eff}=1$ and identical intrinsic frequencies $\omega=1.1\tau_r^{-1}$. (a-d) Two oscillators with coupling strength $k$. (a) Total EPR in units of its uncoupled value $2\sigma_0$. (b,c) Correlation-based coherent cycles $N_{\mathrm{corr}}^{(1)}$ for oscillator 1 and $N_{\mathrm{corr}}^{(Z)}$ for the collective signal $Z$, respectively (Eqs.~(\ref{eq:coherence-correlation}) and (\ref{eq:coherence-observables})). (d) Individual (filled symbols, solid lines) and collective (open symbols, dashed lines) RMS signal amplitudes, each normalized by its uncoupled value. Stars denote the uncoupled pair. (e,f) Population-size dependence for full Cartesian coupling at fixed total incident coupling weight $K=(N-1)k=20\tau_r^{-1}$, compared with uncoupled oscillators: (e) total EPR normalized by $N\sigma_0$ and (f) individual and collective correlation-based coherent cycles. Error bars show one standard error.}}
    \label{fig:coherence}
\end{figure}

\begingroup
\section{Dissipation and oscillatory coherence}\label{sec:coherence}
To relate dissipation to oscillatory performance, we examine how coupling affects coherence and amplitude. We use $m=n=1$ to compare EPR, coherence, and amplitude for an individual oscillator, $z_1=x_1+\mathrm{i}y_1$, and the collective signal $Z=N^{-1}\sum_i z_i$, measured from the same trajectories. Writing $s_1=z_1$ and $s_Z=Z$, we define
\begin{equation}\label{eq:coherence-correlation}
    C_\nu(\tau)=\frac{\langle s_\nu(t+\tau)s_\nu^*(t)\rangle_k}
    {\langle |s_\nu|^2\rangle_k},\qquad \nu\in\{1,Z\}.
\end{equation}
The correlation time $\tau_{c,\nu}$ is the first lag at which $|C_\nu(\tau)|$ decreases to $1/e$. The accumulated correlation phase gives the coherent-cycle count
\begin{equation}\label{eq:coherence-observables}
    N_{\mathrm{corr}}^{(\nu)}=\frac{|\Theta_\nu(\tau_{c,\nu})|}{2\pi},
\end{equation}
where $\Theta_\nu(\tau)$ is the unwrapped phase of $C_\nu(\tau)$, with $\Theta_\nu(0)=0$. The RMS amplitude is $A_{\mathrm{rms},\nu}=\sqrt{\langle |s_\nu|^2\rangle_k}$.

For the identical pairs studied, full Cartesian coupling can reduce the EPR and improve both individual and collective coherence (Fig.~\ref{fig:coherence}(a--d)). Radial coupling also reduces the EPR at this effective temperature, with modest gains in both coherence measures. Phase coupling increases both coherence and EPR, whereas single-coordinate coupling decreases EPR and individual coherence despite increasing collective coherence. Individual RMS amplitudes can decrease while coherence improves. The collective amplitude also reflects phase alignment.

To examine system-size effects on coherence without increasing the total incident coupling weight, we hold $K=(N-1)k$ fixed as $N$ varies for full Cartesian coupling. Figure~\ref{fig:coherence}(e,f) shows decreasing normalized EPR and increasing individual and collective coherence over the population sizes studied, alongside the uncoupled results. These results show that reduced EPR can accompany improved temporal coherence.
\par
\endgroup

\section{Summary and Discussion}\label{sec: 6}
\begin{figure*}[!ht]
    \centering
    \includegraphics[bb=0 0 1133.00000000 450.00000000,width=1.\linewidth]{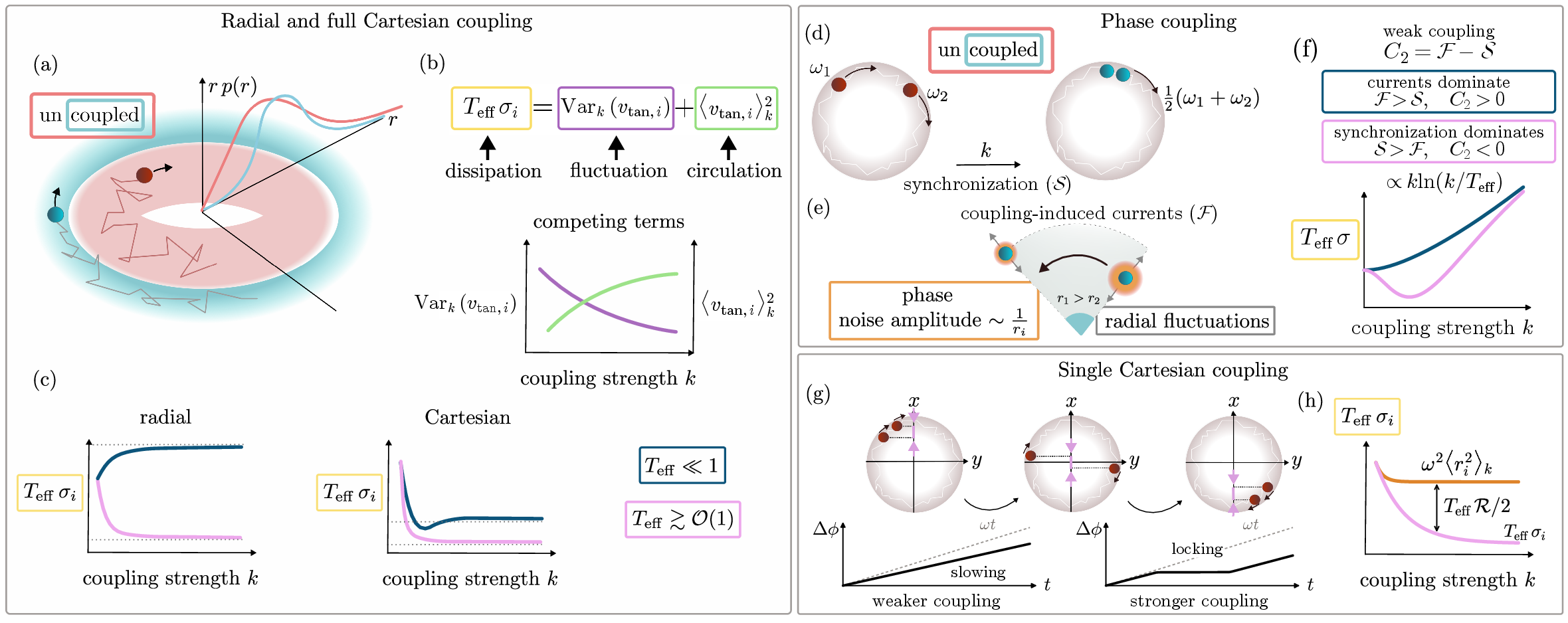}
    \caption{Schematic summary of radial, phase, and Cartesian coupling. (a) Radial and full Cartesian coupling change the radial mean and variance. (b) These determine the variance and squared-mean contributions of the tangential local mean velocity to EPR. (c) Competing changes in these moments yield different EPR responses to coupling. (d) For phase coupling, {synchronization of the radially weighted mean angular velocities lowers the mean-rate EPR contribution by $\mathcal S k^2$ at weak coupling.} (e) {Radial fluctuations leave angular diffusivities unequal and sustain coupling-induced currents; local angular-velocity fluctuations and radial currents add $\mathcal F k^2$ at weak coupling.} (f) {Representative total pair EPR curves with positive and negative initial quadratic coefficient $C_2=\mathcal F-\mathcal S$ (Eq.~(\ref{eq: C_2 pair})). The sign reflects whether additional current dissipation ($\mathcal F$) or the synchronization-related reduction ($\mathcal S$) dominates initially. Both curves ultimately grow as $k\ln(k/T_\mathrm{eff})$.} (g) {Single-coordinate coupling allows slower angular motion and intermittent anti-phase locking to suppress circulation. Upper sketches illustrate intermittent locking at stronger coupling; lower traces show schematic phase accumulation $\Delta\phi$ at weaker and stronger coupling, relative to uncoupled rotation $\omega t$ (dashed).} (h) {For single-coordinate coupling, the radial contribution and the EPR have the same initial slope. Their separation at finite coupling gives the additional reduction due to opposition to rotation, shown per oscillator for the pair in Fig.~\ref{fig:7}(c).}}
    \label{fig:9}
\end{figure*}
We studied how mutual coupling affects the steady-state EPR of stochastic circular limit-cycle oscillators. For a single oscillator, the EPR depends on three key factors: rotational frequency, radial fluctuations, and mean radius. Considering radial, phase Kuramoto-like, and Cartesian couplings, we show how changes in radial statistics and probability currents produce different EPR responses as the coupling strength is varied.

In Section~\ref{sec: 3}, we started with the consideration of radial coupling. At the onset of coupling, suppression of radial fluctuations lowers the variance contribution of the tangential local mean velocity to EPR. This reduction can compete with an increase in mean radius, which raises the contribution from the squared mean velocity (Fig.~\ref{fig:9}(a, b)). {For the evaluated $m=n=1$ model, the} EPR decreases at high effective temperature but increases at sufficiently low effective temperature (Fig.~\ref{fig:9}(c)). The EPR relative to the uncoupled value approaches $86\%$ for a pair when the strong-coupling limit is followed by the high-noise limit. Subsequently taking the number of oscillators to infinity gives $73\%$. At moderate $\beta$, however, the strong-coupling EPR ratio can increase with system size.

{Phase coupling introduces competition between synchronization of the radially weighted mean angular velocities and additional dissipation from coupling-induced currents (Fig.~\ref{fig:9}(d-f)). The exact EPR decomposition separates these mean-rate and fluctuation contributions, together with radial Fisher terms. Strong phase coupling leaves radial fluctuations unsuppressed. The resulting unequal angular diffusivities sustain currents, and the small-radius region produces EPR growth proportional to $k\ln(k/T_\mathrm{eff})$. The leading contribution is carried by the local angular-velocity variance and admits a radially weighted phase-Fisher interpretation in the locked outer region (Appendix~\ref{app sec: strong coupling pair}).}

{For a pair, the leading weak-coupling response is quadratic, with coefficient $C_2=\mathcal F-\mathcal S$ (Eq.~(\ref{eq: C_2 pair})). $\mathcal F$ describes additional dissipation from local angular-velocity fluctuations and radial currents, while $\mathcal S$ describes the synchronization-related reduction of the mean-rate contribution. When $\mathcal S>\mathcal F$, the EPR initially decreases. Its growth at strong coupling then implies a minimum at finite coupling. The network analysis expresses the weak-coupling response as a sum over interacting pairs and establishes the leading growth law for fixed finite all-to-all populations. Thus the energetic response depends on both frequency differences and the coupling structure.}

For identical intrinsic frequencies, full Cartesian coupling reduces the EPR {at the onset of coupling, with the initial slope determined by the variance of the squared radius. Under strong coupling, the collective noise intensity becomes $T_\mathrm{eff}/N$. For $F(r)=r(1-r)$, this gives a reduction relative to the uncoupled system. The reduction persists throughout the intermediate coupling range studied. At sufficiently low noise, competing changes in radial mean and variance produce a finite-coupling minimum below the uncoupled value (Fig.~\ref{fig:9}(a-c)). In the strong-coupling, high-noise limit, the pair EPR approaches} $63\%$ of the uncoupled value, compared with $86\%$ {for radial coupling; the fraction in the full Cartesian case scales as $N^{-2/3}$.} More generally, a finite-coupling minimum below the uncoupled EPR exists throughout the oscillator family at sufficiently low noise, even when the strong-coupling limit lies above the uncoupled value.

{For conservative interactions, Eq.~(\ref{eq:stationary-conservative-epr}) expresses the EPR as the contribution from the coupled radial second moments minus a nonnegative correction $\mathcal R$. This correction vanishes for radial coupling at arbitrary intrinsic frequencies and for full Cartesian coupling at identical frequencies. For single-coordinate Cartesian coupling, $\mathcal R$ measures the coupling force's average opposition to rotation. Since $\mathcal R=\mathcal O(k^2)$, the initial EPR decrease is entirely due to radial redistribution. For example, at $\beta=10/3$ and $k=160\,\tau_r^{-1}$, radial redistribution accounts for a decrease of about $10\%$ of the uncoupled pair EPR, and opposition to rotation accounts for a further $29\%$ (Fig.~\ref{fig:7}(c)). The trajectories illustrate how this opposition can arise through angular slowing and intermittent anti-phase locking (Figs.~\ref{fig:7}(d,e) and~\ref{fig:9}(g)). Over the coupling range studied, the EPR decreases toward a plateau (Fig.~\ref{fig:7}(a)).}

{In addition to the EPR analysis, we examined how coupling affects temporal coherence, finding that reduced EPR can accompany improved coherence (Sec.~\ref{sec:coherence}). For the identical pairs studied, all four coupling types can enhance collective coherence, although their effects on individual coherence and EPR differ. Full Cartesian coupling reduces EPR while improving individual and collective coherence, whereas single-coordinate coupling reduces EPR but lowers individual coherence. Furthermore, we studied how system size affects EPR and coherence, focusing on full Cartesian coupling. Increasing $N$ at fixed total incident coupling weight reduces the normalized EPR and improves both coherence measures over the sizes studied.}

The different radial and full Cartesian responses can also be understood through their Gibbs-reweighted stationary distributions (Sec.~\ref{sec: conservative interaction}). Within this class of distributions with interaction potential $V$, one can enforce sufficient conditions on $V$ such that the EPR decreases (Appendix~\ref{app: gibbs dist}). For arbitrary coupling strength, the EPR gradient can change sign for radial and full Cartesian coupling. At the onset of coupling, however, weaker conditions apply, and these are satisfied by full Cartesian coupling but not by radial coupling. Consequently, full Cartesian coupling is constrained to initially reduce the EPR, whereas radial coupling already admits a competition between opposing contributions. This consideration highlights that even for the canonical coordinate couplings considered in this work, the deformation of the stationary distribution and its effect on the EPR are non-trivial.

{This dependence on interaction type also extends to repulsive coupling (Fig.~\ref{fig:repulsive-coupling}). Reversing $k$ reverses the nonzero leading radial EPR correction near $k=0$. The phase-pair EPR is exactly even in $k$, while repulsive full Cartesian coupling increases the EPR for a pair with identical nonzero intrinsic frequencies (Appendix~\ref{app:repulsive-coupling}).}

{These coupling mechanisms can be investigated in biological oscillator models, with mechanically coupled hair bundles providing a concrete example. Hair bundles can undergo a supercritical Hopf bifurcation, near which their dynamics can be represented by noisy Stuart-Landau oscillators~\cite{roongthumskul2021dynamics}. In the oscillatory regime, the isochronous case belongs to the oscillator family considered here, with $m=2$ and $n=1$ (Appendix~\ref{app: SL}). Mechanical interactions couple the bundle displacement, represented by one Cartesian coordinate~\cite{dierkes2012mean}. Coupling-induced EPR reduction has also been demonstrated in a model of mechanically coupled hair bundles in the stable fixed-point regime~\cite{kharbanda2026residualdynamicsdeterminedissipation}. This connection motivates investigating how displacement coupling changes EPR, coherence, and sensory response around the Hopf transition, including the effects of amplitude-dependent frequency.}

{Driven colloidal rotors provide a complementary experimental setting for investigating coupling and dissipation~\cite{izumida2016energetics,kotar2013optimal}. Existing phase descriptions treat motion on fixed circular trajectories; finite radial compliance introduces amplitude fluctuations whose energetic effects can be examined alongside synchronization. Applying the present framework to such systems requires specifying the driving force, radial confinement, and hydrodynamic mobility.}

{Coupled stochastic oscillator models, including networks with full Cartesian coupling, also provide descriptions of macroscopic brain activity~\cite{deco2017dynamics,kringelbach2024thermodynamics,NARTALLOKALUARACHCHI20261}, where EPR characterizes irreversibility of the modeled dynamics~\cite{PhysRevE.104.014411}. A further direction is to investigate how coupling structure changes this irreversibility and collective dynamics.}

{In applying these results, the calculated EPR characterizes irreversibility within the effective oscillator dynamics. Coarse-graining can conceal dissipation associated with unresolved degrees of freedom~\cite{roldan2021quantifying,ghosal2022inferring}. For example, the correspondence with hair-bundle oscillator models identifies effective displacement dynamics and coupling, but does not explicitly resolve the molecular processes supplying the active driving and noise. $T_{\mathrm{eff}}$ is a dimensionless timescale ratio; relating the calculated EPR to total biological dissipation or physical power, including metabolic expenditure, requires a thermodynamic description of those processes. Establishing biological efficiency additionally requires accounting for any energetic costs of physically implementing and maintaining the coupling, which is a direction for future work.}

Further directions include exploring coupling topologies and coupling between distinct systems~\cite{bera2017chimera}, and developing analytical tools for more general coupled limit cycles.

Overall, this circular limit-cycle model identifies how coupling modifies EPR through radial redistribution and probability currents, and how these energetic changes accompany distinct effects on oscillatory coherence.

\section{Methods}\label{sec:7}
\begingroup
\subsection{Numerical simulation details}\label{app: sim details}
The trajectory calculations are performed in Cartesian coordinates using stochastic Heun integration.
The trajectory EPR is evaluated from the Stratonovich drift integral~\cite{roldan2021quantifying},
\begin{equation}\label{eq:numerical-epr-integral}
 \sigma=\lim_{\tau\to\infty}\frac{\beta}{\tau}
 \left\langle\int_0^\tau \mathbf{A}(\mathbf{X}(t))\cdot{\circ}\,\mathrm{d}\mathbf{X}(t)\right\rangle,
\end{equation}
where $\mathbf{X}=(x_1,y_1,\ldots,x_N,y_N)$, $\mathbf{A}$ is the full Cartesian drift, and $\circ$ denotes Stratonovich integration. Numerically, the drift is evaluated at the arithmetic midpoint of successive Heun states. For Fig.~\ref{fig:7}(e), we calculate the accumulated phase change of each oscillator by summing the signed angular increments between successive integration steps.

Statistical uncertainties are estimated across independently simulated replicas of the complete coupled system. Error bars show one standard error of the mean (SEM) for ordinary means and jackknife standard errors for the phase decomposition and radial variances. Error bars in Fig.~\ref{fig:coherence} show one standard error estimated by whole-replica bootstrap resampling, retaining uncertainty in the measured uncoupled references.

Simulation settings and checks of timestep, equilibration, measurement duration, and replica count are documented in Ref.~\cite{burnet2026revisioncode}.

\subsection{Evaluating analytical expressions}
Radial and full Cartesian stationary expectations and minima are evaluated by numerical quadrature. For unnormalized all-to-all coupling with $k>0$, the finite-$N$ calculations in Figs.~\ref{fig:3 N} and \ref{fig:5N} use exact Gaussian integral identities. For radial coupling,
\begin{equation}\label{eq:gaussian-radial-identity}
 e^{-\frac{\beta k}{2}\sum_{i<j}(r_i-r_j)^2}
 =\sqrt{\frac{\beta kN^2}{2\pi}}
 \int_{\mathbb R}\prod_{i=1}^{N}
 e^{-\frac{\beta kN}{2}(r_i-u)^2}\,\mathrm{d}u.
\end{equation}
For full Cartesian coupling, writing $\mathbf z_i=(x_i,y_i)$,
\begin{equation}\label{eq:gaussian-cartesian-identity}
 e^{-\frac{\beta k}{2}\sum_{i<j}|\mathbf z_i-\mathbf z_j|^2}
 =\frac{\beta kN^2}{2\pi}
 \int_{\mathbb R^2}\prod_{i=1}^{N}
 e^{-\frac{\beta kN}{2}|\mathbf z_i-\mathbf u|^2}\,\mathrm{d}^2\mathbf u.
\end{equation}
At fixed auxiliary variable $u$ or $\mathbf u$, the stationary integrals factor into single-oscillator integrals. These exact identities therefore reduce the high-dimensional stationary averages to nested one-dimensional quadratures. For Cartesian coupling, writing the auxiliary vector as $\mathbf u=s(\cos\psi,\sin\psi)$, each oscillator's phase dependence is $e^{\beta kNr_i s\cos(\theta_i-\psi)}$. Integration over $\theta_i$ gives $2\pi\mathrm{I}_0(\beta kNr_i s)$, independently of $\psi$. The auxiliary phase therefore contributes another $2\pi$, leaving numerical integration over the oscillator radius and auxiliary magnitude $s$.

For phase coupling, Eq.~(\ref{eq:78}) is solved using a Fourier expansion in relative phase and a reversible finite-volume discretization of both radial coordinates, including the origin. Domain, resolution, truncation, and precision checks, together with solver residuals and independent EPR checks, are documented in Ref.~\cite{burnet2026revisioncode}.

OpenAI Codex (GPT Sol. 6.1) assisted with portions of code implementation and mathematical derivations under the authors' direction. The authors reviewed and verified the outputs through analytical and numerical checks and retain responsibility for the results.
\par
\endgroup

\section{Acknowledgments}
We acknowledge funding by the European Research Council (ERC) under the European Union's Horizon 2020 research and innovation program (BacForce, G.A.No. 852585) and by the Deutsche Forschungsgemeinschaft (DFG, German Research Foundation), Project no 492014049. The authors declare that they have no competing interests. 

\section{Data availability}
{The simulation, analysis, and figure-generation code, saved figure data, and numerical validation records are publicly available in Ref.~\cite{burnet2026revisioncode} at \url{https://github.com/AFBurnet/Energetics-of-coupled-stochastic-circular-limit-cycle-oscillators}.}

\section{Appendix}

\subsection{Two coupled Kuramoto oscillators}\label{app:kuramoto}
Consider the coupled stochastic phase oscillators~\cite{kuramoto2003chemical}
\begin{equation}
    \begin{split}
    \dot{\theta}_1 =& \ \,  -\omega_1 + k \sin(\theta_2-\theta_1) + \sqrt{2D}\xi_1 \\
    \dot{\theta}_2 =& \ \, -\omega_2 + k \sin(\theta_1-\theta_2) + \sqrt{2D}\xi_2. 
    \end{split}
\end{equation}
Introducing the relative phase $\varphi=\theta_2-\theta_1$ and center-of-mass phase $\psi=(\theta_2+\theta_1)/2$ gives
\begin{equation}
    \begin{split}
    \dot{\varphi} =& \ \,  -\Delta\omega -2k \sin\varphi + \sqrt{4D}\,\eta_1 \\
    \dot{\psi} =& \ \, -\bar{\omega} + \sqrt{D}\,\eta_2, 
    \end{split}
\end{equation}
where $ \Delta\omega:= \omega_2-\omega_1$, $\bar{\omega}:=(\omega_2+\omega_1)/2$, $\eta_1=(\xi_2-\xi_1)/\sqrt{2}$, and $\eta_2=(\xi_1+\xi_2)/\sqrt{2}$. Since in these coordinates the degrees of freedom decouple, they each contribute to the entropy production independently. 
\\ \\
\noindent
Let us first consider the relative phase $\varphi$. The corresponding Langevin equation corresponds to the well-known ``tilted washboard'' problem for Brownian motion in the potential $V(\varphi)= \Delta\omega \, \varphi -2k\cos\varphi$, where $\Delta\omega$ plays the role of a driving force. The probability current obeys
\begin{equation}
    J_\varphi = (-\Delta\omega-2k \sin\varphi)p_\mathrm{s} - 2D\partial_\varphi p_\mathrm{s}.
\end{equation}
At steady state, the probability current $J_\varphi$ is constant. The stationary solution follows by enforcing periodicity and normalization~\cite{risken1989fpe}. {With $\alpha=-k/D$ and $\rho=\Delta\omega/(2D)$, expansion in modified Bessel functions} $\mathrm{I}_n$ gives
\begin{equation}
    J_\varphi = -\frac{2D}{2\pi\rho \sum\limits_{n=-\infty}^\infty\frac{\mathrm{I}_n(\alpha)\mathrm{I}_n(-\alpha)}{n^2+\rho^2}},
\end{equation}
and 
\begin{equation}
    p_\mathrm{s}(\varphi) = e^{-\alpha \cos\varphi}\frac{\sum\limits_{n=-\infty}^\infty\frac{\mathrm{I}_n(\alpha)}{n^2+\rho^2}[\rho \cos n\varphi + n\sin n \varphi]}{2\pi \rho \sum\limits_{n=-\infty}^\infty\frac{\mathrm{I}_n(\alpha)\mathrm{I}_n(-\alpha)}{n^2+\rho^2}}.
\end{equation}
{For $\Delta\omega=0$, these expressions reduce to $J_\varphi=0$ and $p_\mathrm{s}(\varphi)=e^{(k/D)\cos\varphi}/[2\pi\mathrm{I}_0(k/D)]$.}
\subsubsection{Entropy production rate}
Periodicity gives $\int_0^{2\pi} J_\varphi/p_\mathrm{s}\,\mathrm{d}\varphi=-2\pi\Delta\omega$, and hence
\begin{equation}
    \sigma_\varphi =  \int_0^{2\pi} \frac{ J_\varphi^2}{2D p_\mathrm{s}}\mathrm{d}\varphi =-\frac{\pi \Delta\omega}{D} J_\varphi.
\end{equation}
{For fixed radii $r_i=1$, define the mean local angular velocities $\Omega_i:=\langle v_{\theta_i}\rangle_\mathrm{s}$. Since $v_{\theta_1}+v_{\theta_2}=-2\bar\omega$, the relative-phase contribution can be written as}
\protect\leavevmode{\begin{equation}
\begin{split}
\sigma_\varphi
&=\underbrace{\frac{(\Omega_2-\Omega_1)^2}{2D}}_{\sigma_{\Delta\Omega}}\\
&\quad+\underbrace{\frac{1}{D}\sum_{i=1}^{2}
\left\langle(v_{\theta_i}-\Omega_i)^2\right\rangle_\mathrm{s}}_{\sigma_{\mathrm{var}}}.
\end{split}
\end{equation}
These terms correspond to the fixed-radius limit of the mean-variance decomposition in Eq.~(\ref{eq: EPR mean-variance}). Using $\Omega_2-\Omega_1=\langle v_\varphi\rangle_\mathrm{s}=2\pi J_\varphi$, the two contributions become
\begin{equation}
\begin{aligned}
\sigma_{\Delta\Omega}
&=\frac{2\pi^2}{D}J_\varphi^2,\\
\sigma_{\mathrm{var}}
&=-\frac{\pi\Delta\omega}{D}J_\varphi
-\frac{2\pi^2}{D}J_\varphi^2\\
&=\frac{(\Delta\omega)^2}{2D}
\frac{J_\varphi}{J_\varphi(0)}
\left(1-\frac{J_\varphi}{J_\varphi(0)}\right),
\end{aligned}
\end{equation}
where $J_\varphi(0)=-\Delta\omega/(2\pi)$ is the uncoupled current and the final equality applies for $\Delta\omega\neq0$.
Although the stationary current is independent of phase, the local mean velocity $v_\varphi=J_\varphi/p_\mathrm{s}$ varies with the nonuniform stationary density. For $\Delta\omega\neq0$, the factorized form shows that, as coupling increases, $\sigma_{\mathrm{var}}$ rises from zero at $k=0$, reaches its maximum when $J_\varphi=J_\varphi(0)/2$, and then vanishes as $k\to\infty$. Its initial increase is outweighed by the decrease in $\sigma_{\Delta\Omega}$, so their sum decreases. For $\Delta\omega=0$, both contributions vanish at every coupling strength.
}

The EPR associated with the center-of-mass coordinate $\psi$ is $\sigma_\psi=2\bar{\omega}^2/D$, so that the total EPR is given by
\begin{equation}\label{eq:totEP}
    \sigma =  \sigma_\psi +  \sigma_\varphi =  \frac{2\bar{\omega}^2}{D} - \frac{\pi \Delta\omega}{D} J_\varphi.
\end{equation}
For zero coupling $k=0$, the total entropy production rate is given by the sum of the individual oscillator contributions
\begin{equation}
    \sigma(k=0) = \frac{\omega_1^2 + \omega_2^2}{D},
\end{equation}
whereas in the strong-coupling limit we are left with the center-of-mass contribution
\begin{equation}
    \sigma(k\to\infty) = \frac{2\bar{\omega}^2}{D}.
\end{equation}
The limiting EPR reduction is therefore ${\sigma(0)-\sigma(\infty)}=\Delta\omega^2/(2D)$.

\begingroup
Note that Eq.~(\ref{eq:totEP}) varies smoothly with $k$.
For $\Delta\omega\neq 0$, noise smooths the deterministic locking threshold
$2k=|\Delta\omega|$. For $\Delta\omega=0$, the
relative current vanishes, while
\begin{equation}\label{app eq: kuramoto coherence}
\begin{split}
\sigma(k)&=\frac{2\omega^2}{D},\\
\langle\cos\varphi\rangle
&=\frac{\mathrm{I}_1(k/D)}{\mathrm{I}_0(k/D)},
\end{split}
\end{equation}
where the cosine expectation captures the phase coherence, which increases from $0\to1$ as $k\to\infty$. Thus, increasing phase coherence does not change the EPR in this case.

For comparison, consider $N$ identical noisy Kuramoto oscillators with
the normalized interaction $(K/N)\sum_j\sin(\theta_j-\theta_i)$.
Writing $U=-(K/N)\sum_{i<j}\cos(\theta_i-\theta_j)$, their stationary
distribution is $p_{\mathrm{s}}\propto e^{-U/D}$.
For this Gibbs distribution, the interaction and diffusive contributions
to the current cancel, leaving $J_i=-\omega p_{\mathrm{s}}$;
invariance under common rotation ensures stationarity. This gives
\begin{equation}\label{app eq: kuramoto collective epr}
\frac{\sigma_N(K)}{N}=\frac{\omega^2}{D}.
\end{equation}
Nevertheless, this identical-frequency model synchronizes for
$K>2D$ in the $N\to\infty$ limit~\cite{gupta2014kuramoto}.
Hence, the EPR is not generally a synchronization order parameter.
For nonidentical frequencies, the relative currents and EPR can change
with coupling; their behavior at a collective transition requires
a separate analysis.
\par
\endgroup

\subsection{Free system moments}\label{app:moments}
{For $m=n=1$, the integrals $\mathcal{I}_q$ defined in} Eq.~({\ref{eq:general-radial-moments}}) admit the following closed-form expression:
\begin{equation}\label{eq: free moments}
\resizebox{0.85\columnwidth}{!}{$
\begin{aligned}
\mathcal{I}_{q}&(\beta)
  := \int_0^\infty r\,dr\, e^{\beta(\frac{r^2}{2}-\frac{r^3}{3})}\, r^{q} \\
  &= \frac{1}{8}\,3^{\frac{-1+q}{3}}\,
     \beta^{-\frac{2}{3}-\frac{q}{3}}
     \Biggl(\,8\,\Gamma\!\Bigl(\tfrac{2+q}{3}\Bigr)\,
     {}_{2}F_2\!\Bigl(\{\tfrac{1}{3}+\tfrac{q}{6},\tfrac{5}{6}+\tfrac{q}{6}\};\{\tfrac{1}{3},\tfrac{2}{3}\};\tfrac{\beta}{6}\Bigr) \\
  &\quad + 4\,3^{\tfrac{2}{3}}\beta^{\tfrac{1}{3}}\,
     \Gamma\!\Bigl(\tfrac{4+q}{3}\Bigr)\,
     {}_{2}F_2\!\Bigl(\{\tfrac{2}{3}+\tfrac{q}{6},\tfrac{7}{6}+\tfrac{q}{6}\};\{\tfrac{2}{3},\tfrac{4}{3}\};\tfrac{\beta}{6}\Bigr) \\
  &\quad + 3\,3^{\tfrac{1}{3}}\beta^{\tfrac{2}{3}}\,
     \Gamma\!\Bigl(2+\tfrac{q}{3}\Bigr)\,
     {}_{2}F_2\!\Bigl(\{\!1+\tfrac{q}{6},\tfrac{3}{2}+\tfrac{q}{6}\};\{\tfrac{4}{3},\tfrac{5}{3}\};\tfrac{\beta}{6}\Bigr)
     \Biggr),
\end{aligned}
$}
\end{equation}
${}_{{2}}F_2$ denotes the generalized hypergeometric function, and the moments of the free system are $\langle r^{{q}}\rangle_0=\mathcal{I}_{{q}}/\mathcal{I}_0$.

\begingroup
For the general family, $F'(1)=-1$ and $F''(1)=-(m+2n-1)$. At low noise, expanding the exponent in Eq.~(\ref{eq:general-radial-moments}) through cubic order around $r=1$, and including the prefactor $r^{q+1}$, gives, at fixed $q$,
\begin{equation}\label{app eq: general moment low noise}
\frac{\mathcal I_{q+2}(\beta)}{\mathcal I_q(\beta)}
=1+\frac{2q+4-m-2n}{\beta}
+\mathcal{O}(\beta^{-2}).
\end{equation}
In the high-noise limit, rescaling $r=[m(m+n+1)/\beta]^{1/(m+n+1)}u$ reduces the leading integrals to Gamma functions:
\begin{equation}\label{app eq: general moment high noise}
\begin{split}
\frac{\mathcal I_{q+2}(\beta)}{\mathcal I_q(\beta)}
&=\left[\frac{m(m+n+1)}{\beta}\right]^{2/(m+n+1)}
\frac{\Gamma\!\left(\frac{q+4}{m+n+1}\right)}
{\Gamma\!\left(\frac{q+2}{m+n+1}\right)}\\
&\quad\times\left[1+\mathcal{O}\!\left(\beta^{m/(m+n+1)}\right)\right].
\end{split}
\end{equation}
The strong-coupling second moments for radial and full Cartesian coupling correspond to $q=N-1$ and $q=0$, respectively, evaluated at $N\beta$; the uncoupled second moment uses $q=0$ at $\beta$. These noise limits are taken after $k\to\infty$, at fixed finite $N$.
\endgroup

\begingroup
\subsection{Stationary EPR for interacting oscillators}
\label{app:conservative-epr}

For a coupling force $\boldsymbol f_{\mathrm c}$, the Cartesian drift is
$\boldsymbol A=\boldsymbol f_{\mathrm{rot}}+\boldsymbol f_{\mathrm c}-\nabla U_0$,
where $(\boldsymbol f_{\mathrm{rot}})_i=(\omega_i y_i,-\omega_i x_i)$ and the
intrinsic radial potential is
\begin{equation}
U_0=-\sum_{i=1}^N\int_0^{r_i}F(s)\,\mathrm{d}s.
\end{equation}
Let $p_k$ denote the stationary Cartesian density and
$\boldsymbol v=\boldsymbol A-\beta^{-1}\nabla\ln p_k$ its local mean velocity.
Then $\boldsymbol f_{\mathrm{rot}}+\boldsymbol f_{\mathrm c}-\boldsymbol v
=\nabla(U_0+\beta^{-1}\ln p_k)$.
Since the stationary current $p_k\boldsymbol v$ is divergence-free,
integration by parts with vanishing boundary terms gives
$\langle\boldsymbol v\cdot\nabla\chi\rangle_k=0$ for scalar functions $\chi$.
Using Eq.~(\ref{eq: epr}) with $D=\beta^{-1}I$, we obtain
\begin{equation}\label{app eq: single stationary identity}
\begin{split}
\sigma&=\beta\langle|\boldsymbol v|^2\rangle_k\\
&=\beta\langle(\boldsymbol f_{\mathrm{rot}}+\boldsymbol f_{\mathrm c})\cdot\boldsymbol v\rangle_k\\
&=\beta\langle(\boldsymbol f_{\mathrm{rot}}+\boldsymbol f_{\mathrm c})\cdot\boldsymbol A\rangle_k
 +\langle\nabla\cdot\boldsymbol f_{\mathrm c}\rangle_k.
\end{split}
\end{equation}
The last equality follows by substituting $\boldsymbol v=\boldsymbol A-\beta^{-1}\nabla\ln p_k$
and integrating by parts, using $\nabla\cdot\boldsymbol f_{\mathrm{rot}}=0$.

For conservative coupling, $\boldsymbol f_{\mathrm c}=-\nabla V$, so
$\langle\boldsymbol f_{\mathrm c}\cdot\boldsymbol v\rangle_k=0$.
Equation~(\ref{app eq: single stationary identity}) then gives
$\sigma=\beta\langle\boldsymbol f_{\mathrm{rot}}\cdot\boldsymbol v\rangle_k
=\beta\langle\boldsymbol f_{\mathrm{rot}}\cdot\boldsymbol A\rangle_k$, since
$\nabla\cdot\boldsymbol f_{\mathrm{rot}}=0$.
The intrinsic radial force is perpendicular to the rotational drift, giving
\begin{equation}
\begin{split}
\sigma&=\beta\sum_{i=1}^N\omega_i^2\langle r_i^2\rangle_k-\mathcal R,\\
\mathcal R&=-\beta\langle\boldsymbol f_{\mathrm{rot}}\cdot\boldsymbol f_{\mathrm c}\rangle_k.
\end{split}
\end{equation}

To establish the sign of $\mathcal R$, write $U=U_0+V$ and
$\Psi=U+\beta^{-1}\ln p_k$. Then $\boldsymbol f_{\mathrm{rot}}=\boldsymbol v+\nabla\Psi$,
and $\langle\boldsymbol v\cdot\nabla\Psi\rangle_k=0$ gives
\begin{equation}
\langle|\boldsymbol f_{\mathrm{rot}}|^2\rangle_k
=\langle|\boldsymbol v|^2\rangle_k+\langle|\nabla\Psi|^2\rangle_k.
\end{equation}
Hence
\begin{equation}\label{app eq: single nonnegative correction}
\begin{split}
\mathcal R&=\beta\langle|\nabla\Psi|^2\rangle_k
=\beta\langle|\boldsymbol f_{\mathrm{rot}}-\boldsymbol v|^2\rangle_k\geq0,\\
&0\leq\sigma\leq\beta\sum_{i=1}^N\omega_i^2\langle r_i^2\rangle_k.
\end{split}
\end{equation}
Equation~(\ref{app eq: single nonnegative correction}) gives the upper
bound of Ref.~\cite{dechant2023upper} for the rotational driving force
$\boldsymbol f_{\mathrm{rot}}$. The difference between this bound and
the stationary EPR is $\mathcal R$. The correction vanishes for radial coupling
at arbitrary intrinsic frequencies and for full Cartesian coupling at identical
intrinsic frequencies. In these cases, $p_k\propto e^{-\beta U}$ renders
$\Psi$ constant.
\par
\endgroup

\subsection{Gibbs-reweighted joint probability distributions}\label{app: gibbs dist}

For the steady-state distribution in Eq.~(\ref{eq: dist gen}) {with $\mathbf{J}_0\cdot\nabla V=0$}, the steady-state probability current is given by
\begin{equation}
    \mathbf{J} = \frac{1}{\mathcal{Z}}\mathbf{J}_0 e^{-\beta V},
\end{equation}
{Thus $J_{r_i}=0$ and $J_{\theta_i}=-\omega_i p$.} The EPR is then
\begin{equation}
    \begin{split}
        \sigma &= \beta\sum_{i=1}^N \omega_i^2\int r_i^2 p\,\mathrm{d}\mathbf{x}\\
        &=\beta\sum_{i=1}^N\omega_i^2\langle r_i^2\rangle_k,
    \end{split}
\end{equation}
as encountered for the radial and full Cartesian coupling cases. For the full Cartesian results, we use $\omega_i=\omega$ for all $i$. From Eq.~(\ref{eq: deriv k}), the gradient reads
\begin{equation}
     \frac{\mathrm{d}}{\mathrm{d}k}\sigma = -\beta^2\sum_{i=1}^N\omega_i^2\mathrm{Cov}_k(r_i^2,\partial_k V).
\end{equation}
The EPR decreases when the weighted covariance sum is positive. For radial and full Cartesian coupling, this sum can change sign as coupling increases.

{At $k=0$, suppose}
\begin{equation}
\begin{split}
    \partial_k V\big|_{k=0} &= \sum_{i=1}^N g_i(r_i)+h(\{r_j\},\{\theta_j\}),\\
    \mathrm{Cov}_0(r_i^2,h)&=0\qquad\text{for all }i.
\end{split}
\end{equation}
Independence of the uncoupled radii gives $\left.\mathrm{d}\sigma/\mathrm{d}k\right|_{k=0}{\leq}0$ when the functions $g_i$ are non-decreasing{. The inequality is strict if $\omega_i\ne0$ and $g_i$ is nonconstant for at least one $i$.} For full Cartesian coupling, $\partial_k V\big|_{k=0}=\frac{1}{2}\sum_{i<j}^N c_{ij}[r_i^2+r_j^2-2r_i r_j\cos(\theta_i-\theta_j)]$. {For $c_{ij}\geq0$, the radial terms are non-decreasing, while the uniform uncoupled phases make the cosine terms' covariance with $r_i^2$ vanish.}

\subsection{Radial coupling}

\subsubsection{Expansion of EPR}\label{app:radial-expansion}
For two coupled oscillators, expanding the interaction exponential and applying the binomial theorem gives the EPR in terms of moments of the free system:
\begin{equation}\label{eq: epr expansion rad}
\begin{split}
    \sigma &= \frac{2\omega^2}{T_\mathrm{eff}\mathcal{Z}}
    \sum_{n=0}^\infty\frac{1}{n!}\left(-\frac{k}{2T_\mathrm{eff}}\right)^n\\
    &\quad\times\sum_{m=0}^{2n}(-1)^m\binom{2n}{m}
    \langle r^{2n-m+2}\rangle_0\langle r^m\rangle_0,
\end{split}
\end{equation}
where the normalization term $\mathcal{Z}$ is given by
\begin{equation}
\begin{split}
    \mathcal{Z} &= \sum_{n=0}^\infty\frac{1}{n!}\left(-\frac{k}{2T_\mathrm{eff}}\right)^n\\
    &\quad\times\sum_{m=0}^{2n}(-1)^m\binom{2n}{m}
    \langle r^{2n-m}\rangle_0\langle r^m\rangle_0.
\end{split}
\end{equation}

\subsubsection{EPR gradient}\label{app:EPR grad zero rad}
For $N$ oscillators coupled through the interaction $V(\{r_j\})=\frac{k}{2}\sum_{i<j}^Nc_{ij}(r_i-r_j)^2$, the EPR gradient reads
\begin{equation}\label{eq: deriv rad 0 app}
    \begin{split}
        \frac{\mathrm{d}}{\mathrm{d}k}\sigma &= \omega^2\beta\sum_{i={1}}^N{\Bigl[}\frac{\mathrm{d}}{\mathrm{d}k}\mathrm{Var}_k(r_i)+2\langle r_i\rangle_k\frac{\mathrm{d}}{\mathrm{d}k}\langle r_i\rangle_k{\Bigr]}\\
        &= -\frac{\omega^2\beta^2}{2}\sum_{i={1}}^N [\mathrm{Cov}_k((r_i-\left\langle r_i \right \rangle_k)^2 ,\sum_{l<j}^Nc_{lj}(r_l-r_j)^2) \\
        & \quad\quad\quad + 2\left\langle r_i \right \rangle_k\mathrm{Cov}_k(r_i,\sum_{l<j}^Nc_{lj}(r_l-r_j)^2)].
    \end{split}
\end{equation}
{At $k=0$, the uncoupled radii are independent.} We define $R_i:=r_i-\langle r\rangle_0$ so that, {for $l\ne i$,}
\begin{equation}
\begin{split}
    \mathrm{Cov}_0(R_i^2,(r_i-r_l)^2)&=\mathrm{Var}_0(R_i^2),\\
    \mathrm{Cov}_0(r_i,(r_i-r_l)^2)&=\langle R_i^3\rangle_0=\mu_3,
\end{split}
\end{equation}
where we used that $\langle R_i\rangle_0=0$. {Only edges incident on oscillator $i$ contribute to the covariance sums, giving}
\begin{equation}
\begin{split}
    {\left.\frac{\mathrm{d}\langle r_i\rangle_k}{\mathrm{d}k}\right|_{k=0}
    =-\frac{\beta s_i}{2}}\mu_3,\\
    {\left.\frac{\mathrm{d}\,\mathrm{Var}_k(r_i)}{\mathrm{d}k}\right|_{k=0}
    =-\frac{\beta s_i}{2}}\mathrm{Var}_0(R_i^2),
\end{split}
\end{equation}
where ${s_i}=\sum_jc_{ij}$ is the weighted degree of oscillator $i$. Therefore, bringing these equations together
\begin{equation}
\begin{split}
    \left.\frac{\mathrm{d}\sigma}{\mathrm{d}k}\right|_{k=0}
    &=-\omega^2\beta^2{W}\bigl[\mathrm{Var}_0((r-\langle r\rangle_0)^2)\\
    &\qquad\qquad+2\langle r\rangle_0\mu_3\bigr].
\end{split}
\end{equation}
{$W=\sum_{i<j}c_{ij}=\frac12\sum_i s_i$ is the total edge weight.}

\subsubsection{Strong-coupling limit}\label{app: radial-strong}
\begingroup
For fixed finite $N$, consider radial coupling with $k_{ij}=kc_{ij}$, where the fixed symmetric nonnegative weights form a connected graph. The stationary density is
\begin{equation}
p(\{r_i\})=\frac{1}{\mathcal Z}\prod_{i=1}^N p_0(r_i)
\exp\!\left[-\frac{\beta k}{2}\sum_{i<j}c_{ij}(r_i-r_j)^2\right].
\end{equation}
As $k\to\infty$, the relative radial modes vanish and all radii approach a common value $r$. Integrating out the relative modes leaves the factor $r^N$ from the product radial measure. The common-radius density with respect to $\mathrm{d}r$ is therefore
\begin{equation}\label{app eq: radial common density}
q_\infty(r)=\frac{r^N}{\mathcal I_{N-1}(N\beta)}
\exp\!\left[N\beta\int_0^r F(s)\,\mathrm{d}s\right].
\end{equation}
The graph-dependent factors from integrating the relative modes cancel in the normalization. Hence
\begin{equation}
\begin{split}
\sigma(k\to\infty)
&=\frac{N\omega^2}{T_\mathrm{eff}}\int_0^\infty r^2q_\infty(r)\,\mathrm{d}r\\
&=\frac{N\omega^2}{T_\mathrm{eff}}
\frac{\mathcal I_{N+1}(N\beta)}{\mathcal I_{N-1}(N\beta)}.
\end{split}
\end{equation}

\endgroup\vspace{0pt}

\subsubsection{Distribution under coupling}\label{app: dist radial}

\begin{figure}
    \centering
    \includegraphics[bb=0 0 351.00000000 283.00000000,width=1\linewidth]{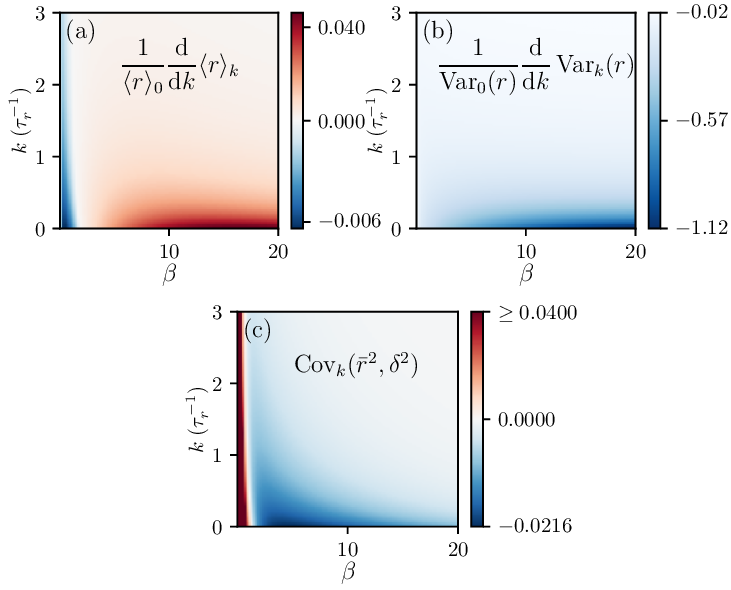}
    \caption{Radial distribution deformation under radial coupling. (a) Heat map of the derivative of the average radius $\langle r\rangle_k$ (in units of $\langle r\rangle_0$) with respect to coupling strength $k$ versus $k$ and the inverse effective temperature $\beta$. (b) Heat map of the derivative of the radial variance $\mathrm{Var}_k(r)$ (in units of $\mathrm{Var}_0(r)$) with respect to $k$ versus $k$ and $\beta$. (c) The covariance {$\mathrm{Cov}_k(\bar r^2,\delta^2)$} (Eq.~(\ref{eq:17})) versus $k$ and $\beta$.}
    \label{fig:app 1}
\end{figure}

Figure~\ref{fig:app 1}(a) shows that, for $\beta\gtrsim1.8$, $\langle r\rangle_k$ increases with coupling strength. For $\beta\lesssim1.8$, $\langle r\rangle_k$ initially decreases then increases with coupling strength. Over the displayed parameter range, the radial variance $\mathrm{Var}_k(r)$ {decreases with $k$, with its derivative approaching zero} (Fig.~\ref{fig:app 1}(b)).

The term $\mathrm{Cov}_k(\bar{r}^2,\delta^2)$ contributes to the EPR gradient in Eq.~(\ref{eq:17}). Coupling suppresses large values of $\delta^2$, so $\mathrm{Cov}_k(\bar{r}^2,\delta^2)>0$ and $\mathrm{Cov}_k(\bar{r}^2,\delta^2)<0$ correspond to ${\langle}\bar{r}^2{\rangle_k}$ decreasing and increasing, respectively. Figure~\ref{fig:app 1}(a, c) shows the sign of $\mathrm{Cov}_k(\bar{r}^2,\delta^2)$ reflects the average radius $\langle r\rangle_k$ increasing or decreasing with $k$.

\subsection{Phase coupling}
\subsubsection{Entropy production rate}\label{app: EPR phase}
{For brevity, define the local mean velocity $v_\alpha:=J_\alpha/p=j_\alpha/p_\varphi$, where $p=p_0(r_1)p_0(r_2)p_\varphi$ and $j_\alpha=J_\alpha/[p_0(r_1)p_0(r_2)]$. For the pair, $\langle\cdot\rangle_0$ denotes the radial average and $\langle\cdot\rangle_\varphi$ is the conditional phase average, so that $\langle\cdot\rangle_k=\langle\langle\cdot\rangle_\varphi\rangle_0$.
}

\paragraph{Mean local velocity decomposition.}
{In the phase and radial coordinates $\{\theta_i,r_i\}$, the total EPR reads
\begin{equation}
    \sigma=\frac{1}{\Teff}\sum_{i=1}^2
    \left[\langle r_i^2v_{\theta_i}^2\rangle_k+\langle v_{r_i}^2\rangle_k\right],
\end{equation}
where
\begin{equation}
\begin{split}
    v_{r_i}&=-\Teff\partial_{r_i}\ln p_\varphi,\\
    v_{\theta_i}&=-\omega_i+k\sin(\theta_j-\theta_i)
    +(-1)^{i+1}\frac{\Teff}{r_i^2}\partial_\varphi\ln p_\varphi.
\end{split}
\end{equation}
Thus $\Teff^{-1}\langle v_{r_i}^2\rangle_k=\Teff\langle(\partial_{r_i}\ln p_\varphi)^2\rangle_k=\sigma_{r_i\varphi}$. Defining $\Omega_i=\langle r_i^2v_{\theta_i}\rangle_k/\langle r^2\rangle_0$, we have
\begin{equation}
    \langle r_i^2v_{\theta_i}^2\rangle_k
    =\langle r^2\rangle_0\Omega_i^2
    +\langle r_i^2(v_{\theta_i}-\Omega_i)^2\rangle_k.
\end{equation}
The conditional stationary equation is symmetric under exchange of $r_1, r_2$ at fixed $\varphi$. Therefore $\langle(r_1^2-r_2^2)\sin\varphi\rangle_k=0$, and $\Omega_1+\Omega_2=-2\bar\omega$. Splitting $\Omega_1^2+\Omega_2^2$ into its mean and difference contributions then gives Eq.~(\ref{eq: EPR mean-variance}).
}

\paragraph{Center-of-mass/relative-coordinate decomposition.}
{Writing $a=r_1^2+r_2^2$ and $b=(r_2^2-r_1^2)/2$, the angular entries of the inverse diffusion matrix are
\begin{equation}
\begin{split}
    (D^{-1})_{\psi\psi}&=\frac{a}{\Teff},\qquad
    (D^{-1})_{\psi\varphi}=\frac{b}{\Teff},\\
    (D^{-1})_{\varphi\varphi}&=\frac{a}{4\Teff}.
\end{split}
\end{equation}
\leavevmode{} The angular local mean velocities are $v_\psi=A_\psi-D_{\psi\varphi}\partial_\varphi\ln p_\varphi$ and $v_\varphi=A_\varphi-D_{\varphi\varphi}\partial_\varphi\ln p_\varphi$, where $A_\psi=-\bar\omega$ and $D_{\psi\varphi}=(\Teff/2)(r_2^{-2}-r_1^{-2})$. The relative-phase coefficients $A_\varphi$ and $D_{\varphi\varphi}$ are given below Eq.~(\ref{eq:78}). Retaining the mixed-current term gives
\begin{equation}\label{app eq: epr rel com}
\begin{split}
    \sigma&=\sum_{i=1}^2\sigma_{r_i\varphi}
    +\frac{1}{\Teff}\left\langle
    a v_\psi^2+\frac{a}{4}v_\varphi^2+2b v_\psi v_\varphi
    \right\rangle_k\\
    &=\sigma_\psi+\sum_{i=1}^2\sigma_{r_i\varphi}
    +\left\langle(D^{-1})_{\psi\psi}
    \frac{D_{\psi\varphi}^2}{D_{\varphi\varphi}^2}
    \langle A_\varphi^2\rangle_\varphi\right\rangle_0\\
    &\quad+\left\langle\frac{v_\varphi^2}{D_{\varphi\varphi}}\right\rangle_k\\
    &=\sigma_\psi+\sum_{i=1}^2\sigma_{r_i\varphi}
    +\sigma_{\psi\varphi}+\sigma_\varphi.
\end{split}
\end{equation}
The second equality follows by completing the square and the term proportional to $bA_\psi A_\varphi$ vanishes under radial exchange. Consequently, $\sigma_{\Delta\Omega}+\sigma_{\mathrm{var}}=\sigma_\varphi+\sigma_{\psi\varphi}$.
}

\paragraph{EPR in terms of observables.}
{In polar coordinates, a conservative interaction potential $V$ contributes $-\partial_{r_i}V$ to the radial drift. Its tangential drift is $-r_i^{-1}\partial_{\theta_i}V$, giving an angular drift of $-r_i^{-2}\partial_{\theta_i}V$ upon division by $r_i$. Since the prescribed coupling leaves every radial drift unchanged, $V$ would have to be independent of all radii. Such a potential cannot generate the radius-independent sine coupling when the radii fluctuate.
For phase coupling, the combined rotational and coupling drift of oscillator $i$ is
$(\boldsymbol f_{\mathrm{rot}}+\boldsymbol f_{\mathrm c})_i=r_iA_{\theta_i}\boldsymbol e_{\theta_i}$,
where $\boldsymbol e_{\theta_i}=(-\sin\theta_i,\cos\theta_i)$ is the unit vector in the direction of increasing $\theta_i$.
The coupling force has divergence $\nabla\cdot\boldsymbol f_{\mathrm c}=\sum_{i=1}^2\partial_{\theta_i}A_{\theta_i}=-2k\cos\varphi$.
The stationary EPR identity of Sec.~\ref{sec: conservative interaction} therefore gives
\begin{equation}\label{eq: radial-phase EPR}
\begin{split}
    \sigma&=\frac{1}{\Teff}\left\langle
    r_1^2A_{\theta_1}^2+r_2^2A_{\theta_2}^2\right\rangle_k
    -2k\langle\cos\varphi\rangle_k\\
    &=\sigma(0)+\frac{\Delta\omega k}{\Teff}
    \langle a\langle\sin\varphi\rangle_\varphi\rangle_0\\
    &\quad+\frac{k^2}{\Teff}
    \langle a\langle\sin^2\varphi\rangle_\varphi\rangle_0
    -2k\langle\cos\varphi\rangle_k,
\end{split}
\end{equation}
where $A_{\theta_1}=-\omega_1+k\sin\varphi$ and $A_{\theta_2}=-\omega_2-k\sin\varphi$. The second equality uses symmetry under exchange of $r_1$ and $r_2$. This expression will be useful in the following derivations.
}

\paragraph{Fisher term balance.}
{Multiplying the full stationary equation, Eq.~(\ref{eq:78}), by $\ln p_\varphi$, integrating over the relative phase and taking the radial average, then integrating by parts, gives
\begin{equation}\label{app eq: fisher balance}
\begin{split}
    &\sum_{i=1}^2\sigma_{r_i\varphi}
    +\left\langle D_{\varphi\varphi}
    \langle(\partial_\varphi\ln p_\varphi)^2\rangle_\varphi\right\rangle_0\\
    &\qquad=-\langle\partial_\varphi A_\varphi\rangle_k
    =2k\langle\cos\varphi\rangle_k.
\end{split}
\end{equation}
The boundary terms vanish by phase periodicity and the stationary radial boundary conditions. Since both Fisher contributions are nonnegative,
\begin{equation}
    0\leq\sum_{i=1}^2\sigma_{r_i\varphi}
    \leq2k\langle\cos\varphi\rangle_k\leq2k
\end{equation}
for $k\geq0$.  The same bound applies to the phase Fisher term weighted by $D_{\varphi\varphi}$.
}

\subsubsection{Weak coupling perturbation}\label{app: phase weak coupling}
{The conditional stationary equation reads
\begin{equation}\label{eq: radial-phase current balance}
    0=\mathcal L_0p_\varphi+2k\partial_\varphi(\sin\varphi\,p_\varphi),
\end{equation}
where $\mathcal L_0:=\mathcal L_r+\Delta\omega\partial_\varphi+D_{\varphi\varphi}\partial_\varphi^2$. Expanding
\begin{equation}\label{eq: p phi fourier}
    p_\varphi=\frac{1}{2\pi}\sum_{\ell=-\infty}^{\infty}
    c_\ell(r_1,r_2)e^{\mathrm{i}\ell\varphi},
\end{equation}
normalization and reality of $p_\varphi$ give $c_0=1$ and $c_{-\ell}=c_\ell^*$. Substitution gives the Fourier recurrence
\begin{equation}
    (\mathcal L_r-\ell^2D_{\varphi\varphi}+\mathrm{i}\ell\Delta\omega)c_\ell
    +k\ell(c_{\ell-1}-c_{\ell+1})=0.
\end{equation}
Expanding the Fourier coefficients as $c_\ell=\delta_{\ell0}+k c_\ell^{(1)}+\mathcal O(k^2)$ and collecting terms of order $k$ shows that only the modes $\ell=\pm1$ have a nonzero source term. The other nonzero modes satisfy homogeneous equations whose solutions vanish under the stationary boundary conditions, while normalization gives $c_0^{(1)}=0$. Setting $h=c_1^{(1)}$ and using $c_{-1}^{(1)}=h^*$, we obtain
\begin{equation}\label{eq: perturb p phi}
\begin{split}
    p_\varphi&=\frac{1}{2\pi}
    \left[1+2k\mathrm{Re}\,[h e^{\mathrm{i}\varphi}]\right]
    +\mathcal{O}(k^2),\\
    &(D_{\varphi\varphi}-\mathcal L_r-\mathrm{i}\Delta\omega)h=1.
\end{split}
\end{equation}
Thus
\begin{equation}
\begin{split}
    \langle\cos\varphi\rangle_\varphi&=k\mathrm{Re}\,h+\mathcal{O}(k^3),\\
    \langle\sin\varphi\rangle_\varphi&=-k\mathrm{Im}\,h+\mathcal{O}(k^3),\\
    \langle\sin^2\varphi\rangle_\varphi&=\frac12+\mathcal{O}(k^2).
\end{split}
\end{equation}
Substituting into Eq.~(\ref{eq: radial-phase EPR}) gives $\sigma(k)=\sigma(0)+C_2k^2+\mathcal{O}(k^4)$, with
\begin{equation}
    C_2=\frac{\langle r^2\rangle_0}{\Teff}
    -2\langle\mathrm{Re}\,h\rangle_0
    -\frac{\Delta\omega}{\Teff}\langle a\,\mathrm{Im}\,h\rangle_0,
\end{equation}
as in Eq.~(\ref{eq: C_2 pair}).
}

{We can interpret this expansion using Eq.~(\ref{eq: EPR mean-variance}). The weighted mean rates satisfy $\Omega_i=-\omega_i\mp k^2\langle r_i^2\mathrm{Im}\,h\rangle_0/\langle r^2\rangle_0+\mathcal{O}(k^4)$, with the upper sign for $i=1$. By radial-exchange symmetry of $h$, this gives
\begin{equation}
\begin{split}
    \sigma_\psi+\sigma_{\Delta\Omega}
    &=\sigma(0)-\underbrace{\frac{\Delta\omega}{\Teff}
    \langle a\,\mathrm{Im}\,h\rangle_0}_{=:\mathcal S}k^2+\mathcal{O}(k^4),\\
    \sigma_{\mathrm{var}}+\sum_{i=1}^2\sigma_{r_i\varphi}
    &=\underbrace{\left[\frac{\langle r^2\rangle_0}{\Teff}
    -2\langle\mathrm{Re}\,h\rangle_0\right]}_{=:\mathcal F}k^2+\mathcal{O}(k^4).
\end{split}
\end{equation}
The coefficient in the second line is nonnegative, since these contributions are nonnegative and vanish at zero coupling. This establishes $\mathcal F\geq0$ in Eq.~(\ref{eq: C_2 pair}).
}

{To establish $\mathcal S\geq0$, write $H:=D_{\varphi\varphi}-\mathcal L_r$, so that $(H-\mathrm{i}\Delta\omega)h=1$. Expanding $\mathcal L_r|h|^2$ and using the imaginary part of this equation yields
\begin{equation}
\begin{split}
    H&\left(2\Delta\omega\,\mathrm{Im}\,h-(\Delta\omega)^2|h|^2\right)\\
    &=(\Delta\omega)^2\left[D_{\varphi\varphi}|h|^2
    +2\Teff\sum_{i=1}^2|\partial_{r_i}h|^2\right]\geq0.
\end{split}
\end{equation}
Multiplying by the negative part of the expression on which $H$ acts and integrating over the stationary radial measure shows that this negative part must vanish. Consequently,
\begin{equation}
\begin{split}
    \Delta\omega\,\mathrm{Im}\,h&\geq\frac{(\Delta\omega)^2}{2}|h|^2\geq0,\\
    \Delta\omega\langle a\,\mathrm{Im}\,h\rangle_0
    &\geq\frac{(\Delta\omega)^2}{2}\langle a|h|^2\rangle_0\geq0,
\end{split}
\end{equation}
which establishes $\mathcal S\geq0$.
}

\subsubsection{Strong coupling limit}\label{app sec: strong coupling pair}
{Multiplying Eq.~(\ref{eq: radial-phase current balance}) by $\cos\varphi$ and integrating over $\varphi$ gives
\begin{equation}\label{eq: integrated pde}
\begin{split}
    0&=\mathcal L_r\langle\cos\varphi\rangle_\varphi
    +\Delta\omega\langle\sin\varphi\rangle_\varphi
    +2k\langle\sin^2\varphi\rangle_\varphi\\
    &\quad-D_{\varphi\varphi}\langle\cos\varphi\rangle_\varphi.
\end{split}
\end{equation}
By detailed balance of the radial dynamics, $\langle f\mathcal L_rg\rangle_0=\langle(\mathcal L_rf)g\rangle_0$. Substituting into Eq.~(\ref{eq: radial-phase EPR}) and using $aD_{\varphi\varphi}/\Teff-4=(r_1/r_2-r_2/r_1)^2$, we find
\begin{equation}
\begin{split}
    \sigma(k)-\sigma(0)
    &=\frac{k}{2}\left\langle
    \left(\frac{r_1}{r_2}-\frac{r_2}{r_1}\right)^2
    \langle\cos\varphi\rangle_\varphi\right\rangle_0\\
    &\quad+\frac{\Delta\omega k}{2\Teff}
    \langle a\langle\sin\varphi\rangle_\varphi\rangle_0\\
    &\quad-\frac{k}{2\Teff}
    \langle(\mathcal L_ra)\langle\cos\varphi\rangle_\varphi\rangle_0.
\end{split}
\end{equation}
At fixed positive radii, $\langle\cos\varphi\rangle_\varphi\to1$ and $\langle\sin\varphi\rangle_\varphi\to0$. Since $\langle\mathcal L_ra\rangle_0=0$, dominated convergence implies that the last two terms are $o(k)$. The first term requires more care: replacing the conditional cosine by unity gives a divergent inverse-radius moment. For $r_1\to0$ with $r_2$ of order one, the integrand is proportional to $p_0(0)r_2^2/r_1$, since $r_1p_0(r_1)=r_1p_0(0)+\mathcal{O}(r_1^{n+2})$.
}

{The condition $D_{\varphi\varphi}\ll k$ in this region requires
\begin{equation}
    r_1\gg\sqrt{\frac{\Teff}{k}}=: \varepsilon_k.
\end{equation}
For $r_2$ of order one, the outer solution describes the region $r_1\gg\varepsilon_k$, where phase relaxation is fast compared with radial diffusion. The inner solution resolves the small-radius region $r_1=\varepsilon_kR$, with $R=\mathcal{O}(1)$, where these timescales are comparable. Since $F(r)=r^n/m+\mathcal{O}(r^{m+n})$ as $r\to0$, the deterministic radial drift is subleading on the scale $r_1=\varepsilon_kR$. The leading equation for the inner solution $p_\varphi^{\mathrm{in}}(R,\varphi)$ therefore reads
\begin{equation}\label{app eq: U dynamics}
    0=\left(\partial_R^2+\frac1R\partial_R\right)p_\varphi^{\mathrm{in}}
    +\partial_\varphi[2\sin\varphi\,p_\varphi^{\mathrm{in}}]
    +\frac1{R^2}\partial_\varphi^2p_\varphi^{\mathrm{in}}.
\end{equation}
The inner and outer solutions must agree in the overlap region $\varepsilon_k\ll r_1\ll1$. Defining $U(R):=\int_0^{2\pi}\cos\varphi\,p_\varphi^{\mathrm{in}}\,\mathrm{d}\varphi$, its moment equation reads
\begin{equation}
    U''+\frac1R U'-\frac1{R^2}U+1-\langle\cos2\varphi\rangle_{\mathrm{in}}=0.
\end{equation}
As $R\to0$, $p_\varphi^{\mathrm{in}}\to1/(2\pi)$, one can find $U(R)=\alpha R-R^2/3+\ldots=\mathcal{O}(R)$, where $\alpha$ is finite. Thus $U(R)/R$ is integrable at the origin. For $R\gg1$, the phase is confined around $\varphi=0$, with $\langle\varphi^2\rangle_{\mathrm{in}}=\mathcal{O}(R^{-2})$, so that $U(R)=1+\mathcal{O}(R^{-2})$. These two limits regularize the small-radius divergence and determine the logarithmic contribution.
}

\begin{figure}
    \centering
    \includegraphics[bb=0 0 375.00000000 274.00000000,width=.9\linewidth]{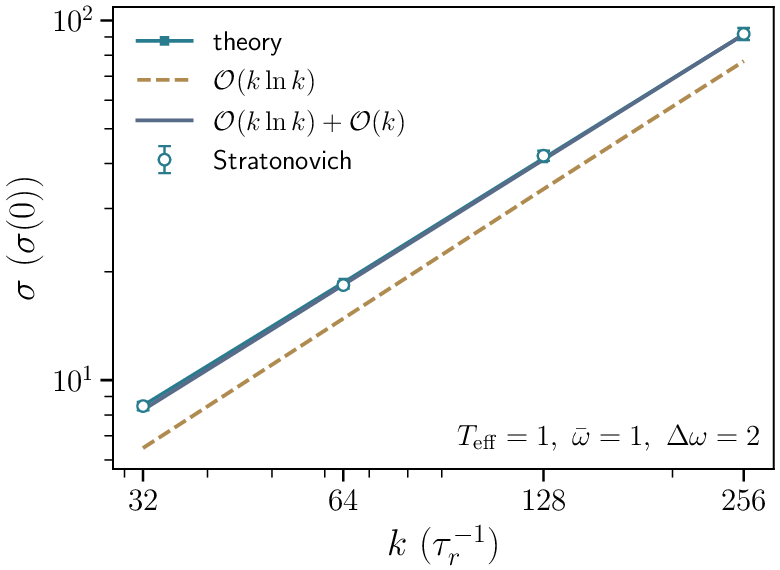}
    \caption{{Strong-coupling EPR for two phase-coupled oscillators at $T_\mathrm{eff}=1$, $\bar\omega=1\tau_r^{-1}$, and $\Delta\omega=2\tau_r^{-1}$, in units of the uncoupled value $\sigma(0)$ at the same intrinsic frequencies and effective temperature. Stratonovich estimates (open circles) are compared with EPR evaluated from numerical solutions of the stationary PDE (squares) and the leading and two-term approximations in Eq.~(\ref{app eq: strong coupling k}). Both approximation curves retain the constant contribution $\sigma_\psi$. Error bars show one standard error.}}
    \label{fig app: epr strong}
\end{figure}

\paragraph{Matching and subleading correction.}
{Writing $u_k(r_1,r_2):=\langle\cos\varphi\rangle_\varphi$, radial-exchange symmetry gives
\begin{equation}
\begin{split}
    \sigma(k)&=k[I_k-\langle u_k\rangle_0]+o(k),\\
    I_k&:=\left\langle\frac{r_2^2}{r_1^2}u_k\right\rangle_0.
\end{split}
\end{equation}
We split the radial integral at $\varepsilon_k\ll\delta\ll1$. The remainders below refer to the matched limit $\varepsilon_k/\delta\to0$ and $\delta\to0$. In the outer region $u_k\to1$, while in the inner region $u_k\approx U(r_1/\varepsilon_k)$, giving
\begin{equation}
\begin{split}
    I_k^{\mathrm{out}}&=\langle r^2\rangle_0
    [-p_0(0)\ln\delta+\mathcal H]+o(1),\\
    I_k^{\mathrm{in}}&=p_0(0)\langle r^2\rangle_0
    \left[\ln\frac{\delta}{\varepsilon_k}+\mathcal C\right]+o(1),
\end{split}
\end{equation}
where
\begin{equation}
\begin{split}
    \mathcal H&=\int_0^1\frac{p_0(r)-p_0(0)}{r}\,\mathrm{d}r
    +\int_1^\infty\frac{p_0(r)}{r}\,\mathrm{d}r,\\
    \mathcal C&=\int_0^1\frac{U(R)}{R}\,\mathrm{d}R
    +\int_1^\infty\frac{U(R)-1}{R}\,\mathrm{d}R.
\end{split}
\end{equation}
The integrals defining $\mathcal C$ converge by the small- and large-$R$ behavior above. Solving the inner equation using Fourier modes on a logarithmic radial grid gives $\mathcal C\approx0.58592$. Numerical refinement checks and code are provided in Ref.~\cite{burnet2026revisioncode}. The matching scale $\delta$ cancels, while $\ln\varepsilon_k^{-1}=\tfrac12\ln(k/\Teff)$. Since $\langle u_k\rangle_0\to1$, the matched expansion yields
\begin{equation}\label{app eq: strong coupling k}
\begin{split}
    \sigma(k)&=\frac{\langle r^2\rangle_0p_0(0)}{2}
    k\ln\frac{k}{\Teff}\\
    &\quad+k\left[\langle r^2\rangle_0
    \bigl(\mathcal H+p_0(0)\mathcal C\bigr)-1\right]+o(k).
\end{split}
\end{equation}
The logarithmic contribution accumulates over $\varepsilon_k\ll r_i\ll1$, while the inner region supplies a finite correction to the radial integral. For moderate-to-strong coupling, the linear correction can be significant, as shown in Fig.~\ref{fig app: epr strong}.
}

\paragraph{Phase-Fisher form of the leading contribution.}
{The angular contribution $\sigma_{\psi\varphi}$ in Eq.~(\ref{app eq: epr rel com}) reads
\begin{equation}\label{app eq: phase weighted current}
\sigma_{\psi\varphi}
=\left\langle(D^{-1})_{\psi\psi}
\frac{D_{\psi\varphi}^2}{D_{\varphi\varphi}^2}
\langle A_\varphi^2\rangle_\varphi\right\rangle_0.
\end{equation}
In the locked outer region $r_i\gg\varepsilon_k$, the leading angular drift balances phase diffusion, $A_\varphi\approx D_{\varphi\varphi}\partial_\varphi\ln p_\varphi$. Its conditional integrand therefore becomes a radially weighted Fisher information for phase shifts (Appendix~\ref{app:FI}). In the outer region, radial evolution is subleading on the phase relaxation timescale, so the linearized angular drift-diffusion balance gives a Gaussian conditional phase peak with variance $D_{\varphi\varphi}/(2k)$ to leading order. Writing $I_\varphi:=\langle(\partial_\varphi\ln p_\varphi)^2\rangle_\varphi$, this gives $I_\varphi\approx2k/D_{\varphi\varphi}$. Hence
\begin{equation}\label{app eq: phase weighted fisher}
\begin{split}
(D^{-1})_{\psi\psi}D_{\psi\varphi}^2 I_\varphi
&\approx2k(D^{-1})_{\psi\psi}
\frac{D_{\psi\varphi}^2}{D_{\varphi\varphi}}\\
&=\frac{k}{2}\left(\frac{r_1}{r_2}-\frac{r_2}{r_1}\right)^2.
\end{split}
\end{equation}
At fixed radii, stronger phase confinement increases this contribution linearly with $k$. The radial factor $(r_1/r_2-r_2/r_1)^2$ grows when one radius becomes small relative to the other. The radial expectation thereof then produces the logarithmic integral cut off at $\varepsilon_k$, yielding the $k\ln k$ growth in Eq.~(\ref{app eq: strong coupling k}).
}

{To establish that $\sigma_{\psi\varphi}$ carries the leading EPR growth, we bound the remaining terms in Eq.~(\ref{app eq: epr rel com}). Put $f:=1/D_{\varphi\varphi}$ and $s_k:=\langle\sin\varphi\rangle_\varphi$. Combining the Fisher balance with the cosine moment equation, Eq.~(\ref{eq: integrated pde}), gives
\begin{equation}\label{app eq: phase remainder bound}
\begin{split}
\sigma_\varphi+\sum_i\sigma_{r_i\varphi}
&=(\Delta\omega)^2\langle f\rangle_0
+2\Delta\omega k\langle f s_k\rangle_0\\
&\quad-2k\langle(\mathcal L_rf)u_k\rangle_0.
\end{split}
\end{equation}
The radial averages $\langle f\rangle_0$ and $\langle|\mathcal L_rf|\rangle_0$ are finite and independent of $k$. Since $|s_k|,|u_k|\leq1$, the right-hand side of Eq.~(\ref{app eq: phase remainder bound}) is $\mathcal{O}(k)$. Moreover, $\sigma_\psi$ is independent of coupling and $\sigma_{\Delta\Omega}\to0$ as $k\to\infty$, while the radial Fisher sum is bounded by $2k$. Equations~(\ref{eq: EPR mean-variance}) and (\ref{app eq: epr rel com}) therefore give
\begin{equation}\label{app eq: phase leading contributions}
\sigma=\sigma_{\psi\varphi}+\mathcal{O}(k)
=\sigma_{\mathrm{var}}+\mathcal{O}(k).
\end{equation}
Thus the leading strong-coupling dissipation admits a radially weighted phase-Fisher interpretation.
}

\subsubsection{N coupled systems}\label{app: EPR phase N}
{The radial dynamics remain autonomous, so the stationary distribution factorizes as $P_k=P_0q_k$, where $P_0=\prod_i p_0(r_i)$ and $\int_{[0,2\pi)^N}q_k\,\mathrm{d}^N\theta=1$. $\langle\cdot\rangle_0$ denotes the $N$-radius average with measure $\prod_i r_i p_0(r_i)\,\mathrm{d}r_i$. Let $D_i=\Teff/r_i^2$ and $b_i=-\omega_i+kg_i$, with $g_i=\sum_jc_{ij}\sin(\theta_j-\theta_i)$ and $c_{ij}=c_{ji}$. The currents are
\begin{equation}
\begin{split}\label{eq: N currents}
    J_{r_i}&=-\Teff P_k\partial_{r_i}\ln q_k,\\
    J_{\theta_i}&=P_k\left(b_i-D_i\partial_{\theta_i}\ln q_k\right).
\end{split}
\end{equation}
Multiplying the stationary equation for $q_k$ in Sec.~\ref{sec: 4}\,B by $\ln q_k$,
averaging over the radii, and integrating by parts in the radial and angular coordinates gives
\begin{equation}
\begin{split}
    &\Teff\sum_{i=1}^N\langle(\partial_{r_i}\ln q_k)^2\rangle_k
    +\sum_{i=1}^N\langle D_i(\partial_{\theta_i}\ln q_k)^2\rangle_k\\
    &\qquad=-\sum_{i=1}^N\langle\partial_{\theta_i}b_i\rangle_k
    =2k\sum_{i<j}c_{ij}\langle\cos(\theta_j-\theta_i)\rangle_k.
\end{split}
\end{equation}
Applying the stationary EPR identity of Sec.~\ref{sec: conservative interaction} to the angular drifts $b_i$ gives
$\sigma_N=T_{\mathrm{eff}}^{-1}\sum_i\langle r_i^2 b_i^2\rangle_k+\sum_i\langle\partial_{\theta_i}b_i\rangle_k$.
Substituting $b_i=-\omega_i+kg_i$ yields
\begin{equation}\label{eq: EPR phase expanded}
\begin{split}
    \sigma_N&=\sum_{i=1}^N\frac{\omega_i^2}{\Teff}\langle r^2\rangle_0
    -\frac{2k}{\Teff}\sum_{i=1}^N\omega_i\langle r_i^2g_i\rangle_k\\
    &\quad+\frac{k^2}{\Teff}\sum_{i=1}^N\langle r_i^2g_i^2\rangle_k
    -2k\sum_{i<j}c_{ij}\langle\cos(\theta_j-\theta_i)\rangle_k.
\end{split}
\end{equation}
This expression will be of use for considering the weak and strong coupling limits.
\newline
The weighted mean-variance decomposition of Eq.~(\ref{eq: EPR mean-variance N}) follows from direct substitution of the currents in Eq.~(\ref{eq: N currents}) into steady-state EPR expression in Eq.~(\ref{eq: epr}) and using $\sum_i\Omega_i^2=N\bar\Omega^2+N^{-1}\sum_{i<j}(\Omega_i-\Omega_j)^2$.
}

\paragraph{Weak coupling.}
{Expanding $q_k=(2\pi)^{-N}(1+kf_1)+\mathcal{O}(k^2)$ gives
\begin{equation}\label{eq: N phase first order}
    \mathcal L_0f_1=-2\sum_{i<j}c_{ij}\cos(\theta_j-\theta_i),
\end{equation}
where $\mathcal L_0=\sum_i[\mathcal L_{r_i}+\omega_i\partial_{\theta_i}+D_i\partial_{\theta_i}^2]$. Each pair is governed by the same first-harmonic response equation as above, with $\Delta_{ij}=\omega_j-\omega_i$ and $D_{ij}=D_i+D_j$. Thus, writing $h_{\Delta_{ij}}=u_{\Delta_{ij}}+\mathrm{i}w_{\Delta_{ij}}$, linearity gives
\begin{equation}
    f_1=2\sum_{i<j}c_{ij}\left[
    u_{\Delta_{ij}}\cos\varphi_{ij}-w_{\Delta_{ij}}\sin\varphi_{ij}\right],
\end{equation}
where $\varphi_{ij}=\theta_j-\theta_i$. The normalization $\int f_1\,\mathrm{d}^N\theta=0$ fixes the additive constant to zero and uniqueness follows by the same nonnegative diffusion argument as for the pair.
\mbox{}}

{For a purely radial function $g$, the first-order averages are
\begin{equation}
\begin{split}
    \langle g\cos\varphi_{ij}\rangle_k
    &=kc_{ij}\langle g u_{\Delta_{ij}}\rangle_0+\mathcal{O}(k^2),\\
    \langle g\sin\varphi_{ij}\rangle_k
    &=-kc_{ij}\langle g w_{\Delta_{ij}}\rangle_0+\mathcal{O}(k^2).
\end{split}
\end{equation}
Distinct pair harmonics are orthogonal under the uncoupled phase average, so the mixed-edge terms do not contribute at this order. Substituting into Eq.~(\ref{eq: EPR phase expanded}) and applying the pair calculation to each edge gives
\begin{equation}
    \sigma_N(k)-\sigma_N(0)
    =k^2\sum_{i<j}c_{ij}^2C_2(\Delta_{ij})+\mathcal{O}(k^3),
\end{equation}
where $C_2(\Delta_{ij})$ is the pair coefficient in Eq.~(\ref{eq: C_2 pair}), evaluated at detuning $\Delta_{ij}$.
}

{The mean-rate and fluctuation contributions follow in the same way from $\Omega_i=-\omega_i+k\langle r_i^2g_i\rangle_k/\langle r^2\rangle_0$:
\begin{equation}
\begin{split}
    \frac{\langle r^2\rangle_0}{\Teff}\sum_{i=1}^N\Omega_i^2
    &=\sigma_N(0)-\frac{k^2}{\Teff}\sum_{i<j}c_{ij}^2\Delta_{ij}\\
    &\quad\times\langle(r_i^2+r_j^2)w_{\Delta_{ij}}\rangle_0+\mathcal{O}(k^3),\\
    \sigma_{\mathrm{var},N}+\sum_{i=1}^N&\sigma_{r_i,\{\theta_j\}}
    =k^2\sum_{i<j}c_{ij}^2\left[
    \frac{\langle r^2\rangle_0}{\Teff}-2\langle u_{\Delta_{ij}}\rangle_0\right]\\
    &\quad+\mathcal{O}(k^3).
\end{split}
\end{equation}
The signs of the pair contributions follow from the two-oscillator argument above.
}

\paragraph{Strong-coupling limit.}\label{app sec: strong coupling N}
{We specialize to unnormalized all-to-all coupling, $c_{ij}=1$ for $i\neq j$. Since $g_i$ is bounded, Eq.~(\ref{eq: EPR phase expanded}) gives
\begin{equation}
    \sigma_N(k)=\frac{k^2}{\Teff}\sum_{i=1}^N
    \langle r_i^2g_i^2\rangle_k+\mathcal{O}(k).
\end{equation}
For fluctuations relative to the other phases, $\phi_i=\theta_i-(N-1)^{-1}\sum_{j\neq i}\theta_j$, under strong coupling the small-angle approximation gives $g_i\approx-(N-1)\phi_i$ and
\begin{equation}\label{app eq: fluctuations phase}
\begin{split}
    \dot\phi_i&\approx-\delta\bar\omega_i-kN\phi_i\\
    &\quad+\sqrt{2\Teff}\left[
    \frac{\eta_{\theta_i}}{r_i}-\frac1{N-1}\sum_{j\neq i}\frac{\eta_{\theta_j}}{r_j}\right],
\end{split}
\end{equation}
where $\delta\bar\omega_i=\omega_i-(N-1)^{-1}\sum_{j\neq i}\omega_j$. In the outer region $r_j\gg\varepsilon_k=\sqrt{\Teff/(kN)}$, phase relaxation is fast compared with radial diffusion. The radii can then be treated as frozen to leading order, yielding
\begin{equation}
    \mathrm{Var}(\phi_i\mid\{r_j\})\approx\frac{D_{\phi_i}}{kN}
    =\frac{\Teff}{kN}\left[
    \frac1{r_i^2}+\frac1{(N-1)^2}\sum_{j\neq i}\frac1{r_j^2}\right].
\end{equation}
The mean relative phase is $\langle\phi_i\mid\{r_j\}\rangle\approx-\delta\bar\omega_i/(kN)$, whose squared contribution to the EPR is only $\mathcal{O}(1)$. The leading conditional contribution is therefore
\begin{equation}
    \frac{k^2r_i^2}{\Teff}\langle g_i^2\mid\{r_j\}\rangle
    \approx\frac{k}{N}\left[(N-1)^2+\sum_{j\neq i}\frac{r_i^2}{r_j^2}\right].
\end{equation}
In the inner region, the boundedness of $g_i$ and $P(r_j<\varepsilon_k)=\mathcal{O}(k^{-1})$ bound its contribution to the EPR by $\mathcal{O}(k)$. In the outer region, for $i\neq j$,
\begin{equation}
\begin{split}
    \left\langle\frac{r_i^2}{r_j^2}\right\rangle_{\mathrm{outer}}
    &=\langle r^2\rangle_0\int_{\varepsilon_k}^\infty
    \frac{p_0(r)}{r}\,\mathrm{d}r+\mathcal{O}(1)\\
    &=\frac{\langle r^2\rangle_0p_0(0)}{2}
    \ln\frac{kN}{\Teff}+\mathcal{O}(1).
\end{split}
\end{equation}
There are $N(N-1)$ ordered pairs. Since $N$ is fixed, the $\ln N$ contribution is absorbed into the linear term, and
\begin{equation}
    \sigma_N(k)=(N-1)\frac{\langle r^2\rangle_0p_0(0)}{2}
    k\ln\frac{k}{\Teff}+\mathcal{O}(k).
\end{equation}\mbox{}
}

{The same inner and outer estimates also identify which EPR contribution carries this growth. At fixed radii, Cauchy-Schwarz gives $|\langle g_i\mid\{r_j\}\rangle|\leq\langle g_i^2\mid\{r_j\}\rangle^{1/2}$. Using $g_i\approx-(N-1)\phi_i$ and the outer conditional variance above yields
\begin{equation}\label{app eq: phase N mean bound}
\begin{split}
|\langle r_i^2g_i\rangle_k|
&\lesssim (N-1)\sqrt{\frac{\Teff}{kN}}\\
&\quad\times\left[\langle r\rangle_0+\langle r^2\rangle_0
\langle r^{-1}\rangle_0\right]+\mathcal{O}(k^{-1})\\
&=\mathcal{O}(k^{-1/2}).
\end{split}
\end{equation}
We used radial independence and $\langle r^{-1}\rangle_0=\int_0^\infty p_0(r)\,\mathrm{d}r<\infty$. The $\mathcal{O}(k^{-1})$ term includes the mean angular displacement and the inner contribution, where $|g_i|\leq N-1$ and the probability that any radius is below $\varepsilon_k$ is $\mathcal{O}(k^{-1})$ at fixed $N$. Hence $\langle r^2\rangle_0\sum_i\Omega_i^2/\Teff=\mathcal{O}(k)$ within these asymptotic estimates. The network Fisher balance gives
\begin{equation}
    0\leq\Teff\sum_{i=1}^N\langle(\partial_{r_i}\ln q_k)^2\rangle_k
    \leq N(N-1)k.
\end{equation}
Together with the weighted decomposition, this yields $\sigma_N(k)=\sigma_{\mathrm{var},N}(k)+\mathcal{O}(k)$ at fixed $N$.
}

\subsubsection{Fisher Information}\label{app:FI}
Consider a probability distribution $p(x;\theta)$, where $\theta$ parametrizes $p$. The Fisher information is defined as
\begin{equation}\label{eq:FI}
    \mathcal{J}(\theta)= \int (\partial_\theta\ln p(x;\theta))^2 p(x;\theta)\mathrm{d}x,
\end{equation}
where the integral is performed over the $x$-domain. In Eq.~({\ref{eq: EPR mean-variance}}) we identify the EPR contribution $\sigma_{r_i\varphi}$ as a weighted radial average of a Fisher information term since
\begin{equation}
\begin{split}
    \mathcal{J}_{r_j}(r_i)
    &=\int_0^{2\pi}(\partial_{r_i}\ln p(\varphi|r_1,r_2))^2\\
    &\qquad\times p(\varphi|r_1,r_2)\,\mathrm{d}\varphi\\
    &=\langle(\partial_{r_i}\ln p_\varphi)^2\rangle_\varphi,
\end{split}
\end{equation}
where $r_j$ is kept fixed. At fixed radii, define $p_\varepsilon(\varphi{\mid r_1,r_2}):=p(\varphi-\varepsilon{\mid r_1,r_2})$, where $\varepsilon\in(-\pi,\pi]$. {Since $\partial_\varepsilon\ln p_\varepsilon=-\partial_\varphi\ln p_\varepsilon$, periodicity gives}
\begin{equation}\label{app eq: phase shift fisher}
\begin{split}
    \mathcal{J}(\varepsilon)
    &=\int_0^{2\pi}(\partial_\varepsilon\ln p_\varepsilon)^2
    p_\varepsilon\,\mathrm{d}\varphi\\
    &=\langle(\partial_\varphi\ln p_\varphi)^2\rangle_\varphi
    {=I_\varphi(r_1,r_2).}
\end{split}
\end{equation}
{Thus $I_\varphi$ is} the Fisher information of the phase shift $\varepsilon$ at fixed radii~\cite{makowski2021transactional}.

\subsection{Full Cartesian coupling}
\subsubsection{Expansion of EPR}\label{app:cartesian-expansion}
{Substituting Eq.~(\ref{eq:51}) into Eq.~(\ref{eq: dist gen}) and integrating over the phases gives the joint radial distribution
\begin{equation}\label{eq:54}
 p(r_1,r_2)=\frac{p_0^1p_0^2}{\mathcal{Z}}
 e^{-\frac{\beta k}{2}(r_1^2+r_2^2)}\mathrm{I}_0(\beta kr_1r_2).
\end{equation}}
To expand the normalization in uncoupled moments, we use 
\begin{equation}\label{eq:binomial-powers}
\begin{split}
     |z_1-z_2|^{2n} &= r_1^{2n}(1-\frac{r_2}{r_1}e^{-{\mathrm{i}}\varphi})^{{n}}(1-\frac{r_2}{r_1}e^{{\mathrm{i}}\varphi})^{{n}}\\
     &= r_1^{2n} \sum_{l=0}^n \sum_{m=0}^n \binom{n}{l}\binom{n}{m} \left(-\frac{r_2}{r_1}\right)^{l+m}e^{{\mathrm{i}}\varphi(m-l)}\\
     &= r_1^{2n} \sum_{s=-n}^n C_s\left(\frac{r_2}{r_1}\right)e^{{\mathrm{i}}s\varphi},
\end{split}
\end{equation}
where 
\begin{equation}\label{eq:binomial-bounds}
     C_s\left(\frac{r_2}{r_1}\right) = \sum_{m={\max(0,s)}}^{{\min(n,n+s)}} \binom{n}{m-s}\binom{n}{m} \left(-\frac{r_2}{r_1}\right)^{2m-s}.
\end{equation}
Since {$\int_{0}^{2\pi}\frac{\mathrm{d}\theta_1\mathrm{d}\theta_2}{(2\pi)^2}e^{\mathrm{i}s(\theta_1-\theta_2)}=\delta_{s,0}$}, it follows that
\begin{equation}
\begin{split}
    \mathcal{Z} &= \sum_{n=0}^\infty\frac{1}{n!}\left(-\frac{k}{2T_\mathrm{eff}}\right)^n\\
    &\quad\times\sum_{m=0}^n\binom{n}{m}^2
    \langle r^{2m}\rangle_0\langle r^{2(n-m)}\rangle_0.
\end{split}
\end{equation}
Moving onto the EPR, the currents in Cartesian coordinates are given by $J_{x_i}=\omega y_i p$ and  $J_{y_i}=-\omega x_i p$, so that
\begin{equation}\label{eq:epr expansion cart}
\begin{split}
    \sigma &= \frac{2\omega^2}{T_\mathrm{eff}}\frac{1}{\mathcal{Z}}
    \sum_{n=0}^\infty\frac{1}{n!}\left(-\frac{k}{2T_\mathrm{eff}}\right)^n\\
    &\quad\times\sum_{m=0}^n\binom{n}{m}^2
    \langle r^{2m+2}\rangle_0\langle r^{2(n-m)}\rangle_0.
\end{split}
\end{equation}

\subsubsection{EPR gradient}\label{app: EPR grad zero cart}
For $N$ oscillators coupled through the interaction $V(\{r_j\}{,\{\theta_j\}})=\frac{k}{2}\sum_{i<j}^Nc_{ij}[r_i^2+r_j^2 -2r_ir_j\cos(\theta_i-\theta_j)]$, the gradient of the second radial moment reads
\begin{equation}\label{eq: deriv cart 0 app}
    \begin{split}
        \frac{\mathrm{d}}{\mathrm{d}k}\langle r_i^2  \rangle_k &= -\frac{\beta}{2}\mathrm{Cov}_k(r_i^2 ,\sum_{l<j}^Nc_{lj}[r_l^2+r_j^2\\&\qquad-2r_lr_j\cos(\theta_l-\theta_j)]).
    \end{split}
\end{equation}
At $k=0$, the radii and phases are independent, {and the phases are uniform, so the cosine contribution vanishes. Only edges incident on oscillator $i$ contribute to the remaining covariance.} Therefore the above gradient at $k=0$ simplifies to
\begin{equation}
    \begin{split}
        \frac{\mathrm{d}}{\mathrm{d}k}\langle r_i^2  \rangle_k \bigg|_{k=0}
        &= -\frac{\beta}{2}\left(\sum_{j=1}^Nc_{ij}\right)\mathrm{Var}_0(r^2).
    \end{split}
\end{equation}
The gradient of the total EPR at $k=0$ therefore reads
\begin{equation}\label{eq: cart n derv zero}
    \begin{split}
       \frac{\mathrm{d}}{\mathrm{d}k}\sigma \bigg|_{k=0} &={\omega^2\beta}\sum_{i=1}^N \frac{\mathrm{d}}{\mathrm{d}k}\langle r_i^2  \rangle_k \bigg|_{k=0}\\
         &= -\omega^2\beta^2 {W} \mathrm{Var}_0(r^2).
    \end{split}
\end{equation}

\subsubsection{Strong-coupling limit}\label{app: limit cart}
The steady-state probability distribution of $N$ coupled oscillators with interaction $V=\frac{k}{2}\sum_{i<j} c_{ij}|z_i-z_j|^2$,
{where $c_{ij}=c_{ji}\geq0$ are fixed symmetric weights, $c_{ii}=0$, and $kc_{ij}$ is the link strength}, reads
\begin{equation}\label{eq:pn_def}
p(\{z_i\})=\frac{1}{{\widetilde{\mathcal Z}}}
e^{
\beta\sum\limits_{i=1}^N{\int_0^{|z_i|}F(s)\,\mathrm{d}s}
-\frac{\beta k}{2}\sum\limits_{i<j}c_{ij}|z_i-z_j|^2
}.
\end{equation}
{$\widetilde{\mathcal Z}=(2\pi\mathcal I_0)^N\mathcal Z$ is the normalization for the density written as an unweighted exponential.}
To find the strong-coupling limit, it is convenient to transform the system into the relative and center-of-mass coordinates which diagonalize the interaction term. Let $z=(z_1,\dots,z_N)^T\in\mathbb{C}^N$ and $\mathbf 1=(1,\dots,1)^T$. Using
\begin{equation}\label{eq:pairwise_identity_clean}
\sum_{i<j}c_{ij}|z_i-z_j|^2
= z^\dagger L z,
\end{equation}
we identify the graph Laplacian $L$, whose matrix elements are given by
\begin{equation}
L_{ij}=
\begin{cases}
{s_i}, & i=j,\\
-c_{ij}, & i\neq j,
\end{cases}
\end{equation}
{$s_i=\sum_jc_{ij}$ is the weighted degree of node $i$.} Since $L\mathbf 1=0$, the vector $\mathbf 1$ spans the zero mode of the interaction. If the coupling topology corresponds to a connected graph, then this zero mode is unique and all remaining eigenvalues are strictly positive.

The matrix $L$ can be diagonalized with a unitary matrix $U\in\mathbb{C}^{N\times N}$ such that
$
U^\dagger L U=\mathrm{diag}(0,\lambda_1,\dots,\lambda_{N-1}),
$
where $\lambda_\alpha>0$ for $\alpha\in\{1,\dots,N-1\}$. The first column is defined as
$
e_0=\frac{\mathbf 1}{\sqrt N},
$
and the remaining columns $\{e_\alpha\}_{\alpha=1}^{N-1}$ form an orthonormal basis of the subspace $\mathbf 1^T v=0$. The new coordinates $q := U^\dagger z$ read
\begin{equation}\label{eq:mode_transform}
q_0=e_0^\dagger z=\frac{1}{\sqrt N}\sum_{i=1}^N z_i=\sqrt N\,\bar z,\qquad
q_\alpha=e_\alpha^\dagger z,
\end{equation}
where $\alpha\in\{1,\dots,N-1\}$. In the coordinates $(q_0,\{q_\alpha\})$ the steady state \eqref{eq:pn_def} becomes
\begin{equation}\label{eq:pn_modes}
p(q_0,\{q_\alpha\})=\frac{1}{{\widetilde{\mathcal Z}}}
e^{
\beta\sum\limits_{i=1}^N{\int_0^{|(Uq)_i|}F(s)\,\mathrm{d}s}
-\frac{\beta k}{2}\sum\limits_{\alpha=1}^{N-1}\lambda_\alpha |q_\alpha|^2
}.
\end{equation}
Now
\begin{equation}\label{eq:relative_collapse}
\begin{split}
&\lim_{k\to\infty}\prod_{\alpha=1}^{N-1}
{\frac{\beta k\lambda_\alpha}{2\pi}}
\exp\!\left[-\frac{\beta k\lambda_\alpha}{2}|q_\alpha|^2\right]\\
&\qquad{=}\prod_{\alpha=1}^{N-1}\delta^{(2)}(q_\alpha),
\end{split}
\end{equation}
where $\delta^{(2)}$ denotes the 2d Dirac delta function. Since $\lambda_\alpha>0$ for all $\alpha\geq1$, an integration over the delta functions sets $q_\alpha\to0$. Then $(Uq)_i=(q_0e_0)_i=\bar z$.

The total EPR in complex coordinates can be written as
\begin{equation}\label{eq:sigma_general}
\sigma=\frac{\omega^2}{T_{\rm eff}}\left\langle \sum_{i=1}^N |z_i|^2\right\rangle_k.
\end{equation}
In the strong-coupling limit $z_i=\bar z$ for all $i$, so $\sum_{i=1}^N|z_i|^2=N|\bar z|^2$ and therefore
\begin{equation}\label{eq:sigma_strongN}
\sigma(k\to\infty)
=\frac{N\omega^2}{T_{\rm eff}}\left\langle |\bar z|^2\right\rangle_{p_\infty}.
\end{equation}
The integration over the delta functions in $p_\infty(\bar z)$ reduces the expectation to that of a free single oscillator with distribution $p_0$, but with inverse temperature $N\beta$. Thus
\begin{equation}\label{eq:sigma_strongN_M2}
\sigma(k\to\infty)= \frac{N\omega^2}{T_{\rm eff}}\left\langle r^2\right\rangle_{0,N\beta}
=  \frac{N\omega^2}{T_{\rm eff}}\frac{\mathcal{I}_2}{\mathcal{I}_{0}}(N\beta).
\end{equation}

\begingroup
We now obtain the leading EPR gradient as $k\to\infty$. Let $r=|\bar z|$ and define
\begin{equation}
	\begin{split}
		R&:=\sum_\alpha|q_\alpha|^2,\qquad
		Q:=\sum_\alpha\lambda_\alpha|q_\alpha|^2,\\
		S&:=\sum_i|z_i|^2=Nr^2+R,
	\end{split}
\end{equation}
where sums over $\alpha$ run over the $N-1$ relative modes.
Since $\sigma=\beta\omega^2\langle S\rangle_k$ and the coupling
enters the stationary density through $e^{-\beta kQ/2}$,
\begin{equation}\label{eq:cartesian-mode-gradient}
	\frac{\mathrm{d}\sigma}{\mathrm{d}k}
	=-\frac{\omega^2\beta^2}{2}\mathrm{Cov}_k(S,Q).
\end{equation}

Since $L$ is real and symmetric, $U$ can be taken real and orthogonal,
giving $z_i=\bar z+\sum_\alpha U_{i\alpha}q_\alpha$.
At fixed $\bar z$, $q_{\alpha,\parallel}$ and $q_{\alpha,\perp}$ denote
the components of $q_\alpha$ parallel and perpendicular to $\bar z$,
respectively.
Writing $\Phi(r)=\int_0^rF(s)\,\mathrm{d}s$, expansion to
quadratic order gives
\begin{equation}\label{eq:cartesian-relative-expansion}
	\begin{split}
		\sum_i\Phi(|z_i|)
		&\approx N\Phi(r)\\
		&\quad+\frac12\sum_\alpha
		\left[F'(r)q_{\alpha,\parallel}^2+\frac{F(r)}r q_{\alpha,\perp}^2\right].
	\end{split}
\end{equation}
The two coefficients are the curvatures of $\Phi$ along and
perpendicular to the collective radius.

Together with the coupling term, this gives independent,
zero-mean Gaussian components at quadratic order, with
\begin{equation}
	\begin{split}
		\langle q_{\alpha,\parallel}^2\mid\bar z\rangle_k
		&\approx\frac{1}{\beta[k\lambda_\alpha-F'(r)]},\\
		\langle q_{\alpha,\perp}^2\mid\bar z\rangle_k
		&\approx\frac{1}{\beta[k\lambda_\alpha-F(r)/r]},
	\end{split}
\end{equation}
where $\langle\cdot\mid\bar z\rangle_k$ denotes averaging over
the relative modes at fixed $\bar z$.
Set $h(r):=F'(r)+F(r)/r$. Expanding the variances in $1/k$ and
using $\mathrm{Var}(X^2)=2\langle X^2\rangle^2$ for a zero-mean
Gaussian gives
\begin{equation}
	\begin{split}
		\langle Q\mid\bar z\rangle_k
		&\approx\frac{2(N-1)}{\beta k}
		+\frac{h(r)}{\beta k^2}\sum_\alpha\lambda_\alpha^{-1},\\
		\langle R\mid\bar z\rangle_k
		&\approx\frac{2}{\beta k}\sum_\alpha\lambda_\alpha^{-1}
		+\frac{h(r)}{\beta k^2}\sum_\alpha\lambda_\alpha^{-2},\\
		\mathrm{Cov}_k(R,Q\mid\bar z)
		&\approx\frac{4}{\beta^2k^2}\sum_\alpha\lambda_\alpha^{-1}.
	\end{split}
\end{equation}

The law of total covariance, with conditional averages over the relative coordinates at fixed $\bar z$ and outer averages over the stationary center-of-mass distribution, gives, to leading order,
\begin{equation}
	\begin{split}
		\mathrm{Cov}_k(S,Q)
		&\approx\left\langle\mathrm{Cov}_k(R,Q\mid\bar z)\right\rangle_k\\
		&\quad+N\,\mathrm{Cov}_k\left(
		r^2,\langle Q\mid\bar z\rangle_k\right).
	\end{split}
\end{equation}
The omitted covariance between $\langle R\mid\bar z\rangle_k$
and $\langle Q\mid\bar z\rangle_k$ is $\mathcal{O}(k^{-4})$
in this expansion, since their radius dependence first appears
at order $k^{-2}$. The remaining terms are of order $k^{-2}$, and their coefficient can be evaluated using the limiting center-of-mass radial distribution, equivalent to the free-oscillator radial distribution at inverse temperature $N\beta$. Substitution
into Eq.~(\ref{eq:cartesian-mode-gradient}) therefore yields
\begin{equation}\label{eq:cartesian-graph-strong-gradient}
	\begin{split}
		\frac{\mathrm{d}\sigma}{\mathrm{d}k}
		&\approx-\frac{\omega^2}{k^2}
		\left(\sum_{\alpha=1}^{N-1}\lambda_\alpha^{-1}\right)\\
		&\quad\times\left[
		2+\frac{N\beta}{2}\mathrm{Cov}_{0,N\beta}
		\left(r^2,F'(r)+\frac{F(r)}r\right)\right].
	\end{split}
\end{equation}
The two contributions describe suppression of relative
fluctuations and redistribution of the collective radius.

This is Eq.~(\ref{eq:cartesian-N-strong-gradient}). For $F(r)=r(1-r)$,
$h(r)=2-3r$. Unnormalized all-to-all
coupling has $\lambda_\alpha=N$, with $N=2$ recovering
Eq.~(\ref{eq:cartesian-pair-strong-gradient}).
At fixed $N$, $\beta$, and radial drift, other connected
symmetric weights change the positive graph factor but not
the sign of the leading coefficient. 

For the general family, $h'(1)=-(m+2n)$. Expanding the covariance in Eq.~(\ref{eq:cartesian-graph-strong-gradient}) at low noise therefore gives
\begin{equation}\label{app eq: cartesian strong gradient low noise general}
\begin{split}
\frac{\mathrm d\sigma}{\mathrm dk}
&\approx\frac{\omega^2}{k^2}\left(\sum_{\alpha=1}^{N-1}\lambda_\alpha^{-1}\right)\\
&\quad\times\left[m+2n-2+\mathcal O((N\beta)^{-1})\right].
\end{split}
\end{equation}
Since $m+2n-2>0$, the strong-coupling gradient is positive at sufficiently low noise throughout this family. Together with the negative initial gradient, this implies a finite-coupling EPR minimum.
\endgroup

\subsubsection{Distribution under coupling}\label{app: dist cartes}
\begingroup
For two Cartesian-coupled oscillators, differentiating the normalized radial density in Eq.~(\ref{eq:54}) gives the initial changes in the radial mean and variance:
\begin{equation}\label{eq:  m v deriv 0}
 \begin{split}
 \left.\frac{\mathrm{d}}{\mathrm{d}k}\mathrm{Var}_k(r_i)\right|_{k=0}
 &=-\frac{\beta}{2}\Bigl[\mathrm{Var}_0((r_i-\langle r_i\rangle_0)^2)\\
 &\qquad\quad+2\langle r_i\rangle_0\mu_3\Bigr],\\
 \left.\frac{\mathrm{d}}{\mathrm{d}k}\langle r_i\rangle_k\right|_{k=0}
 &=-\frac{\beta}{2}\mathrm{Cov}_0(r_i,r_i^2),
 \end{split}
\end{equation}
where $\mu_3=\langle(r_i-\langle r_i\rangle_0)^3\rangle_0$ is the third central moment. Since $r_i$ and $r_i^2$ are increasing functions of $r_i$, their covariance is nonnegative and the mean radius initially decreases. For the free distribution in Eq.~(\ref{eq:general-radial-density}) with $m=n=1$, $\mu_3$ changes from positive to negative near $\beta=1.8$, and the initial variance gradient becomes slightly positive for $\beta\gtrsim8.6$.
\endgroup

\begin{figure}
    \centering
    \includegraphics[bb=0 0 362.00000000 287.00000000,width=1.\linewidth]{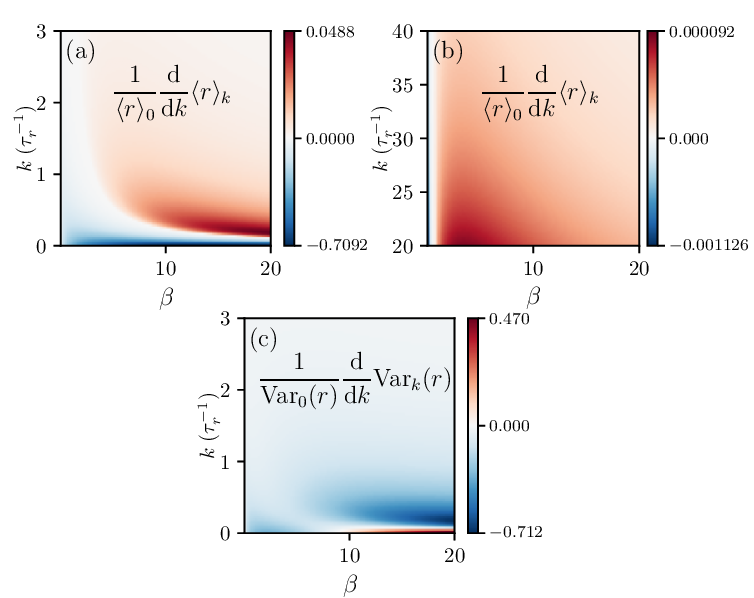}
    \caption{Radial distribution deformation under full Cartesian coupling. Heat maps versus coupling strength $k$ and inverse effective temperature $\beta$: (a, b) derivative of the average radius $\langle r\rangle_k$, normalized by $\langle r\rangle_0$, with respect to $k$ {for small and large $k$, respectively}; (c) derivative of the radial variance $\mathrm{Var}_k(r)$, normalized by $\mathrm{Var}_0(r)$, with respect to $k$.}
    \label{fig:app 2}
\end{figure}

Figure~\ref{fig:app 2}(a,b) shows the gradient of the average radius $\langle r\rangle_k$ at small and large coupling strengths, respectively\kern0pt. Over the plotted range, for $\beta\gtrsim0.5$, $\langle r\rangle_k$ initially decreases then increases with coupling strength. For $\beta\lesssim0.5$, $\langle r\rangle_k$ decreases with coupling strength. Figure~\ref{fig:app 2}(c) shows that the radial variance $\mathrm{Var}_k(r)$ decreases with increasing coupling strength for $\beta\lesssim{8.6}$. For $\beta\gtrsim{8.6}$, $\mathrm{Var}_k(r)$ initially increases (Eq.~(\ref{eq:  m v deriv 0})) and then decreases.

\begingroup
\subsection{Single-coordinate coupling: initial EPR gradient}
\label{app:single-coordinate-gradient}

To compare the initial EPR slopes for identical oscillators, assign independent
coupling strengths $k_x$ and $k_y$ to the two Cartesian coordinates. Exchanging $k_x$ and $k_y$ is equivalent to rotating the system
by $90^\circ$, which leaves the EPR unchanged. Consequently, the
first-order coefficients of $k_x$ and $k_y$ in the expansion of
$\sigma(k_x,k_y)$ about $(0,0)$ are equal. For full Cartesian coupling
($k_x=k_y=k$), both coefficients contribute to the initial slope. For
single-coordinate coupling ($k_x=k$, $k_y=0$), only one contributes,
giving half the full Cartesian slope. Using
Eq.~(\ref{eq: cart n derv zero}) gives
\begin{equation}
\left.\frac{\mathrm{d}\sigma}{\mathrm{d}k}\right|_{k=0}
=-\frac{\omega^2\beta^2W}{2}\mathrm{Var}_0(r^2),
\end{equation}
where $W=\sum_{i<j}c_{ij}$, establishing
Eq.~(\ref{eq:single-coordinate-weak}).
\par
\endgroup

\begingroup
\subsection{Single-coordinate coupling: oscillation death}\label{app: single-od}
We consider the deterministic limit of Eq.~(\ref{eq:64}) with
$F(r)=r(1-r)$ and $c_{12}=c_{21}=1$ for two identical oscillators,
with $k>0$ and $\omega>0$. In the invariant
antisymmetric subspace $(x_2,y_2)=-(x_1,y_1)$, write $(x,y)=(x_1,y_1)$.
Writing $x=r\cos\phi$ and $y=-r\sin\phi$, the deterministic anti-phase dynamics become
\begin{equation}\label{app eq: anti-phase dynamics}
\begin{split}
    \dot r&=r[1-r-2k\cos^2\phi],\\
    \dot\phi&=\omega+k\sin(2\phi).
\end{split}
\end{equation}
For $0<k<\omega$, one angular revolution takes time $2\pi/\sqrt{\omega^2-k^2}$.
The time average of $\cos(2\phi)$ over this period vanishes. For a nonzero
periodic radial solution, averaging $\dot r/r$ over one period therefore gives
\begin{equation}\label{app eq: anti-phase mean radius}
    \langle r\rangle_{\mathrm{period}}=1-k.
\end{equation}
Sustained anti-phase oscillation thus requires $k<1$. For $1\leq k<\omega$,
the radius instead decays to zero while the phase continues winding.

Defining $\chi:=1-r$, a nonzero stationary state satisfies
\begin{equation}\label{app eq: od balance}
    0=(\chi-2k)x+\omega y,\qquad 0=\chi y-\omega x,
\end{equation}
so that $\chi^2-2k\chi+\omega^2=0$. The branch relevant at strong coupling is
\begin{equation}\label{app eq: od states}
\begin{split}
    \chi&=k-\sqrt{k^2-\omega^2},\qquad r_*=1-\chi,\\
    (x_*,y_*)&=\pm\frac{1-\chi}{\sqrt{\chi^2+\omega^2}}(\chi,\omega),
\end{split}
\end{equation}
with the other oscillator at $(-x_*,-y_*)$. These states require
$k\geq\omega$ and $\chi<1$. At fixed $\omega$ and large $k$,
$\chi\approx\omega^2/(2k)$, so $r_*\to1$ and
$x_*/y_*\approx\omega/(2k)$, so the stationary states approach opposite sides
of the $y$-axis.

Stability must also hold for perturbations away from the antisymmetric
subspace. Let $\boldsymbol q_i=(x_i,y_i)^\top$ and
$\boldsymbol n=(\chi,\omega)^\top/\sqrt{\chi^2+\omega^2}$, where $\boldsymbol n$ is the radial unit vector at the fixed point. The single-oscillator
Jacobian at either fixed point is
\begin{equation}
    M=\begin{pmatrix}\chi&\omega\\-\omega&\chi\end{pmatrix}
    -(1-\chi)\boldsymbol n\boldsymbol n^\top.
\end{equation}
The symmetric and antisymmetric perturbations
$\delta\boldsymbol q_\pm=(\delta\boldsymbol q_1\pm\delta\boldsymbol q_2)/2$
therefore evolve with $M_+=M$ and
$M_-=M-2k\,\mathrm{diag}(1,0)$, respectively. Their eigenvalues are
\begin{equation}\label{app eq: od eigenvalues}
\begin{split}
    \lambda_{\mathrm{sym},\pm}
    &=\frac{3\chi-1\pm\sqrt{(1-\chi)^2-4\omega^2}}{2},\\
    \lambda_{\mathrm{asym},1}&=-(1-\chi),\qquad
    \lambda_{\mathrm{asym},2}=-2\sqrt{k^2-\omega^2}.
\end{split}
\end{equation}
For $k>\omega$ and $\chi<1$, the antisymmetric modes are stable. Since
$\mathrm{tr}\,M_+=3\chi-1$ and
$\det M_+=2\chi^2-\chi+\omega^2$, full linear stability additionally requires
\begin{equation}\label{app eq: od stability}
    \chi<\frac13,\qquad 2\chi^2-\chi+\omega^2>0.
\end{equation}
The other root $\chi=k+\sqrt{k^2-\omega^2}$, when it gives a nonzero state,
has a positive antisymmetric eigenvalue for $k>\omega$.
Existence and stability thus give different thresholds. For example,
at $\omega=1$ the nonzero branch in Eq.~(\ref{app eq: od states}) exists for
$k>1$, but is stable only for $k>5/3$.

For the parameters in Fig.~\ref{fig:7}(d), $k=50$ and $\omega=0.05$
in the dimensionless equations, the fixed points have
$(x_*,y_*)\approx\pm(0.000500,0.999975)$. These states satisfy
Eq.~(\ref{app eq: od stability}) and are therefore stable. 
\par
\endgroup

\begingroup
\subsection{Attractive and repulsive coupling}\label{app:repulsive-coupling}
Figure~\ref{fig:repulsive-coupling} compares stationary EPRs for two
oscillators with $m=n=1$ at finite positive and negative coupling. 

For radial and full Cartesian coupling, the stationary densities remain
normalizable at each finite negative $k$: the confining term proportional to
$r^{m+n+1}$ in Eq.~(\ref{eq:general-radial-density}) dominates the quadratic
interaction because $m+n+1>2$. For radial
coupling, $A:=\sigma'(0)$ is evaluated from Eq.~(\ref{eq:18}), giving
$\sigma(k)=2\sigma_0+Ak+\mathcal{O}(k^2)$.
Thus weak repulsion reverses a nonzero linear correction. 

For phase coupling of a pair,
$A_\varphi(\varphi+\pi;-k)=A_\varphi(\varphi;k)$.
Reversing $k$ therefore translates the stationary density and probability
currents by $\pi$ in relative phase while preserving the EPR integral,
giving $\sigma(-k)=\sigma(k)$.
The equality holds at finite coupling, including detuning and radial
fluctuations. The negative-$k$ branch in
Fig.~\ref{fig:repulsive-coupling}(b) is obtained from this exact pair symmetry.

For the full Cartesian pair with identical nonzero intrinsic frequency
$\omega$, set $\kappa=-k>0$. Integrating the stationary radial density,
Eq.~(\ref{eq:54}), over the other radius gives a marginal radial density
proportional to $rp_0(r)H_\kappa(r)$, where
\begin{equation}\label{app eq: repulsive cartesian factor}
    H_\kappa(r)=e^{\beta\kappa r^2/2}
    \int_0^\infty s p_0(s)e^{\beta\kappa s^2/2}
    \mathrm I_0(\beta\kappa rs)\,\mathrm ds.
\end{equation}
For $\kappa>0$, $H_\kappa(r)$ is strictly increasing for
$r>0$, so its covariance with $r^2$ is positive.
Since the EPR is $2\beta\omega^2\langle r^2\rangle_{-\kappa}$,
\begin{equation}\label{app eq: repulsive cartesian epr}
    \sigma(-\kappa)-2\sigma_0
    =\frac{2\beta\omega^2\operatorname{Cov}_0(r^2,H_\kappa)}
    {\langle H_\kappa\rangle_0}>0,
\end{equation}
where the average is with respect to $rp_0(r)$. Thus repulsion increases
the EPR above its uncoupled value for every finite negative $k$ in this
identical-frequency pair model (Fig.~\ref{fig:repulsive-coupling}(c)).

\begin{figure}[t]
    \centering
    \includegraphics[bb=0 0 245.00000000 353.00000000,width=\columnwidth]{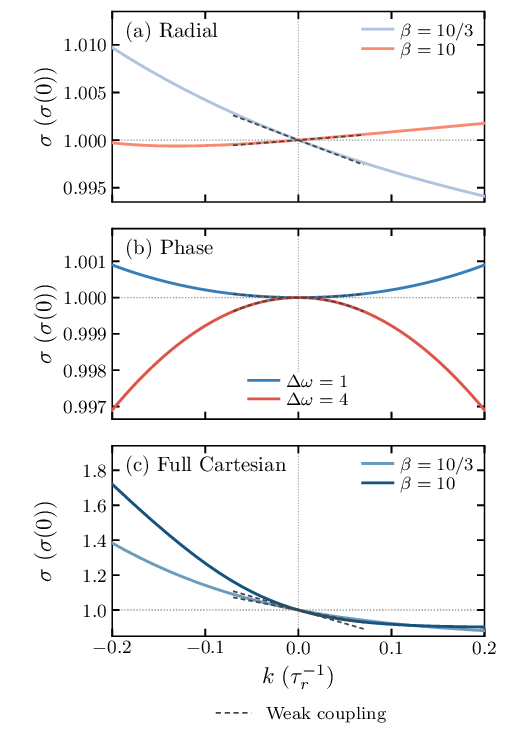}
    \caption{{Attractive and repulsive coupling of two oscillators.
    The EPR is expressed in units of its uncoupled value $\sigma(0)$
    at the same intrinsic frequencies and effective temperature.
    (a) Radial and (c) full Cartesian coupling at
    $\beta\in\{10/3,10\}$ with identical intrinsic frequencies.
    (b) Phase coupling at $\beta=1$,
    $\bar\omega=1.1\tau_r^{-1}$ and
    $\Delta\omega\in\{1,4\}\tau_r^{-1}$.
    Solid curves show EPR evaluated from the stationary densities in (a,c) and
    numerical solutions of the stationary PDE in (b); the negative-$k$ phase branch follows
    the exact pair symmetry.
    Dashed curves show the weak-coupling predictions of
    Eqs.~(\ref{eq:18}), (\ref{eq: C_2 pair}) and
    (\ref{eq:cartesian-pair-weak-gradient}) for (a), (b) and (c), respectively.}}
    \label{fig:repulsive-coupling}
\end{figure}
\par
\endgroup

\subsection{Stochastic Stuart-Landau oscillator}\label{app: SL}
The Stuart-Landau oscillator (SL) is the normal form of a supercritical Hopf bifurcation~\cite{kuramoto2003chemical}. In Cartesian coordinates, the corresponding Langevin equations read
\begin{equation}\label{eq:112}
    \begin{split}
        \dot{x} &= \lambda x - \alpha y - (x-\zeta y)(x^2+y^2) + \sqrt{2D}\xi_x\\
        \dot{y} &= \lambda y + \alpha x - (\zeta x +y)(x^2+y^2) + \sqrt{2D}\xi_y,\\
    \end{split}
\end{equation}
{A concrete biological correspondence is provided by the model of mechanically coupled hair bundles in Ref.~\cite{roongthumskul2021dynamics}. Writing $z_i=x_i+\mathrm{i}y_i$, $x_i$ represents bundle displacement and $y_i$ an unobserved effective coordinate. In the notation used here, its dynamics take the form}
\begingroup
\begin{equation}\label{app eq: hair bundle mapping}
\begin{split}
    \dot z_i &= (\lambda+\mathrm{i}\alpha_i)z_i
    -(1+\mathrm{i}\zeta)|z_i|^2z_i \\
    &\quad +k(x_j-x_i)
    +\sqrt{2D}\,(\xi_{x_i}+\mathrm{i}\xi_{y_i}),
    \qquad i\neq j.
\end{split}
\end{equation}
\endgroup
{The elastic interaction therefore couples a single Cartesian coordinate. The parameters $\mu$, $\omega_{0,i}$, $\beta$, and $K$ in Ref.~\cite{roongthumskul2021dynamics} correspond to $\lambda$, $\alpha_i$, $\zeta$, and $k$, respectively; its noise variance is $2D$.}

{For an uncoupled oscillator, the polar It\^o equations read}
\begin{equation}\label{app eq: sl polar ito}
    \begin{split}
        \dot{r} &= \lambda r - r^3 {+\frac{D}{r}} + \sqrt{2D}\xi_r\\
        \dot{\phi} &= \alpha - \zeta r^2+ \frac{\sqrt{2D}}{r}\xi_\phi.\\
    \end{split}
\end{equation}
In the deterministic limit, when $\lambda<0$, the intrinsic dynamics are given by a fixed point and the particle undergoes damped oscillations. The system undergoes a Hopf bifurcation at $\lambda=0$, whereby for $\lambda>0$, the intrinsic dynamics admit a stable limit cycle. Without noise, the underlying limit cycle has radius $r_c=\sqrt{\lambda}$ and intrinsic driving frequency $\omega_0 = \alpha - \zeta \lambda$. {For $\lambda>0$ and $\zeta=0$, scaling length by $r_c=\sqrt{\lambda}$ and time by $\tau_r=(2\lambda)^{-1}$ gives the dimensionless model with $m=2$, $n=1$, and $T_\mathrm{eff}=D/(2\lambda^2)=\tau_r/\tau_D$, where $\tau_D=\lambda/D$, up to the orientation of the angular coordinate. Nonzero $\zeta$ introduces amplitude-dependent angular frequency.} Rotational symmetry and vanishing radial current give $p(r,\phi)=\frac{1}{2\pi}{p_0(r)}$, with
\begin{equation}\label{eq:115}
    p_0(r) = \frac{1}{\mathcal{Z}}e^{\frac{\lambda r^2}{2D}- \frac{ r^4}{4D}},
\end{equation}
where 
\begin{equation}
    \mathcal{Z} = \frac{\sqrt{D\pi}}{2}e^{\lambda^2/4D}\left(1 + \mathrm{erf}\left[\frac{\lambda}{2\sqrt{D}}\right]\right).
\end{equation}
{As $\lambda$ becomes positive, the minimum of the radial potential $V(r)=-\lambda r^2/(2D)+r^4/(4D)$ moves from $r=0$ to $r=\sqrt{\lambda}$.}

\subsubsection{Entropy production rate}
{With $p(r,\phi)=p_0(r)/(2\pi)$, the nonzero angular current is $J_\phi=(\alpha-\zeta r^2)p$.} The EPR is then~\cite{cao2015free, lee2018thermodynamic}
\begin{equation}\label{eq:117}
\begin{split}
\sigma
&= {\int_0^{2\pi}\mathrm{d}\phi \int_0^\infty \frac{r^2J_\phi^2}{Dp}\,r\,\mathrm{d}r} \\
&= \int_0^\infty \frac{r^2(\alpha-\zeta r^2)^2}{D}p_0(r)\,rdr .
\end{split}
\end{equation}
Using the moments of $p_0$, this becomes
\begin{equation}
\sigma
= \frac{\alpha^2}{D}\langle r^2\rangle_0
+ \frac{\zeta^2}{D}\langle r^6\rangle_0
- \frac{2\alpha\zeta}{D}\langle r^4\rangle_0 .
\end{equation}
where the $m$th moment is given by
\begin{equation}
    \resizebox{\columnwidth}{!}{$
\begin{aligned}
   \langle r^m \rangle_0 =  \frac{2^{m/2} D^{\frac{m-2}{4}} \left(\lambda  \Gamma \left(\frac{m}{4}+1\right)
   \, _1F_1\left(\frac{2-m}{4};\frac{3}{2};-\frac{\lambda ^2}{4 D}\right)+\sqrt{D} \Gamma
   \left(\frac{m+2}{4}\right) \, _1F_1\left(-\frac{m}{4};\frac{1}{2};-\frac{\lambda ^2}{4
   D}\right)\right)}{\sqrt{\pi } \left(\text{erf}\left(\frac{\lambda }{2
   \sqrt{D}}\right)+1\right)}.
\end{aligned}
$}
\end{equation}

\subsubsection{Radial coupling}
{Radial coupling preserves the stationary reweighting construction of Eq.~(\ref{eq: dist gen}) for $\zeta\neq0$. For $N$ oscillators with common $\alpha$ and $\zeta$ and symmetric coupling weights $c_{ij}=c_{ji}$, the EPR is}
\begin{equation}\label{eq:119}
     \sigma=\frac{1}{D}\sum_{i=1}^{{N}}
       \langle r_i^2\tilde\omega^2(r_i)\rangle_k,
\end{equation}
{where $\tilde\omega(r)=\alpha-\zeta r^2$ is the amplitude-dependent angular frequency. The coupled expectation of a function $f$ is}
\begin{equation}
    \begin{split}
     \left\langle f \right \rangle_k &:= \frac{\left \langle f \, e^{-\frac{k}{2D}\sum_{i<j}{c_{ij}}(r_i-r_j)^2} \right \rangle_0}{\left \langle e^{-\frac{k}{2D}\sum_{i<j}{c_{ij}}(r_i-r_j)^2} \right \rangle_0},\\
    \end{split}
\end{equation}
where the expectation $\langle\ldots\rangle_0$ is with respect to {the product of the $N$ uncoupled radial distributions $r_i p_0(r_i)$, with} $p_0$ given in Eq.~(\ref{eq:115}).

{For $\zeta=0$, the moment expansions in Appendix~\ref{app:moments} apply with $m=2$, $n=1$. For a pair, the strong-coupling EPR approaches approximately $0.927$ of the uncoupled value in the high-noise limit.}
\subsubsection{Cartesian coupling}
{For full Cartesian coupling, amplitude-dependent angular frequencies generally invalidate the stationary-density construction used here. When $\zeta=0$ and the intrinsic frequencies are identical, Eq.~(\ref{eq:54}) applies to a pair with $\beta$ replaced by $1/D$ and the stationary radial density in Eq.~(\ref{eq:115}).}

\begingroup
Using the dimensionless variables introduced above for $\zeta=0$, $F(r)=r(1-r^2)/2$, so Eq.~(\ref{eq:cartesian-graph-strong-gradient}) becomes
\begin{equation}\label{app eq: sl cartesian strong gradient}
\begin{split}
\frac{\mathrm d\sigma}{\mathrm dk}
&\approx\frac{\omega^2}{k^2}
\left(\sum_{\alpha=1}^{N-1}\lambda_\alpha^{-1}\right)\\
&\quad\times\left[N\beta\,\mathrm{Var}_{0,N\beta}(r^2)-2\right].
\end{split}
\end{equation}
The bracket changes sign at $N\beta\approx1.213$. Above this threshold, the positive strong-coupling gradient implies a finite-coupling minimum. The $1/\beta$ correction to the limiting EPR ratio vanishes for this model. In the high-noise limit, the ratio approaches $N^{-1/2}$, approximately $0.707$ for a pair.
\endgroup
\subsubsection{Phase coupling}
{For $\zeta=0$, the analysis in Sec.~\ref{sec: 4} applies with the Stuart-Landau radial density and generator. Nonzero $\zeta$ introduces a radius-dependent frequency difference into the conditional phase equation.}

\let\LCorigbibhref\href
\let\LCorigbiburl\url
\renewcommand{\href}[2]{#2}
\let\url\nolinkurl
\bibliography{bibliography}
\let\href\LCorigbibhref
\let\url\LCorigbiburl

\end{document}